\documentclass[prb,twocolumn,showpacs,preprintnumbers,amsmath,amssymb]{revtex4-1}

\usepackage{graphicx,color}
\usepackage{bm}
\usepackage{soul}
\usepackage{dcolumn}
\usepackage{lipsum}
\usepackage{epstopdf}
\usepackage{ulem}

\newcommand{\hatm}{\mbox{$\hat{m}$}}
\newcommand{\HATn}{\hat{{m}}}

\begin{document}

\preprint{APS/123-QED}

\title{First-principles calculation of the parameters used by atomistic
  magnetic simulations}

\author{Sergiy Mankovsky}
\affiliation{%
Department of Chemistry/Phys.\ Chemistry, LMU Munich,
Butenandtstrasse 11, D-81377 Munich, Germany \\
}%
\author{Hubert Ebert}
\affiliation{%
Department of Chemistry/Phys.\ Chemistry, LMU Munich,
Butenandtstrasse 11, D-81377 Munich, Germany \\
}%
\date{\today}

\begin{abstract}
While the ground state of magnetic
materials is in general well described on the
basis of spin density functional theory (SDFT), the theoretical description of
finite-temperature and non-equilibrium properties require an extension
beyond the standard SDFT. Time-dependent SDFT (TD-SDFT), which
give for example access to dynamical properties are
computationally very demanding and can currently be hardly applied to
complex solids. 
Here we focus on the alternative approach based on the combination of a
parameterized phenomenological spin Hamiltonian and SDFT-based electronic
structure calculations, giving access to the dynamical and finite-temperature
properties for example 
via spin-dynamics simulations using the
Landau-Lifshitz-Gilbert (LLG) equation
or Monte Carlo simulations. 
We present an overview on the various methods
to calculate the parameters of the various phenomenological Hamiltonians
with an emphasis on the KKR Green function method as one of the 
most flexible band structure methods giving access to practically all
relevant parameters.
Concerning these, 
it is crucial to account for the spin-orbit coupling (SOC) 
by performing relativistic
SDFT-based calculations as it plays a key role for magnetic anisotropy
and chiral exchange interactions represented by the DMI parameters in
the spin Hamiltonian. 
This concerns also  the Gilbert damping parameters characterizing
magnetization dissipation in the LLG equation, chiral multispin
interaction parameters of the extended Heisenberg Hamiltonian,
as well as spin-lattice interaction parameters 
describing the interplay
of spin and lattice dynamics processes,
for which an efficient computational scheme
has been developed recently by the present authors.
\end{abstract}

\maketitle


\section{Introduction}

Density functional theory (DFT) is a 'formally exact approach to the
static electronic many-body problem' for the electron gas in the
equilibrium, which was adopted for a huge number of investigations
during the last decades to describe 
the ground state of solids, both magnetic and non-magnetic, as well
as various ground state properties \cite{ED11}. 

However, dealing with real systems, the
properties in an out-of-equilibrium situation are of great interest. 
An example for this is the presence of
external perturbation varying in time, which could be accounted for
by performing time-dependent 
first-principles electronic structure calculations. The
time-dependent extension of density functional theory (TD-DFT)
\cite{KDE+15a} is used successfully to study various dynamical
processes in atoms  and  molecules, in particular, giving access to the
time evolution of the electronic structure in a system affected by a
femtosecond laser pulse. However, TD-DFT can be hardly applied
to complex solids because of the lack of universal parameter-free
approximations for 
the exchange-correlation kernel.
Because of this, an approach
based on the combination of simulation methods for spin- and lattice
dynamics, using model spin and lattice Hamiltonians is more popular for
the moment.  
A great progress with this approach has been achieved during last decade 
due to the availability of parameters for the model Hamiltonians
calculated on a first principles level, that is a central issue of
the present contribution. 
  As it was pointed out in Ref.\ \onlinecite{ED11}, this approach has the
  advantage, that the spin-related many-body effects in this case are
  much simpler to be taken into account when compared to the ab-initio
  approach. Thus, the isotropic exchange coupling parameters $J_{ij}$ for
  the classical Heisenberg Hamiltonian worked out Liechtenstein et
al. \cite{LKG84,LKAG87} have been successfully used by many authors to
predict the ground state magnetic structure of material and to
investigate its finite-temperature properties. Depending on the materials,
the isotropic $J_{ij}$ can exhibit only spatial anisotropy.
Extension of the Heisenberg Hamiltonian accounting for anisotropy in spin
subspace is often done by adding the so-called Dzyaloshinskii-Moriya
interactions (DMI) and the magnetic anisotropy term,
\begin{eqnarray}
  H_{H,rel} & = & - \sum_{i,j} {J}_{ij} (\hat{e}_i \cdot \hat{e}_j) -
                   \sum_{i,j} \vec{D}_{ij} (\hat{e}_i \times \hat{e}_j)
                  + \sum_{i} \hat{e}_i \underline{K}_{ii} \hat{e}_i \,. \nonumber\\
\label{Eq_Heisenberg-rel1}
\end{eqnarray}
with $\hat{e}_{i(j)}$ the orientation of the spin magnetic moment at
site $i (j)$.
Alternatively, one may describe exchange interactions in the more
general tensorial form, $\underline{J}_{ij}$, leading to:
\begin{eqnarray}
H_{H,rel} & = & - \sum_{i,j} \hat{e}_i  \underline{J}_{ij}  \hat{e}_j +
                \sum_{i} \hat{e}_i  \underline{K}_{ii}  \hat{e}_i \,,
\label{Eq_Heisenberg-rel2}
\end{eqnarray}
In the second case the DMI is represented as the
antisymmetric part of the exchange tensor, i.e. $D^{\alpha}_{ij} =
\frac{1}{2}(J^{\beta\gamma}_{ij} -
J^{\gamma\beta}_{ij})\epsilon_{\alpha\beta\gamma}$.
It should be stressed, that calculations of the spin-anisotropic
exchange interaction parameters as well as of the
magnetic anisotropy parameters require a relativistic treatment of the
electronic 
structure in contrast to the case of the isotropic exchange parameters
which can be calculated on a non-relativistic level.
Various schemes to map the dependence of the electronic energy on the
magnetic configuration were suggested in the literature to calculate the
parameters of the spin Hamiltonians\cite{USPW03,EM09a,HBB08,HBB09},
depending of its form given in Eqs.\
(\ref{Eq_Heisenberg-rel1}) or (\ref{Eq_Heisenberg-rel2}).

Despite of its simplicity, the spin Hamiltonian gives access to a 
reasonable description of the temperature dependence of magnetic
properties of materials when combined with Monte Carlo (MC)
simulations \cite{RBKT06}, or non-equilibrium spin dynamics simulations
based on the phenomenological Landau-Lifshitz-Gilbert equations
\cite{AKH+96,EBBH22}   
\begin{eqnarray}
\frac{1}{\gamma}\frac{d\vec{M}}{d\tau}  &= & 
-\vec{M}\times\vec{H}_{\rm eff}
+ \vec{M} \times  \left[ \frac{\tilde{G}(\vec{M})}{\gamma^2 M_s^2}
\frac{d\vec{M}}{d\tau} \right] \;.
\label{eq:LLG_0}
\end{eqnarray}
Here $\vec{H}_{\rm eff}$ is the effective magnetic field defined as
$\vec{H}_{\rm eff} = -\frac{1}{M} \frac{\partial F}{\partial \hat{m}}$,
where $F$ is the free energy of the system and $\hat{m} =
\frac{\vec{M}}{\vec{M}_s}$ with $ M_s$ the saturation magnetization
treated at first-principles level, and $\gamma$ is the gyromagnetic 
ratio and $ \tilde G $ is the Gilbert damping parameter.
Alternatively, the effective magnetic field can be represented in terms
of the spin Hamiltonian in Eq.\ (\ref{Eq_Heisenberg-rel2}), i.e.
  $\vec{H}_{\rm eff} = -\frac{1}{M} \frac{\partial \langle
    {\cal H}_{H,rel} \rangle_T}{\partial \hat{m}}$,
with $\langle ... \rangle_T$ denoting the thermal average for the
extended Heisenberg Hamiltonian ${\cal H}_{H,rel}$.

The first-principles calculation of the parameters for the Heisenberg
Hamiltonian as well as for the LLG equation for spin dynamics 
have been reported in the literature by various groups who applied
different approaches based on ab-initio methods. 
Here we will focus on calculations based on the Green function
multiple-scattering formalism being a rather powerful tool to supply all
parameters for the extended Heisenberg Hamiltonian as well as for the LLG
equation.

\subsection{Magnetic anisotropy}

Let's first consider the magnetic anisotropy term in spin Hamiltonian,
characterized by parameters (written in tensorial form in Eqs.\
(\ref{Eq_Heisenberg-rel1}) and (\ref{Eq_Heisenberg-rel2})) deduced from
the total energy dependent on the orientation of the magnetization $\hat{{m}}$.
The latter is traditionally split into the magneto-crystalline anisotropy
(MCA) energy, 
$E_{\rm MCA}(\hat{{m}})$, induced by spin-orbit coupling (SOC) 
and the shape anisotropy energy, $E_{\rm shape}(\hat{{m}})$,
caused by magnetic dipole interactions, 
%
\begin{eqnarray}
  E_{\rm A}(\hat{{m}})  &=&  E_{\rm MCA}(\hat{{m}}) + E_{\rm shape}(\hat{{m}}) \; .
\label{eq:Anisotropy}
\end{eqnarray}
%
%
Although a quantum-mechanical description of the magnetic shape anisotropy
deserves separate discussion \cite{BMB+12} this contribution can be
reasonably well estimated based on classical magnetic dipole-dipole
interactions.
Therefore, we will focus on the MCA contribution which is fully
determined by the electronic structure of the considered system. In the
literature the focus is in general on the MCA energy of the ground
state, which can 
be estimated straightforwardly from the total energy calculated for
different orientations of the magnetization followed by a mapping onto a
model spin 
Hamiltonian, given e.g. by an expansion in terms of  spherical harmonics
$Y_{lm}(\hatm)$ \cite{Blu99}
%
%
\begin{eqnarray}
   E_{\rm MCA}(\hat{{m}})  &=& \sum_{ l \, {\rm even}}
 \sum_{m = -l}^{m = l} 
 \kappa_l^m \,   Y_{lm}(\hat{{m}})  \,.
\label{eq:MCA-YLM}
\end{eqnarray}
%
Alternative approach to calculate the MCA parameters is based on magnetic torque calculations, using the
definition 
\begin{eqnarray}
 T^{\hat{m}}(\theta^{\hat{u}}) &=& - \frac{\partial E(\hat{m})}{\partial \theta^{\hat{u}}}
\; ,
\label{eq:Torque_3}
\end{eqnarray}
avoiding the time-consuming total energy calculations.
This scheme is based on the  
so-called magnetic force theorem that allows to represent the MCA energy
in terms of a corresponding electronic single-particle energies change
under rotation of magnetization, as follows \cite{RSP97}:
%
%
\begin{eqnarray}
\label{eq:FORCE-THEOREM}
\Delta {\cal E}_{\rm SOC}(\HATn, \HATn') &=&
- \int^{ E_{\rm F}^{\HATn}  } dE \, 
\left[ N^{\HATn}(E) - N^{\HATn'}(E) \right]
\nonumber \\
&&
- \frac{1}{2} n^{\HATn'}(E_{\rm F}^{\HATn'}) \, 
 (E_{\rm F}^{\HATn} - E_{\rm F}^{\HATn'})^2 \nonumber \\
&& + {\cal O}(E_{\rm F}^{\HATn} - E_{\rm F}^{\HATn'})^3 \; 
\end{eqnarray}
%
%
with $ N^{\HATn}(E) = \int^{E} dE' \, n^{\HATn}(E')$ the integrated DOS for the 
magnetization along the direction $\HATn$, and $n^{\HATn}(E)$  the density
of states (DOS) represented in terms of the Green function as follows
%
\begin{eqnarray}
\label{eq:DOS-vs-GF}
n^{\HATn}(E) &=& -\frac{1}{\pi} \mbox{Im Tr} \, G^{\HATn}(E) \;. 
\end{eqnarray}
%
%
This expression can be used in a very efficient way within the framework
of the multiple-scattering formalism. In this case the Green function is
given in terms of the 
scattering path operator $\underline{\underline{\tau}}(E)^{nn'}$
connecting the sites $n$ and $n'$ as follows
\begin{eqnarray}
G_0(\vec{r},\vec{r}\,',E) & = &
\sum_{\Lambda \Lambda'} 
Z^{n}_{\Lambda}(\vec{r},E)\,
                              {\tau}^{n n'}_{\Lambda\Lambda'}(E)\,
Z^{n' \times}_{\Lambda'}(\vec{r}\,',E)
 \nonumber \\
 & & 
-  \sum_{\Lambda} \Big[ 
Z^{n}_{\Lambda}(\vec{r},E)\, J^{n \times}_{\Lambda}(\vec{r}\,',E)\,
\Theta(r'-r)
 \nonumber \\
&& + J^{n}_{\Lambda}(\vec{r},E) \, Z^{n \times}_{\Lambda}(\vec{r}\,',E)\, \Theta(r-r')
\Big] \delta_{nn'} \; ,
\label{Eq:GF_KKR}
\end{eqnarray}
%
where the combined index $\Lambda =(\kappa,\mu)$ represents
the relativistic spin-orbit and magnetic quantum numbers
 $\kappa$ and $\mu$, respectively  \cite{Ros61};
$Z^{n}_{\Lambda}(\vec{r},E)$ and $J^{n}_{\Lambda}(\vec{r},E)$ are the
regular and irregular solutions of the single-site Dirac equation
(\ref{Eq:Dirac}) \cite{SPR-KKR8.5,EKM11,EBKM16}.
The scattering path operator is given by the expression
\begin{eqnarray}
  \underline{\underline{\tau}}(E) &=& [\underline{\underline{m}}(E) - \underline{\underline{G}}_0(E)]^{-1}
\label{eq:tau}
\end{eqnarray}
with  $\underline{\underline{m}}(E) = \underline{\underline{t}}^{-1}(E)$
and  $\underline{\underline{G}}_0(E)$ the inverse single-site
scattering and structure constant matrices, respectively.
The double underline used here indicates matrices with respect to site
and angular momentum indices \cite{EKM11}. 

Using the Lloyd's formula that gives the integrated DOS 
in terms of the scattering path operator, Eq.\ (\ref{eq:FORCE-THEOREM}) can
be transformed to the form
\begin{eqnarray}
 \Delta {\cal E}_{\rm SOC}(\HATn, \HATn') &=&  -\frac{1}{\pi} \mbox{Im\, Tr\,}
                                       \int^{E_F}dE\,\nonumber \\
                   &&\times   \left(\mbox{ln}\,
                                       \underline{\underline{\tau}}(\HATn,E)
                                       - \mbox{ln}\,
                                       \underline{\underline{\tau}}(\HATn',E)\right)  
\label{eq:Free_energy-2}
\end{eqnarray}
with the scattering path operator evaluated for the magnetization along
$\HATn$ and $\HATn'$, respectively.

With this, the magnetic torque $T(\theta)$
can be expressed by means of multiple scattering theory leading for the
torque component with respect to a rotation of the magnetization around
an axis $\hat{{u}}$, to the expression \cite{SSB+06}
%
\begin{equation}
T^{\hat{m}}(\theta^{\hat{u}}) = - \frac{1}{\pi} \, {\Im}
                \int^{E_{\rm F}} dE \; \frac{\partial}{\partial \theta^{\hat{{u}}}}
                \left[\ln \det \left(\underline{\underline{t}}(\hatm)^{-1}  
                -\underline{\underline{G}}^0    \right)   \right]
\; .
\label{eq:TORQUE-MST}
\end{equation}
%
Mapping the resulting torque onto a corresponding parameterized
expression as for example 
Eq.\ (\ref{eq:MCA-YLM}), one obtains the corresponding parameters of the
spin Hamiltonian.

However, one should note that the magnetic anisotropy of materials
changes when the temperature 
increases. This occurs first of all due to the increasing amplitude of
thermally induced spin fluctuations responsible for a modification
of the electronic structure. A corresponding expression for
magnetic torque st finite temperature was worked out by Staunton et al.\
\cite{SSB+06}, on the basis of the 
relativistic generalization of the disordered local moment (RDLM) theory
\cite{GPS+85}. 
To perform the necessary thermal averaging over different orientational
configurations 
of the local magnetic moments it uses a technique similar to the one used
to calculate the configurational average in the case of random 
metallic alloys, so-called  Coherent Potential Approximation (CPA) alloy
theory \cite{Sov67,SPG+00}. Accordingly, 
the free energy difference for two different orientations of the
magnetization is given by 
\begin{eqnarray}
 \Delta {\cal F}(\HATn,\HATn')  &=&  - \int dE\,
                                    f_{FD}(E,\HATn) \\
  && \bigg[
                                    \langle N^{\HATn} \rangle(E) -
                                    \langle N^{\HATn'} \rangle(E)
                                    \bigg]. 
\label{eq:Free_energy-T}
\end{eqnarray}

By using in this expression the configurational averaged integrated density
of states \cite{FS80,GPS+85} given by Lloyd's formula, the corresponding
expression for the magnetic torque at temperature $T$
\begin{eqnarray}
 T^{\hat{m},T}(\theta^{\hat{u}}) &=& - \frac{\partial
                                     }{\partial
                                     \theta^{\hat{u}}} \bigg(\sum_i\int
                                     P^{\HATn}_i(\hat{e}_i)  \langle
                                     \Omega^{\HATn} \rangle_{\hat{e}_i}
                                     d \hat{e}_i \bigg)\,.
\label{eq:Torque_T1}
\end{eqnarray}
can be written explicitly as:
\begin{eqnarray}
 T^{\hat{m},T}(\theta^{\hat{u}}) &=& - \frac{1}{\pi} \mbox{Im} \int^{E_F}dE\,
                                     f_{FD}(E,\HATn) \nonumber \\
  &&  \bigg(\sum_i\int
                                     \frac{\partial P^{\HATn}_i(\hat{e}_i)
                                     }{\partial
                                     \theta^{\hat{u}}} \mbox{ln det} M^{\HATn}_i(\hat{e}_i,E)
                                     d \hat{e}_i \bigg)\,.
\label{eq:Torque_T2}
\end{eqnarray}
where 
\begin{eqnarray}
 \underline{M}^{\HATn}_i(\hat{e}_i,E) &=&  1 +
                                                          ([\underline{t}_i(\hat{e}_i)]^{-1}
                                                           -
                                                           \underline{t}^{\HATn}_{i,c}(\hat{e}_i)]^{-1})
                                                           \underline{\tau}^{\HATn}_{ii,c}\,,
\label{eq:Torque_T3}
\end{eqnarray}
and
\begin{eqnarray}
\underline{\underline{\tau}}^{\HATn}_{ii,c} &=&([\underline{\underline{t}}^{\HATn}_{i,c}(\hat{e}_i)]^{-1}
                                               - \underline{\underline{G}}_0)^{-1}\,.
\label{eq:Torque_T4}
\end{eqnarray}
where the index $c$ indicates quantities related to the CPA medium.

\begin{figure}[hbt]
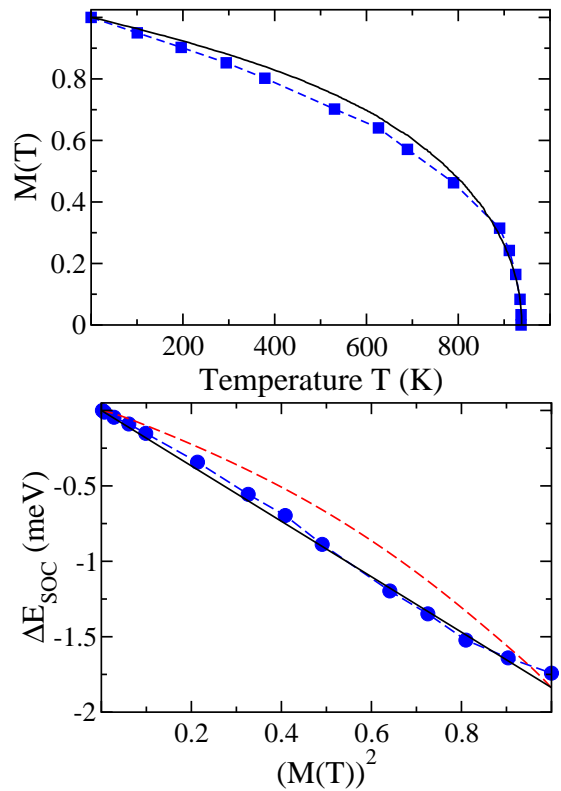

 \begin{center}
\includegraphics[width=0.4\textwidth,angle=0,clip]{RDLM_Staunt_1_mod.eps}\,\,\,\,\,
\includegraphics[width=0.4\textwidth,angle=0,clip]{RDLM_Staunt_2_mod.eps}
 \end{center}
\caption{\label{fig:RDLM-FePt} 
RDLM calculations on FePt.
Top: the magnetization $M(T)$  versus $T$ for the 
  magnetization along the easy  [001] axis (filled squares). 
The full line shows the mean field approximation to a classical
Heisenberg model for comparison.
Bottom: the magnetic anisotropy  energy $\Delta E_{\rm SOC}$  
as a function of the square of the magnetization $M(T)$. 
The filled circles show the RDLM-based results, 
the full line give $K(T)\sim[M(T)/M(0)]^2$, and the dashed line is based on the single-ion model function. 
All data taken from \cite{SOR+04}.
 } 
\end{figure}

Fig.~\ref{fig:RDLM-FePt} (top) shows as an example the results for the
temperature-dependent 
magnetization ($M(T)$) calculated within the RDLM calculations for
$L1_0$-ordered FePt \cite{SOR+04}.
Fig.~\ref{fig:RDLM-FePt} (bottom) gives the corresponding parameter $K(T)$
for a uni-axial magneto-crystalline anisotropy, which is obviously in
good agreement with  experiment. 

\subsection{Inter-atomic bilinear exchange interaction parameters}

\begin{figure}
\includegraphics[width=0.35\textwidth,angle=0,trim= 0 0.0cm 0 0.0cm,clip]{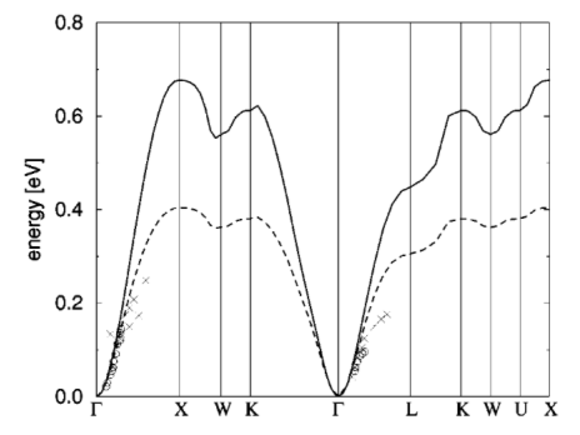}\;
\caption{\label{fig:Ni_magnon} Adiabatic  spin-wave  dispersion
  relations  along  high-symmetry  lines  of  the  Brillouin  zone  for
  Ni.  Broken  line:  frozen-magnon-torque method, full line: transverse
  susceptibility method \cite{GEF01}. All data are taken from Ref.\
  \onlinecite{GEF01}.  }    
\end{figure}

Most first-principles calculations on the bilinear exchange coupling
parameters reported in the literature, 
are based on the magnetic force theorem (MFT) by evaluating the energy change
due to a perturbation on the spin subsystem with respect to a suitable
reference configuration \cite{SKS+22}. Many results are based on calculations of the
spin-spiral energy 
$\epsilon(\vec{q})$, giving access to the exchange
parameters in the momentum space, $J_{\vec{q}}$
\cite{USK94,HEPO98,SB02,HBB08},  
followed by a Fourier transformation to the real space representation
$J_{ij}$. Alternatively, the real space exchange parameters 
are calculated directly by evaluating the energy change due to the
tilting of spin moments of interacting atoms. The corresponding
non-relativistic 
expression (so-called Liechtenstein or LKAG formula) has been
implemented based on the KKR as well as LMTO Green function (GF)
\cite{LKG84,LKAG87,PKT+00,SKS+22} 
band structure methods.
 It should be noted that the magnetic force theorem provides a
 reasonable accuracy for the
exchange coupling parameters in the case of infinitesimal rotations of the
spins close to some equilibrium state, that can be justified only in the long
wavelength and strong-coupling limits \cite{Sol21}. Accordingly, 
calculations of the exchange coupling parameters beyond the magnetic
force theorem, represented in terms of the inverse transverse susceptibility,
were discussed in the literature by various 
authors \cite{GEF01,Ant03,Bru03,Sol21,SKS+22}. Grotheer et
al., for example, have demonstrated\cite{GEF01} a deviation of the
spin-wave dispersion 
curves away from $\Gamma$ point in the BZ, calculated for fcc Ni using 
the exchange parameters $\underline{J}_{\vec{q}} \sim  
\underline{\chi}^{-1}_{\vec{q}}$,  from the 
MFT-based results for $\underline{J}_{\vec{q}}$. On the other hand, the
results are close to each other in the long-wavelength limit (see Fig. \ref{fig:Ni_magnon}). 
The calculations beyond the standard DFT are done by making use of the
so-called constrained-field DFT. The latter theory was also used by Bruno
\cite{Bru03} who suggested the 'renormalization' of the exchange
coupling parameters  
expressed in terms of non-relativistic transverse magnetic
susceptibility, according to $\underline{J} = \frac{1}{2} \underline{M}
\underline{\chi}^{-1} 
\underline{M} = \frac{1}{2} \underline{M} (\underline{\tilde{\chi}}^{-1}
- \underline{I}^{xc})
\underline{M}$, with the various quantities defined as follows
\begin{eqnarray}
  \tilde{\chi}^{-1}_{ij}  &=& \frac{2}{\pi} \int^{E_F}dE\,
                               \int_{\Omega_i} d^3r  \int_{\Omega_j}
                               d^3r'\\
  &&\times
                               \mbox{Im\,}[G^\uparrow(\vec{r},\vec{r}',E)G^\downarrow(\vec{r}',\vec{r},E)]\,,
\label{eq:Bru1}
\end{eqnarray}
\begin{eqnarray}
  M_{i}  &=&   \int_{\Omega_i} d^3r \, m(\vec{r})\,,
\label{eq:Bru2}
\end{eqnarray}
and
\begin{eqnarray}
 \tilde{I}^{xc}_{ij}  &=&  \delta_{ij} \frac{\Delta_i}{2M_i}\,,
\label{eq:Bru3}
\end{eqnarray}
with $\Delta_i = \frac{4}{M_i}\sum_j \tilde{J}_{ij}$, where
\begin{eqnarray}
 \tilde{J}_{ij}  &=&  \frac{1}{\pi}  \mbox{Im\,} \int^{E_F}dE\,
                               \int_{\Omega_i} d^3r  \int_{\Omega_j}
                      d^3r' \\
  &&\times  [B_{xc}(\vec{r}) G^\uparrow(\vec{r},\vec{r}',E)B_{xc}(\vec{r}')G^\downarrow(\vec{r}',\vec{r},E)]\,.
\label{eq:Bru4}
\end{eqnarray}
This approach results in a Curie temperature of 634 K for fcc Ni 
(vs. 350 K based on the MFT) which is in good agreement with the
experimental value of ($621-631$ K). As was pointed out by Solovyev
\cite{Sol21}, such a corrections can be significant only for a certain
class of materials, while, for instance, the calculations of spin-wave
energies \cite{GEF01} and $T_C$ \cite{Bru03} for bcc Fe demonstrate that 
these corrections are quite small.
As most results in the literature were
obtained using the exchange parameters based on the magnetic force
theorem,  we restrict below to this approximation.

Similar to the case of the MCA discussed above, application of the magnetic
force theorem gives the energy change due to tilting of two spin moments
represented in terms of the integrated DOS \cite{LKAG87}. Within the 
multiple scattering formalism, this energy can be transformed using the
Lloyd's formula leading to the expression
\begin{eqnarray}
 \Delta {\cal E} &=&  -\frac{1}{\pi} \mbox{Im\, Tr\,} \int^{E_F}dE\,
                      \left(\mbox{ln}\, \underline{\underline{\tau}}(E) - \mbox{ln}\, \underline{\underline{\tau}}^{0}(E)\right) 
\label{eq:Free_energy-22}
\end{eqnarray}
with 
$\underline{\underline{\tau}}^{(0)}(E)$ and
$\underline{\underline{\tau}}(E)$ the scattering path
operators for non-distorted and distorted systems, respectively.

As reported in Ref.\ \onlinecite{LKAG87}, the expression for $J_{ij}$
representing the exchange interaction between the spin moments on
sites $i$ and $j$,  is given by the expression
\begin{eqnarray}
 J_{ij}  &=&  -\frac{1}{4\pi}  \mathrm{Im Tr_L} \int^{E_F} dE 
\underline{\Delta}_i \underline{\tau}^\uparrow_{ij}
\underline{\Delta}_j \underline{\tau}^\downarrow_{ji}   \; ,
\label{Eq:Jij_parameter_LKAG}
\end{eqnarray}
with $\underline{\Delta}_{i(j)} = ([\underline{t}^\uparrow]^{-1}_{i(j)} -
[\underline{t}^\downarrow]^{-1}_{i(j)})$, where $t^\uparrow_{i(j)}$ and
$\underline{t}^\downarrow_{i(j)}$ are the spin-up and spin-down single-site
scattering matrices, respectively, while
$\underline{\tau}^\uparrow_{ij}$ and $\underline{\tau}^\downarrow_{ji}$
are the spin-up and spin-down, respectively, scattering path operators.
As relativistic effects are not taken into account, the 
exchange interactions are isotropic with respect to the orientation of
the magnetization as well as with respect to the direction of the spin
tilting. On the other hand, spin-orbit coupling gives rise to an  
anisotropy for exchange interactions requiring a
representation in the form of the exchange tensor
$\underline{J}_{ij}$ with its antisymmetric part
giving access to the Dzyaloshinskii-Moriya (DM) interaction  
$\vec{D}_{ij}$.


 Udvardi et al. \cite{USPW03} and later Ebert and Mankovsky \cite{EM09a}
 suggested an extension of the classical 
 Heisenberg Hamiltonian by accounting for relativistic effects for the
 exchange coupling (see also Ref.\ \onlinecite{SKS+22}). These calculations are based on a fully relativistic
 treatment of the electronic structure obtained by use of of the Dirac
 Hamiltonian  
 \begin{eqnarray}
 {\cal H}_{\rm D}   & =&
 - i c \vec{\alpha} \cdot \vec{\nabla}  
+ \frac{1}{2} \, c^{2} (\beta - 1) 
\nonumber \\ 
&&\qquad 
+ \bar V(\vec {r}) + \beta \, \vec{\sigma} \cdot  \vec { B}(\vec {r})  
 + e \vec{\alpha} \cdot \vec {A}(\vec {r})
\;.
\label{Eq:Dirac}
\end{eqnarray}
Here, $ {\alpha}_i $ and $ \beta $ are the standard
 Dirac matrices \cite{Ros61} while $\bar V(\vec {r}) $ and 
$   \vec { B}(\vec {r}) $ are the spin independent and spin dependent
parts of the electronic potential. 

Considering a ferromagnetic (FM) state as a reference state with the
magnetization along the $z$ direction, a tilting of the magnetic moments on
sites $i$ and $j$ leads to a modification of the scattering path
operator implying the relation
\begin{eqnarray}
  \mbox{ln} \,\underline{\underline{\tau}}
- \mbox{ln} \,\underline{\underline{\tau}}^0
  &=& - \mbox{ln}\left(1 +
       \underline{\underline{\tau}}\,[\Delta
      {\underline{m}}_i + \Delta
      {\underline{m}}_j       
      +  ... ] \right) \; ,
\end{eqnarray}
with ${\underline{m}}_i =  {\underline{t}}^{-1}_i$.
This allows to write down the expression for the energy change due to a
spin tilting on sites $i$ and $j$ as follows
\begin{eqnarray}
{\cal E}_{ij} &=& -\frac{1}{\pi} \mbox{Im\,Tr} \int^{E_F} dE\,  \Delta m_i\, 
                               \tau_{ij}\,  \Delta m_j \,  \tau_{ji}
\label{eq:Delta_energy_ij}
\end{eqnarray}

Within the approach of Udvardi et al. \cite{USPW03}, the
dependence of the single-site inverse scattering matrix
${\underline{m}}_i$ on the orientation of magnetic moment $\hat{e}_i$
is accounted for by performing a corresponding rotation operation using
the rotation matrix $\underline{R}(\theta,\phi)$, i.e., one has
${\underline{m}}_i(\theta,\phi) = \underline{R}(\theta,\phi)
{\underline{m}}^0_i \underline{R}^+(\theta,\phi)$.
The change of the scattering matrix ${\underline{m}}_i$ under spin rotation,
$\Delta{\underline{m}}_i$, linearized with respect to the rotation
angles, is given by the expression
\begin{eqnarray}
\Delta {\underline{m}}_i&=& \underline{R}(\theta_i,\phi_i)  \underline{m}^0_i
\underline{R}^+(\theta_i,\phi_i) - \underline{m}^0_i  \nonumber \\
                &=&  \underline{m}^\theta_i \delta \theta_i + \underline{m}^\phi_i \delta \phi_i
\label{eq:Delta_tau-1}
\end{eqnarray}
with 
\begin{eqnarray}  
\underline{m}^\theta_i  &=& \frac{\partial}{\partial \theta}
                            \underline{m}_i =  \frac{\partial R}{\partial \theta}
                            \underline{m}_i R^+ + R  \underline{m}_i
                            \frac{\partial R^+}{\partial \theta}  \,,
                             \nonumber \\
\underline{m}^\phi_i&=& \frac{\partial}{\partial \phi}
                            \underline{m}_i =  \frac{\partial R}{\partial \phi}
                            \underline{m}_i R^+ + R  \underline{m}_i
                            \frac{\partial R^+}{\partial \phi} \,.
\label{eq:Delta_tau-2}
\end{eqnarray}
 To calculate the derivatives of the rotation matrix, the definition
\begin{eqnarray}
\hat{R}(\alpha_{\hat{n}},\hat{n}) &=&e^{i\alpha_{\hat{n}}(\hat{n} \cdot \hat{\vec{{J}}} )}
\label{eq:Rotation_matrix}
\end{eqnarray}
for the corresponding operator is used,
with $\hat{\vec{{J}}}$ the total angular momentum operator.
$\hat{R}(\alpha_{\hat{n}},\hat{n})$ describes a rotation of the magnetic
moment $\hat{m}$  by the 
angle $\alpha_{\hat{n}}$ about the direction $\hat{n} \perp \hat{m}$,
that gives in particular $\underline{R}(\theta,\hat{n})$
for $\hat{n} = \hat{y}$ and $\underline{R}(\phi,\hat{n})$ for $\hat{n} =
\hat{z}$.

This leads to the second derivatives of the total energy with respect to the
titling angles $\alpha_i = \{\theta_i, \phi_i\}$ and  $\beta_j = \{\theta_j, \phi_j\}$
\begin{eqnarray}
\frac{\partial^2 {\cal E}}{\partial \alpha_i \partial \beta_j} &=&
                                                          -\frac{1}{\pi}
                                                          \mbox{Im\,Tr}
                                                          \int^{E_F} dE
                                                          \underline{m}^\alpha_i
                                                           \underline{\tau}_{ij}
                                                          \underline{m}^\beta_j
                                                          \underline{\tau}_{ji}
\end{eqnarray}
As is discussed by Udvardi et al. \cite{USPW03}, these derivatives
give access to all elements $J^{\mu \nu}_{ij}$ of the exchange tensor,
where $\mu(\nu)= \{x,y,z\}$. Note, however, that only the tensor elements
with $\mu(\nu)= \{x,y\}$ can be calculated using the magnetization
direction along the $\hat{z}$ axis, giving access to the $z$ component
$D^z_{ij}$ of the DMI. In order to obtain all other tensor elements, 
an auxiliary rotation of the magnetization towards the $\hat{x}$ and
$\hat{y}$ directions 
of the global frame of reference is required. For example, the
component $D^x_{ij}$ if the DMI vector can be evaluated via the tensor
elements 
\begin{eqnarray}
J^{zy}_{ij} &=& \frac{\partial^2 E}{\partial \theta_i \partial \phi_j}
\,\,\, \mbox{and} \,\,\,
  J^{yz}_{ij} = \frac{\partial^2 F}{\partial \phi_i \partial \theta_j}
\end{eqnarray}
for $\theta = \frac{\pi}{2}$ and $\phi = 0$.


An alternative expression within the KKR multiple scattering formalism
has been worked out by Ebert and Mankovsky \cite{EM09a},
by using the alternative convention for the electronic Green function (GF) as
suggested by Dederichs and coworkers \cite{DDZ92}. 
According to this convention, the off-site part of the GF is given by the
expression:
%
\begin{eqnarray}
G(\vec{r}_i,\vec{r}_j,E) &=& \sum_{\Lambda\Lambda'}
R^{i}_{\Lambda}(\vec{r}_i,E) G^{ij}_{\Lambda\Lambda'}(E)
R^{j\times}_{\Lambda'}(\vec{r}_j,E) \; , 
\label{Eq:GF_KKR_Julich}
\end{eqnarray}
%
where $G^{ij}_{\Lambda\Lambda'}(E)$ is the so-called structural Green's
function, $R^{i}_{\Lambda}$ is a regular solution to the single-site
Dirac equation labeled by the 
combined quantum numbers $\Lambda$ \cite{Ros61}.
The energy change $\Delta {\cal E}_{ij}$ due to a spin
tilting on sites $i$ and $j$ , given by Eq.\ (\ref{eq:Delta_energy_ij}),
transformed to the above mentioned convention is
expressed as follows
\begin{eqnarray}
\Delta {\cal E}_{ij} & = & -\frac{1}{\pi} \mathrm{Im Tr} \int dE \Delta
\underline{t}^i  \underline{G}^{ij} \Delta\underline{t}^j
\underline{G}^{ji} \;,
\end{eqnarray}
where the change of the single-site t-matrix $\Delta \underline{t}^{i}$ 
can be represented in terms of the perturbation $\Delta V^{i}(\vec{r})$ at
site $i$ using the expression
\begin{eqnarray}
\Delta t_{\Lambda'\Lambda}^i & = & \int d^3r
R^{i\times}_{\Lambda'}(r)\Delta V(r)R^{i}_\Lambda(r) = \Delta
V_{\Lambda'\Lambda}^{(R)i}  \; ,  
\end{eqnarray}
where the perturbation caused by the rotation of the spin magnetic moment  
$\hat{e}_i$ is represented by a change of the spin-dependent
potential in Eq.\ (\ref{Eq:Dirac}) (in contrast to the approach used in
Ref.\ \onlinecite{USPW03})
\begin{eqnarray}
\Delta V(r) & = & V_{\hat{n}}(r) -  V_{\hat{n}_0}(r) 
  =  \beta \vec{\sigma}(\hat{n} - \hat{n}_0)B(r) \; .
\label{Eq:Delta_V}
\end{eqnarray}
Using again the frozen potential approximation
   implies that the spatial part of the potential $V_{\hat{n}}(r)$
   does not change upon rotation of spin orientation.

Coming back to the convention for the GF used by Gy\"orffy and coworkers
\cite{Wei90a} according to Eq.\ (\ref{Eq:GF_KKR})
the expression for the elements of the exchange tensor represented in
terms of the scattering path operator $\tau^{ij}_{\Lambda'\Lambda}(E)$
has the form   
\begin{eqnarray}
 J_{ij}^{\alpha_i \alpha_j}  &=&  -\frac{1}{\pi}  \mathrm{Im Tr} \int dE 
 \underline{T}^{\alpha_i} \underline{\tau}^{ij}
\underline{T}^{\alpha_j} \underline{\tau}^{ji} \; ,
\label{Eq:Jij_tensor}
\end{eqnarray}
where
\begin{eqnarray}
T^{\alpha_i}_{\Lambda\Lambda'} & = & \int d^3r
Z^{\times}_{\Lambda}(\vec{r})\beta \sigma_{\alpha}B(r)
Z_{\Lambda'}(\vec{r}) \; .
\label{Eq:Jij_tensor_ME}
\end{eqnarray}

When compared to the approach of Udvardi et al.\cite{USPW03}, the
expression in Eq.\ 
(\ref{Eq:Jij_tensor}) is given explicitly in Cartesian coordinates. 
However, 
auxiliary rotations of the magnetization are still required to
calculate all tensor elements, and as a consequence, all components of
the DMI vector.  
This can be avoided using the approach reported recently \cite{MPE19}
for DMI calculations.


In this case, using the grand-canonical potential in the operator form
%
\begin{eqnarray}
 {\cal  K}   &=&  {\cal H - \mu N }  \;,
\end{eqnarray}
with $\mu$ the chemical potential, the variation of single-particle
energy density $\Delta{\cal E}(\vec{r})$ caused by a perturbation is written in
terms of the electronic Green function for $T = 0$ K as follows 
%
\begin{eqnarray}
\Delta{\cal E}(\vec{r}) &=& 
  -\frac{1}{\pi} \, \mbox{Im}\, \mbox{Tr} \int^\mu dE\, (E - \mu)\, \Delta
G(\vec{r},\vec{r},E) \;.
\label{K1_dens_new2}
\end{eqnarray}
%
Assuming the perturbation $\Delta V$ responsible for the change of the Green
function $\Delta G = G - G_0$ (the index $0$ indicates here the collinear
ferromagnetic reference state) to be small,  
$\Delta G$ can be expanded up to any order w.r.t. the perturbation
\begin{eqnarray}
\Delta G(E) & = & G_0 \Delta V G_0  \nonumber \\
&&+  G_0 \Delta V G_0 \Delta V G_0 \nonumber \\
&&+ G_0 \Delta V G_0 \Delta V G_0 \Delta V G_0 \nonumber \\
&& +  G_0\Delta V G_0\Delta V G_0 \Delta V G_0 \Delta V G_0  + ... \;,
\label{Eq_GF_expansion}
\end{eqnarray}
%
leading to a corresponding expansion for the energy change with respect to
the perturbation as follows 
\begin{eqnarray}
\Delta{\cal E} &=& \Delta{\cal E}^{(1)} + \Delta{\cal E}^{(2)} +
                   \Delta{\cal E}^{(3)} + \Delta{\cal E}^{(4)}  +
                   ... \;, 
\label{K1_dens_new3}
\end{eqnarray}

Here and below we drop the energy argument for the Green function $G(E)$
for the sake of convenience.
This expression is completely general as it gives the energy change as a
response to any type of perturbation. 
When $\Delta V$ is associated with tiltings of the spin magnetic moments,
 it can be expressed within the frozen potential approximation and in
 line with Eq.\ (\ref{Eq:Delta_V}) as follows 
\begin{eqnarray}
 \Delta V(\vec{r}) &=&  \sum_i \beta \big( \vec{\sigma}\cdot\hat{s}_i
  -  \sigma_z\big) B_{xc}(\vec{r}) \;.
\label{Eq_perturb}
\end{eqnarray}
With this, the energy expansion in Eq (\ref{K1_dens_new3}) gives access to the
bilinear DMI as well as to higher order multispin interactions
\cite{MPE20}.
To demonstrate the use of this approach, we start with the $x$ and $y$
components of the DMI vector, which 
can be obtained by setting the perturbation
$\Delta V$ in the form of a spin-spiral described by the configuration
of the magnetic moments 
\begin{equation}
  \hat{m}_i =  \Big(\sin(\vec{q}\cdot\vec{R}_i),0, \cos(\vec{q}\cdot\vec{R}_i)\Big) 
\; ,
\label{spin-spiral_1}
\end {equation}
with the wave vector $\vec{q} = (0,q,0)$.
As it follows from the spin Hamiltonian, the slope of the spin wave energy
dispersion at the $\Gamma$ point is determined by the DMI as follows 
\begin{eqnarray}
  \lim_{q \to 0} \frac{\partial E_{\rm DM}^{(1)}}{\partial q_y}   
  &=& \lim_{q \to 0} \frac{\partial }{\partial q_y}   \sum_{ij} D^y_{ij} 
 \, \sin(\vec{q}\cdot(\vec{ R}_j - \vec{R}_i))  \nonumber \\  
  &=&  \sum_{ij} D^y_{ij} \, (\vec{R}_j - \vec{R}_i)_y    \;.
\label{Slope}
\end{eqnarray}
Identifying this with the corresponding derivative of the energy $\Delta {\cal E}^{(1)}$ in Eq. \ref{K1_dens_new3} 
%
\begin{equation}
 \frac{\partial \Delta {\cal E}^{(1)}}{\partial q_\alpha} \bigg\vert _{q \to 0}  =
 \frac{\partial E_{\rm DM}^{(1)}}{\partial q_\alpha} \bigg\vert _{q \to 0} \; ,
\label{Limit_q0}
\end {equation}
%
and equating the corresponding terms for each atomic pair $(i,j)$, 
one obtains the following expression for the $y$ component of the DMI vector:
%
\begin{eqnarray}
  D^y_{ij} & = &
\left(-\frac{1}{2\pi} \right) \mbox{Im}\,\mbox{Tr} \int^\mu dE \,(E - \mu) 
 \nonumber \\     &\times & 
       \bigg[ { \underline{O}^{j}(E)} 
     \,   \underline{\tau}^{j i}(E)  \, { \underline{T}^{i,x}(E)} \,
                               \underline{\tau}^{i j}(E) \nonumber  \\
  && -  { \underline{O}^{i}(E)} \,  \underline{\tau}^{i
      j}(E) \, {\underline{T}^{j,x}(E)}  \,
        \underline{\tau}^{ji}(E)\bigg] \; , \label{Eq:DMI_Dij}
\end{eqnarray}
In a completely analogous way one can derive the x-component of the DMI
vector, $D^x_{ij}$. The overlap integrals $O^{j}_{\Lambda\Lambda'}$ 
and matrix elements $T^{i,\alpha}_{\Lambda\Lambda'}$ of the operator
${\cal T}^{i,\alpha}  = \beta \sigma_{\alpha} B_{xc}^i(\vec{r})$  (which are 
connected with the components of the torque operator $\beta[\vec{\sigma}
\times \hat{m}] B^i_{xc}(\vec{r})$) are defined as follows:\cite{EM09a}
%
\begin{eqnarray}
 O^{j}_{\Lambda\Lambda'} & = & \int_{\Omega_j} d^3r  \,
 Z^{j \times}_{\Lambda}(\vec{r},E) \, Z^{j}_{\Lambda'}(\vec{r},E)  \label{Eq:ME1}
  \\
 T^{i,\alpha}_{\Lambda\Lambda'} & = & \int_{\Omega_i} d^3r  \, Z^{i \times}_{\Lambda}(\vec{r},E)\, \Big[\beta \sigma_{\alpha} B_{xc}^i(\vec{r})\Big] \, Z^{i}_{\Lambda'}(\vec{r},E)\;.  \label{Eq:ME2}
\end{eqnarray}
%

As is shown in Ref.\ \onlinecite{MPE20}, the $D_{ij}^z$ component of the
DMI, as well 
isotropic exchange parameter $J_{ij}$ can also be obtained on the basis
of Eqs.\ (\ref{Eq_GF_expansion}) and (\ref{K1_dens_new3}) using the
second order term w.r.t. the perturbation, for a spin spiral with the
form   
\begin{eqnarray}
  \hat{s}_i &=&
(\sin\theta \cos(\vec{q}\cdot\vec{R}),\sin\theta
                \sin(\vec{q}\cdot\vec{R}), \cos\theta) \,.
\label{Eq:spin-spiral-stand}
\end {eqnarray}

In this case case,  the DMI component $D^{z}_{ij}$ and  the isotropic
exchange interaction are obtained by taking the first-
and second-order derivatives of the energy 
$\Delta {\cal E}^{(2)}(\vec{q})$ (see Eq.\ (\ref{K1_dens_new3})),
respectively, with respect to $\vec{q}$: 
 \begin{eqnarray}
   \frac{\partial}{\partial \vec{q}}\Delta E_H(\vec{q})\bigg|_{q \to 0} &=&  
    - \sin^2\theta \sum_{i \neq j}^N
         D^{z}_{ij}   \hat{q}\cdot(\vec{R}_i - \vec{R}_j) 
 \label{Eq_deriv_Heisenberg_2-derivativ1}
 \end{eqnarray}
and
 \begin{eqnarray}
   \frac{\partial^2}{\partial \vec{q}^2}\Delta E_H(\vec{q})\bigg|_{q \to 0} &=&  
     \sin^2\theta \sum_{i,j}
      J_{ij} (\hat{q}\cdot(\vec{R}_i - \vec{R}_j))^2
 \label{Eq_deriv_Heisenberg_2-derivativ2}
 \end{eqnarray}
 with $\hat{q} = \vec{q}/|\vec{q}|$ the unit vector giving the direction of
 the wave vector $\vec{q}$.   
Identifying these expressions again with the corresponding derivatives
of $\Delta {\cal 
  E}^{(2)}(\vec{q})$, one obtains the following relations for $D^{z}_{ij}$ 
 \begin{eqnarray}
   D^z_{ij} &=&  \frac{1}{2} ({\cal J}^{xy}_{ij}  - {\cal J}^{yx}_{ij})
 \label{Eq_DMI_z}
 \end{eqnarray}
 and for $J_{ij}$
 \begin{eqnarray}
  J_{ij} &=&  \frac{1}{2} ({\cal J}^{xx}_{ij}  + {\cal J}^{yy}_{ij}) \,,
 \label{Eq_Js}
 \end{eqnarray}
 where the tensor elements ${\cal J}^{\alpha\beta}$ are given by Eqs.\
 (\ref{Eq:Jij_tensor}) and (\ref{Eq:Jij_tensor_ME}).

 Similar to the magnetic anisotropy, the exchange coupling parameters depend
 on temperature, that should be taken into account within the finite
 temperature spin dynamic simulations. An approach that gives access to
 calculations of exchange  coupling parameters for finite temperature has
 been reported in Ref.\ \onlinecite{MPE20}. It 
 accounts for the electronic structure modification due to temperature
 induced lattice vibrations by using the alloy
 analogy model in the adiabatic approximation. This implies 
calculations of the thermal average $\langle ... \rangle_{ T}$ 
as the configurational average over a set of appropriately chosen set of
atomic displacements, using the CPA alloy theory
 \cite{EMKK11,MKWE13,EMC+15}.

To make use of this scheme to account for lattice vibrations, a discrete  
set of $N_{v}$  vectors $\Delta \vec{R}^q_v(T)$ is
introduced for each atom, with the temperature dependent amplitude,
which characterize a rigid displacement of the atomic potential  
in the spirit of the rigid muffin-tin approximation \cite{PZDS97,Lod76}.
The corresponding single-site t-matrix in the common global frame of the
solid is given by the transformation:
%
\begin{equation}
\label{eq:U-trans}
\underline{t}^q_v = \underline{U}(\Delta \vec{R}_v)\,\underline{t}^{q,\rm loc}\,
                     \underline{U}(\Delta \vec{R}_v)^{-1} \;,
\end{equation}
%
with the so-called U-transformation matrix $\underline{U}(\vec{s})$ 
given  in its non-relativistic form by:\cite{Lod76,PZDS97}
%
\begin{equation}
\label{eq:U-trans-matrix}
U_{LL'}(\vec{s}) = 
4\pi \sum_{L''}i^{l+l''-l'}\, C_{LL'L''}\, j_{l''}(|\vec{s}|k)\, Y_{L''}(\hat{s})
\;.
\end{equation}
%
Here $L=(l,m)$ represents the non-relativistic 
angular momentum quantum numbers,
$j_{l}(x)$ is a spherical Bessel function,
$Y_{L}(\hat{r})$ a real spherical harmonics, 
$C_{LL'L''}$ a corresponding  Gaunt number
and $k=\sqrt{E}$ is the electronic wave vector.
The relativistic version of the U-matrix 
is obtained by a standard Clebsch-Gordan transformation.\cite{Ros61}

Every displacement characterized by a displacement vectors
$\Delta \vec{R}_v(T)$ can be treated as a pseudo-component of a pseudo
alloy. Thus, the thermal averaging can be performed as the site diagonal
configurational average for a substitutional alloy,
by solving the multi-component CPA equations within the global frame of
reference \cite{EMC+15}.

The same idea can be used also to take into account thermal spin
fluctuations. A set of representative 
orientation vectors $\hat{e}_f$ 
(with $f=1,...,N_f$) for the local magnetic
moment is introduced. Using the rigid spin approximation, 
the single-site t-matrix 
in the global frame, corresponding to a given orientation vector, 
is determined by:
%
\begin{equation}
\label{eq:R-trans}
\underline{t}^q_f = \underline{R}(\hat{e}_f)\,\underline{t}^{q,\rm loc}\,
                \underline{R}(\hat{e}_f)^{-1} \;,
\end{equation}
%
where $ \underline{t}^{q,\rm loc}$ is the single-site t-matrix in the
local frame.
Here the transformation from the
local to the global frame of reference
is expressed by the rotation matrices $ \underline{R}(\hat{e}_f)$
that are determined by the vectors $\hat{e_f}$
or corresponding Euler angles.\cite{Ros61}
Again, every orientation can be treated as a
pseudo-component of a pseudo alloy, that allows to use the alloy analogy
model to calculate the thermal average over all types of spin fluctuations
\cite{EMC+15}.

\begin{figure}
\includegraphics[width=0.35\textwidth,angle=0,clip]{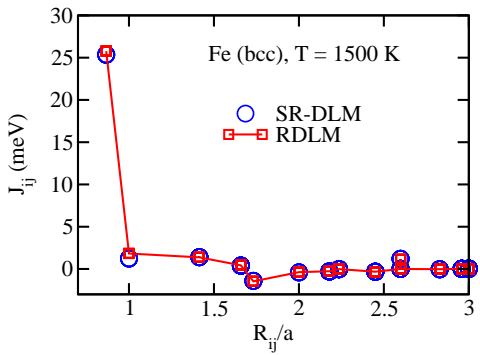}\;
\caption{\label{fig:Fe_J-DLM} Isotropic exchange coupling parameters
  calculated for the disordered magnetic state of bcc Fe within the
  scalar-relativistic approach, using the DLM scheme (circles, SR-DLM)
  and within the fully-relativistic approach, using the RDLM scheme
  \cite{SOR+04,SSB+06} (squares, RDLM). 
    }  
\end{figure}

The alloy analogy for thermal vibrations applied to the temperature
dependent exchange coupling parameters leads to
\begin{eqnarray}
 \bar{J}_{ij}^{\alpha_i \alpha_j}  &=&  -\frac{1}{2\pi} \Im  \int dE \, \mathrm{Trace}
\langle \Delta \underline{V}^{\alpha_i} \underline{\tau}^{ij} \Delta \underline{V}^{\alpha_j} \underline{\tau}^{ji} \rangle_c \; ,
\label{eq:Jij-T_1}
\end{eqnarray}
where $\langle ... \rangle_{\rm c}$ represents  the configurational
average with respect to the set of displacements.
  In case of the exchange coupling parameters one has to distinguish
  between the averaging over thermal lattice 
  vibrations and spin fluctuations. In the first case the
  configurational average is approximated as follows   
  $\langle \Delta \underline{V}^{i} \underline{\tau}^{ij} \Delta
  \underline{V}^{j} \underline{\tau}^{ji} \rangle_{vib} \approx \langle
  \Delta \underline{V}^{i} \underline{\tau}^{ij} \rangle_{vib}
  \langle \Delta 
  \underline{V}^{j}  \underline{\tau}^{ji} \rangle_{vib} $, assuming
  a negligible impact of the so-called vertex corrections
  \cite{But85}. This averaging accounts for the
  impact of thermally induced phonons on the exchange coupling
  parameters for every temperature before their use in MC or spin
  dynamics simulations that deal subsequently with the
  thermal averaging in spin subspace.
   The impact of spin fluctuations can be incorporated as well within the
    electronic structure calculations.
   For a non-polarized paramagnetic reference state, this can be done, e.g.,
   by using the so-called disorder local moment (DLM) scheme
   formulated in general within
  the non-relativistic (or scalar-relativistic) framework. Magnetic
  disorder in this case can be modeled by creating a pseudo alloy
  with an occupation of the atomic sites by two types of atoms with
  opposite spin moments oriented upwards, $M^{\uparrow}$  and
  downwards $M^{\downarrow}$, respectively, i.e. considering the alloy
  $M_{0.5}^{\uparrow}M_{0.5}^{\downarrow}$.
  In the relativistic case the corresponding RDLM scheme
  has to describe the magnetic disorder by a discrete set of 
  $N_f$ orientation vectors, and as a consequence, the average $\langle
  \underline{\tau}^{ij}\rangle_{spin}$ has to be calculated taking into 
  account all these orientations. A comparison of the results
  obtained for the isotropic exchange coupling constants $J_{ij}$ for
  bcc Fe using the DLM and RDLM schemes is shown in 
  Fig. \ref{fig:Fe_J-DLM}, demonstrating close agreement, with the small
  differences to be ascribed to the different account of relativistic
  effects, i.e. in particular the spin-orbit coupling. 

\subsection{Multi-spin expansion of spin Hamiltonian: General remarks}


Despite the obvious success of the classical Heisenberg model for many
applications, higher-order multi-spin expansion $H_{ms}$ of the spin
Hamiltonian $H$, given by the expression
\begin{eqnarray}
  H_{ms} &=& 
 - \frac{1}{3!}\sum_{i,j,k}
                   J_{ijk} \hat{s}_i\cdot (\hat{s}_j \times \hat{s}_k) \; ,  \nonumber \\          
 &&      -  \frac{2}{p!}\sum_{i,j,k,l}  J^{s}_{ijkl} (\hat{s}_i \cdot
        \hat{s}_j)(\hat{s}_k \cdot \hat{s}_l)  \nonumber \\  
 &&  
 - \frac{2}{p!} \sum_{i,j,k,l}  \vec{{\cal D}}_{ijkl} \cdot (\hat{s}_i \times
    \hat{s}_j) (\hat{s}_k \cdot \hat{s}_l) + ... \;, \nonumber \\
  &=& H^{3} +  H^{4,s} +  H^{4,a} + ...
\label{Eq_Heisenberg_general}
\end{eqnarray}
can be of great importance to describe more subtle properties of magnetic
materials
\cite{HO63,HO64,AB67,IU74,IU76,Aks80,Bro84a,IUS14,ALU+08,MKS97,GT09,GST14,FESS15}.

This concerns first of all systems with a non-collinear ground
  state characterized by finite spin tilting angles, that makes 
  multispin contributions to the energy  non-negligible. In
  particular, many reports published recently 
  discuss the impact of the multispin
  interactions on the stabilization of exotic topologically non-trivial
  magnetic textures, e.g. skyrmions, hopfions, etc. \cite
  {MdSDG+21,GHMH21,Hay22}

Corresponding calculations of the
multi-spin exchange parameters have been reported by different groups.
The approach based on the Connolly-Williams scheme has been used to
calculate the four-spin non-chiral (two-site and three-site) and
chiral interactions for Cr trimers \cite{ALU+08} and for a deposited Fe
atomic chain \cite{LRP+19a}, respectively, for  the biquadratic,
three-site four spin and four-site four spin interaction parameters
\cite{PHMH20, GHMH21}. The authors discuss the role of these type of
interactions for the stabilization of different types of non-collinear magnetic
structures as skyrmions and antiskyrmions.
  
A more flexible  mapping scheme using perturbation theory within the KKR
  Green function formalism was only  reported recently by
  Brinker et  al. \cite{BSL19,BSL20}, and by the present authors
  \cite{MPE20}.
Here we discuss the latter approach, i.e. the energy expansion w.r.t.\ $\Delta V$ in Eq.\
(\ref{K1_dens_new3}). One has to point out that a spin tilting in a real
system has a finite amplitude and therefore the higher order 
terms in this expansion might become non-negligible and in general should be
taken into account. Their role obviously depends on the specific material and should
increase with temperature that leads to an increasing
amplitude of the spin fluctuations. 
As these higher-order terms are directly connected to the multispin
terms in the extended Heisenberg Hamiltonian, one has to expect also a
non-negligible role of the multispin interactions for some magnetic
properties.   


Extending the spin Hamiltonian to go beyond the classical Heisenberg model,
we discuss first the four-spin exchange interaction terms  $J_{ijkl}$
and  $\vec{\cal D}_{ijkl}$.
 They can be calculated using the fourth-order term of the Green
 function expansion $\Delta {\cal E}^{(4)}$ given by:
\begin{eqnarray}
\Delta {\cal E}^{(4)} &=& - \frac{1}{\pi} \mbox{Im}\,\mbox{Tr} \int^{E_F}
                    dE\, \nonumber \\
  && \times (E-E_F) \Delta V
     G  \Delta V G \Delta V G \Delta V G  \nonumber\\
 &=& -\frac{1}{\pi} \mbox{Im}\,\mbox{Tr} \int^{E_F}
                          dE\, \Delta V G \Delta V G \Delta V G \Delta V
     G\;. \nonumber \\
\label{Eq_Free_Energy-4}
\end{eqnarray}
where the sum rule for the Green function $\frac{dG}{dE} = -GG$ followed by
integration by parts was used to get a more compact expression. Using
the multiple-scattering representation for the Green function, this
leads to:
\begin{eqnarray}
\Delta {\cal E}^{(4)} &=&  \sum_{i,j,k,l} -\frac{1}{\pi} \mbox{Im}\,\mbox{Tr} \int^{E_F}
                          dE\,  \nonumber \\
                      && \times \Delta \underline{V}_{ii} \underline{\tau}_{ij}
                        \Delta \underline{V}_{jj} \underline{\tau}_{jk}
                         \Delta \underline{V}_{kk}  \underline{\tau}_{kl}
                         \Delta \underline{V}_{ll} \underline{\tau}_{li} \;.
\label{Eq_Free_Energy-5}
\end{eqnarray}
with the matrix elements $\Delta \underline{V}_{ii} = \langle Z_i| \Delta V | Z_i \rangle$.
Using the ferromagnetic state with $\vec{M}||\hat{z}$ as a reference
state, and creating the perturbation $\Delta V$ in the form
of a spin-spiral according to  Eq.\ (\ref {Eq:spin-spiral-stand}), one obtains
the corresponding $\vec{q}$-dependent energy change $\Delta {\cal
  E}^{(4)}(\vec{q})$, written here explicitly as an example
\begin{eqnarray}
\Delta {\cal E}^{(4)} &=&  -\frac{1}{\pi} \sum_{i,j,k,l}\mbox{Im}\,\mbox{Tr} \int^{E_F}
                          dE\, \mbox{sin}^4\theta \nonumber \\
                      && \times \bigg[ I^{xxxx}_{ijkl} \;
                    \mbox{cos}(\vec{q} \cdot \vec{R}_i)\,  \mbox{cos}(\vec{q}
                         \cdot \vec{R}_j )\,  \mbox{cos}(\vec{q} \cdot
                         \vec{R}_k )\,  \mbox{cos}(\vec{q} \cdot \vec{R}_l)\nonumber \\
                      && +  I^{xxyy}_{ijkl}  \;
                    \mbox{cos}(\vec{q} \cdot \vec{R}_i)\,  \mbox{cos}(\vec{q}
                         \cdot \vec{R}_j) \,  \mbox{sin}(\vec{q} \cdot
                         \vec{R}_k ) \, \mbox{sin}\vec{q} \cdot \vec{R}_l)\nonumber\\
                      && + I^{yyxx}_{ijkl}   \;
                    \mbox{sin}(\vec{q} \cdot \vec{R}_i)\,  \mbox{sin}(\vec{q}
                         \cdot \vec{R}_j )\,  \mbox{cos}(\vec{q} \cdot
                         \vec{R}_k)\,   \mbox{cos}(\vec{q} \cdot \vec{R}_l)\nonumber\\
                      && +  I^{yyyy}_{ijkl}   \;
                    \mbox{sin}(\vec{q} \cdot \vec{R}_i)\,  \mbox{sin}(\vec{q}
                         \cdot \vec{R}_j )\,  \mbox{sin}(\vec{q} \cdot
                         \vec{R}_k )\,  \mbox{sin}(\vec{q} \cdot
                         \vec{R}_l) \nonumber\\
                      && +  I^{xyxx}_{ijkl}   \;
                    \mbox{cos}(\vec{q} \cdot \vec{R}_i)\,  \mbox{sin}(\vec{q}
                         \cdot \vec{R}_j )\,  \mbox{cos}(\vec{q} \cdot
                         \vec{R}_k )\,  \mbox{cos}(\vec{q} \cdot \vec{R}_l)\nonumber \\
                      && +  I^{yxyy}_{ijkl}   \;
                    \mbox{sin}(\vec{q} \cdot \vec{R}_i)\,  \mbox{cos}(\vec{q}
                         \cdot \vec{R}_j)\,   \mbox{sin}(\vec{q} \cdot
                         \vec{R}_k )\,  \mbox{sin}\vec{q} \cdot \vec{R}_l)\nonumber\\
                      && +  I^{yxxx}_{ijkl}   \;
                    \mbox{sin}(\vec{q} \cdot \vec{R}_i)\,  \mbox{cos}(\vec{q}
                         \cdot \vec{R}_j )\,  \mbox{cos}(\vec{q} \cdot
                         \vec{R}_k) \,  \mbox{cos}(\vec{q} \cdot \vec{R}_l)\nonumber\\
                      && +  I^{xyyy}_{ijkl}   \;
                    \mbox{cos}(\vec{q} \cdot \vec{R}_i)\,  \mbox{sin}(\vec{q}
                         \cdot \vec{R}_j )\,  \mbox{sin}(\vec{q} \cdot
                         \vec{R}_k )\,  \mbox{sin}(\vec{q} \cdot
                         \vec{R}_l) +  ...  \bigg]    \nonumber\\
\label{Eq_Free_Energy-4spin-2}
\end{eqnarray}
where
\begin{eqnarray}
  I^{\alpha\beta\gamma\delta}_{ijkl}  &=&  
\underline{T}^{i, \alpha}(E)\, \underline{\tau}^{ij}(E)
   \underline{T}^{j, \beta}(E)\, \underline{\tau}^{jk}(E)    \nonumber    \\
       && \times  \underline{T}^{k, \gamma}(E)\, \underline{\tau}^{kl}(E)
\underline{T}^{l, \delta}(E)\, \underline{\tau}^{li}(E) \; .
\label{Eq:J_XYZL} 
\end{eqnarray}

As is shown in Ref.\ \onlinecite{MPE20}, the four-spin isotropic
exchange interaction $J_{ijkl}$ and $z$-component of the DMI-like
interaction ${\cal D}^z_{ijkl}$ can be obtained calculating the energy
derivatives  $\frac{\partial^4}{\partial q^4} 
\Delta {\cal E}^{(4)}$ and 
$ \frac{\partial^3}{\partial q^3}
\Delta  {\cal E}^{(4)}$ in the limit of $q = 0$, and then identified with
the corresponding derivatives of the terms $H^{4,s}$ and $H^{4,a}$ in Eq.\
(\ref{Eq_Heisenberg_general}). 
These interaction terms are given by the expressions
\begin{eqnarray}
J_{ijkl}^{s} &=& \frac{1}{4} \bigg[ {\cal J}_{ijkl}^{xxxx} + {\cal
                         J}_{ijkl}^{xxyy} + {\cal J}_{ijkl}^{yyxx} + {\cal
                         J}_{ijkl}^{yyyy} \bigg]
\label{Eq_SYM-spin-4}
\end{eqnarray}
and
\begin{eqnarray}
D_{ijkl}^{z} &=& \frac{1}{4} \bigg[{\cal J}_{ijkl}^{xyxx} + {\cal
                         J}_{ijkl}^{xyyy} -  {\cal J}_{ijkl}^{yxxx} - {\cal
                         J}_{ijkl}^{yxyy}) \;\bigg]
\label{Eq_DMI-spin-4}\,,
\end{eqnarray}
where the following definition is used:
\begin{eqnarray}
{\cal J}_{ijkl}^{\alpha\beta\gamma\delta} &=&  \frac{1}{2\pi} \mbox{Im}\,\mbox{Tr} \int^{E_F}
                          dE\,  T^\alpha_i \tau_{ij}  T^\beta_j \tau_{jk}  T^\gamma_k
                         \tau_{kl} T^\delta_l \tau_{li} 
\label{Eq_j_ijkl}
\end{eqnarray}


These expression obviously give also access to a special cases, i.e. the four-spin
three-site interactions with $l = j$, and the four spin two-site, so
called  biquadratic exchange interactions  with $k=i$ and $l=j$.

\begin{figure}
\includegraphics[width=0.35\textwidth,angle=0,trim= 0 2.5cm 0 9.5cm,clip]{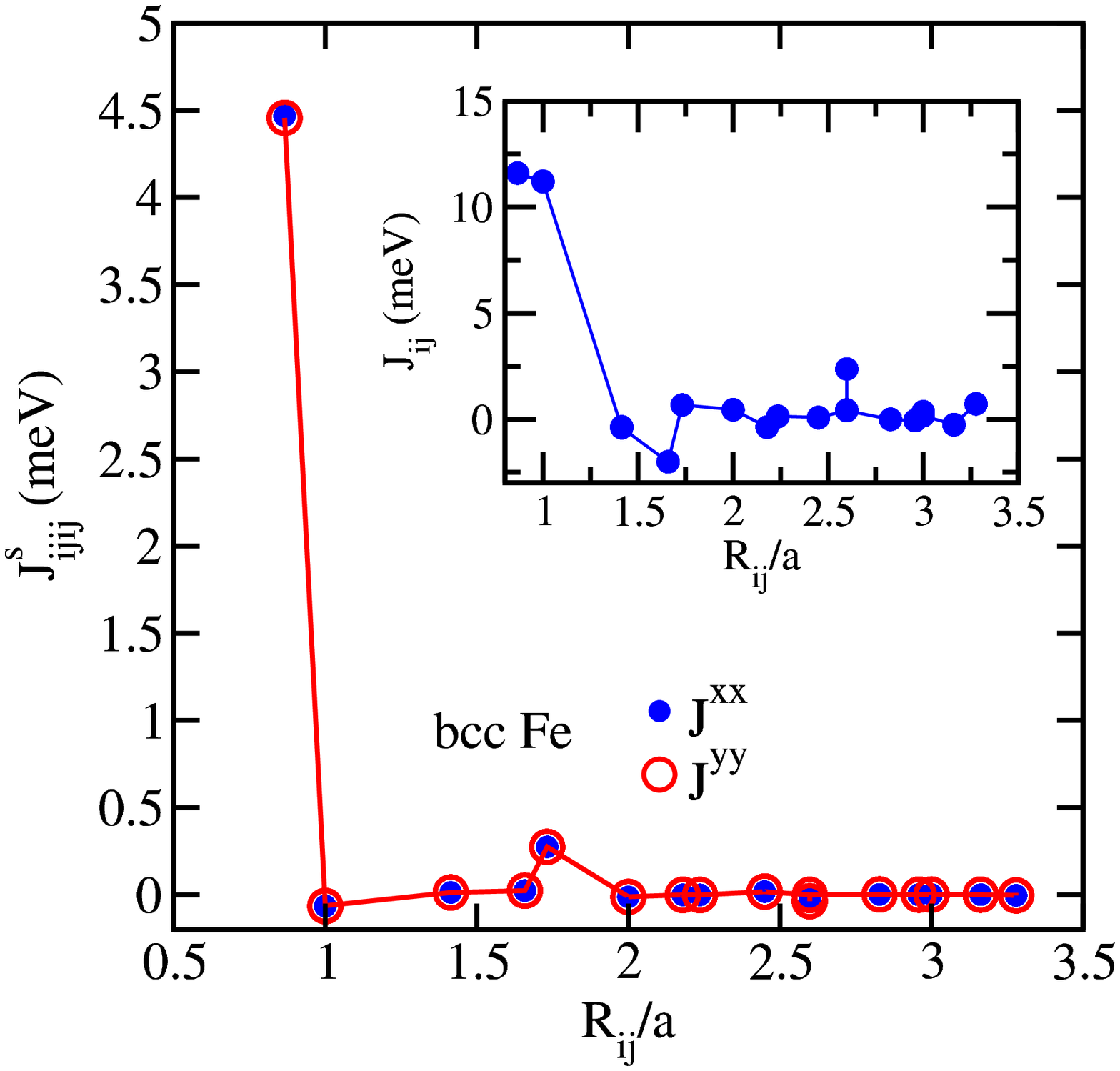}\;(a)
\includegraphics[width=0.35\textwidth,angle=0,trim= 0 2.5cm 0 9.5cm,clip]{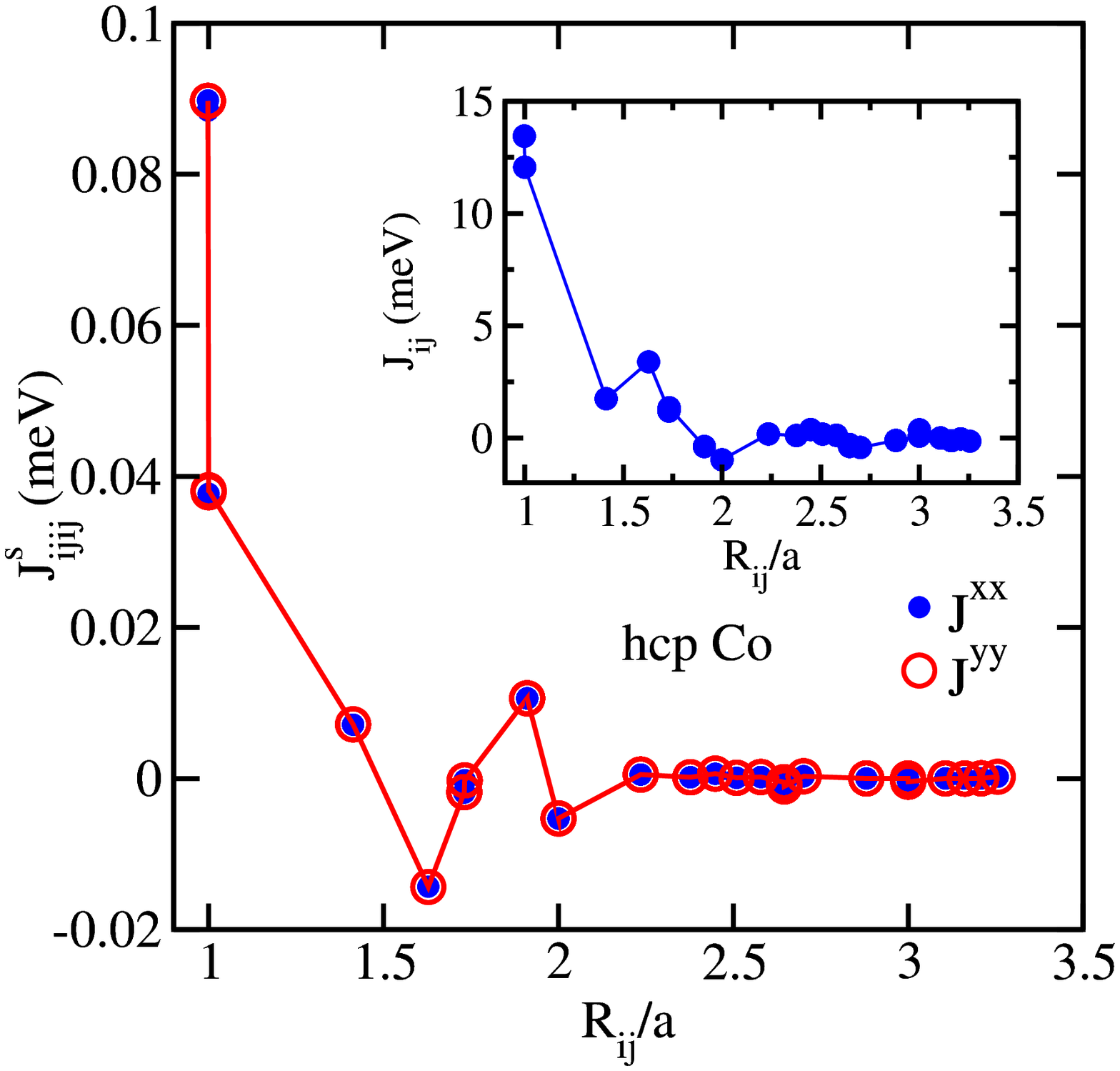}\;(b)
\includegraphics[width=0.35\textwidth,angle=0,trim= 0 2.5cm 0 9.5cm,clip]{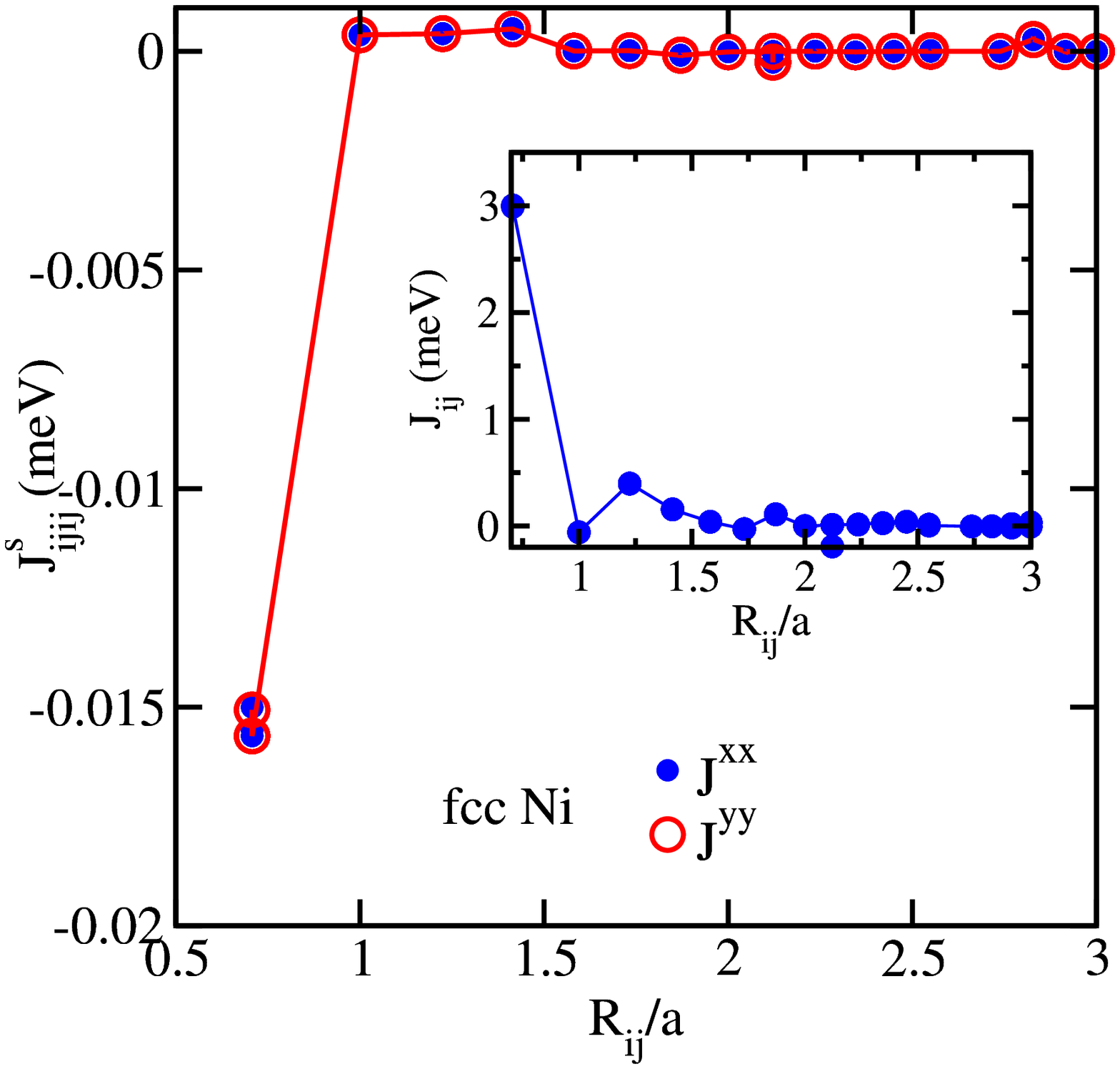}\;(c)
\caption{\label{fig:FeCoNi_biquad}  Scalar biquadratic exchange
  interactions $J^s_{ijij}$ in bcc  Fe (a), hcp Co (b) and fcc Ni
  (Ni). For comparison, the insets 
  show the bilinear exchange interaction parameters calculated for the
  FM state with the magnetization along the $\hat{z}$-axis. All data are taken from Ref.\
  \onlinecite{MPE20}.
    }   
\end{figure}

The scalar biquadratic exchange interaction parameters $J^s_{ijij}$ calculated on the basis
of Eq.\ (\ref{Eq_SYM-spin-4}) for 
the three $3d$ bulk ferromagnetic systems bcc Fe, hcp Co and fcc
Ni have been reported in Ref.\ \onlinecite{MPE20}.
The results are plotted in Fig.\ \ref{fig:FeCoNi_biquad} as a function
of the distance $R_{ij} + R_{jk} + R_{kl} + R_{li}$.
For comparison, the insets give the corresponding bilinear
isotropic exchange interactions for these materials. One can see rather
strong first-neighbor interactions for bcc Fe, demonstrating the
non-negligible character of the biquadratic 
interactions. This is of course a material-specific property, and one notes
as decrease for the biquadratic exchange parameters when going to Co and
Ni as shown in Fig.\ \ref{fig:FeCoNi_biquad} (b) and (c), respectively.


In order to calculate the $x$ and $y$ components of the four-spin and as
a special case the three-site-DMI (TDMI) and biquadratic-DMI (BDMI)
type interactions, the scheme suggested in Ref.\ \onlinecite{MPE20} for
the calculation of the DMI parameters   
\cite{ME17,MPE19} can be used, which exploited the DMI-governed behavior of the
spin-wave dispersion having a finite slope at the $\Gamma$ point of the
Brillouin zone.
Note, however, that a more general form of perturbation is required in this case
 described by a 2D spin modulation field according to the
 expression 
\begin{eqnarray}
  \hat{s}_i &=&
 \big(\mbox{sin}(\vec{q}_1 \cdot \vec{R}_i) \;\mbox{cos}(\vec{q}_2 \cdot \vec{R}_i),
                \mbox{sin}(\vec{q}_2 \cdot \vec{R}_i), \nonumber \\
&&  \mbox{cos}(\vec{q}_1 \cdot
 \vec{R}_i)\mbox{cos}(\vec{q}_2 \cdot \vec{R}_i) \big) \;,
\label{spiral2}
\end {eqnarray}
where the wave vectors $\vec{q}_1$ and $\vec{q}_2$ are
orthogonal to each other, as for example $\vec{q}_1 = q_1\hat{y}$ and
$\vec{q}_2 = q_2\hat{x}$. 

Taking the second-order  derivative with respect to the wave-vector
$\vec{q}_{2}$ and the first-order derivative with respect to the 
  wave-vectors $\vec{q}_{1}$ and $\vec{q}_{2}$, and considering the limit
  $q_{1(2)} \to 0$, one obtains
\begin{eqnarray}
 \frac{\partial^3}{\partial q_2^3}
  \bigg|_{q_2 = 0}  {H}^{4,a}
  &=&
       \sum_{i,j,k,l} {\cal D}^x_{ijkl}\,
   ({\hat{q}_2}\cdot \vec{R}_{ij})\,  ({\hat{q}_2}\cdot \vec{R}_{lk})^2
      \,, \nonumber
\label{Eq_Heisenberg_H4_Deriv-1}
\end{eqnarray}
and
\begin{eqnarray}
\frac{\partial}{\partial q_1}\bigg|_{q_1 = 0} \frac{\partial^2}{\partial
  q_2^2}\bigg|_{q_2 = 0} {H}^{4,a}
  &=&
       \sum_{i,j,k,l} {\cal D}^y_{ijkl}\,
    ({\hat{q}_1}\cdot \vec{R}_{ij}) \, ({\hat{q}_2}\cdot \vec{R}_{lk})^2  \,, \nonumber
\label{Eq_Heisenberg_H4_Deriv-2}
\end{eqnarray}
where $\vec{R}_{ij} = \vec{R}_j -\vec{R}_i$ and $\vec{R}_{lk} = \vec{R}_k -\vec{R}_l$.

The microscopic expressions for the $x$ and $y$ components of
$\vec{\cal D}_{ijkl}$ describing the four-spin interactions is derived 
on the basis of the third-order term in Eq.\ (\ref{Eq_GF_expansion}) 
\begin{eqnarray}
\Delta {\cal E}^{(3)} &=& -\frac{1}{\pi} \mbox{Im}\,\mbox{Tr} \int^{E_F}
                    dE (E - E_F)\, \nonumber \\
  && \times G_0  \Delta V G_0 \Delta V G_0 \Delta V G_0 \; .
\label{Eq_Free_Energy-3}
\end{eqnarray}
%
The final expression for $ {\cal D}^\alpha_{ijkl}$  is achieved by
taking the second-order  derivative 
with respect to the wave-vector  $\vec{q}_{2}$ and the
first-order derivative with respect to the
  wave-vectors $\vec{q}_{1(2)}$, considering the limit $q_{1(2)} \to 0$, i.e.
 equating within the ab-initio and model
expressions the corresponding terms 
proportional to $({\vec R}_i - {\vec R}_j)_y ({\vec R}_k - {\vec
  R}_l)^2_x$ and $({\vec R}_i - {\vec R}_j)_x ({\vec R}_k - {\vec
  R}_l)^2_x$ (we keep a similar form in both cases for the sake of
convenience) gives the elements ${\cal D}^{y,x}_{ijkl}$ and
${\cal D}^{y,y}_{ijkl}$, as well as ${\cal D}^{x,x}_{ijkl}$ and
${\cal D}^{x,y}_{ijkl}$, respectively, of the four-spin chiral interaction 
as follows
\begin{eqnarray}
  {\cal D}^{\alpha,\beta}_{ijkj}  &=&   \epsilon_{\alpha\gamma} \frac {1}{8\pi}\mbox{Im}\, \mbox{Tr} \int^{E_F} dE (E - E_F)\, 
               \nonumber \\       
                     &&\Big[ \underline{O}^{i}\, \underline{\tau}^{ij}
                        \underline{T}^{j, \gamma}\,  \underline{\tau}^{jk}
                        \underline{T}^{k, \beta}\, \underline{\tau}^{kl}
                  \underline{T}^{l, \beta}\, \underline{\tau}^{li}
                        \;\nonumber \\
                     && - \underline{T}^{i, \gamma}\, \underline{\tau}^{ij}
                        \underline{O}^{j}  \underline{\tau}^{jk}
                        \underline{T}^{k, \beta}\, \underline{\tau}^{kl}
                  \underline{T}^{l, \beta}\,
                        \underline{\tau}^{li}\Big] \;          \,\nonumber\\
                     && + \Big[ \underline{O}^{i}\, \underline{\tau}^{ij}
                        \underline{T}^{j, \beta}\,  \underline{\tau}^{jk}
                        \underline{T}^{k, \beta}\, \underline{\tau}^{kl}
                  \underline{T}^{l, \gamma}\, \underline{\tau}^{li}
                        \;\nonumber \\
                     && -   \underline{T}^{i, \gamma}\, \underline{\tau}^{ij}
                       \underline{T}^{j, \beta}  \underline{\tau}^{jk}
                        \underline{T}^{k, \beta}\, \underline{\tau}^{kl}
                \underline{O}^{l} \,  \underline{\tau}^{li}\Big] \;
\label{Eq:D-4-4_XYZ} 
\end{eqnarray}
with $\alpha, \beta = x,y$, and $\epsilon_{\alpha\gamma}$ the elements
of the transverse Levi-Civita tensor $ \underline{\epsilon}
= \begin{bmatrix}  0 & 1 \\  -1 & 0   \end{bmatrix} $.
The TDMI and BDMI parameters can be obtained as the special cases $l =
j$ and $l = j, k = i$, respectively, from Eq.\ (\ref{Eq:D-4-4_XYZ}). 

The expression in  Eq.\ (\ref{Eq:D-4-4_XYZ}) gives access to the $x$ and $y$
components of the DMI-like three-spin interactions
\begin{eqnarray}
 {\cal D}^{\alpha}_{ijkj}= {\cal D}^{\alpha,x}_{ijkj} +
 {\cal D}^{\alpha,y}_{ijkj} \;.
\label{Eq_Heisenberg_D4}
\end{eqnarray}


Finally, 
three-spin chiral exchange interaction (TCI) represented by first term in the
extended spin Hamiltonian has been discussed in Ref.\
\onlinecite{MPE20}. As it follows from this expression, the 
contribution due to this type of interaction is non-zero only in case of
a non-co-planar and non-collinear magnetic structure characterized by
the scalar spatial type product $\hat{s}_i\cdot (\hat{s}_j \times \hat{s}_k)$
involving the spin moments on three different lattice sites.

In order to work out the expression for the $J_{ijk}$ interaction,
one has to use a multi-Q spin modulation
\cite{SMM12,OCK12,BLHK16} which ensure a non-zero scalar spin chirality
for every three atoms.
The energy contribution due to the TCI,
  is non-zero only if $J_{ijk}\neq  J_{ikj}$, etc. 
  Otherwise, the terms $ijk$ and $ikj$
  cancel each other due to the relation $\hat{s}_i\cdot (\hat{s}_j
 \times \hat{s}_k) = - \hat{s}_i\cdot (\hat{s}_k \times \hat{s}_j)$.

Accordingly, the expression for the TCI is derived using the 2Q non-collinear
spin texture described by Eq.\ (\ref{spiral2}), 
which is characterized by two wave 
vectors oriented along two mutually perpendicular directions, 
as for example  $\vec{q}_1 = (0,q_y,0)$
and $\vec{q}_2 = (q_x,0,0)$.
Applying such a spin modulation in Eq.\ (\ref{spiral2}) for the term
$H^3$ associated with the three-spin interaction in the spin Hamiltonian in
Eq.\ (\ref{Eq_Heisenberg_general}), the second-order
derivative of the energy $E^{(3)}(\vec{q}_1, \vec{q}_2)$ with respect to
the wave vectors $q_1$ and $q_2$ is given in the limit $q_1 \to 0$,  $q_2 \to 0$ 
 by the expression
\begin{eqnarray}
  && \frac{\partial^2}{\partial\vec{q}_1\partial\vec{q}_2} H^{(3)} 
                                                                       \nonumber \\
  && = - \sum_{i \neq j\neq k}
J_{ijk} \big( \hat{z} \cdot [(\vec{R}_i - \vec{R}_j) \times(\vec{R}_k -
  \vec{R}_j) ] \big) \; .
\label{Eq_Heisenberg_3-spin_deriv}
\end{eqnarray}

\begin{figure}[h]
\includegraphics[width=0.2\textwidth,angle=0,clip]{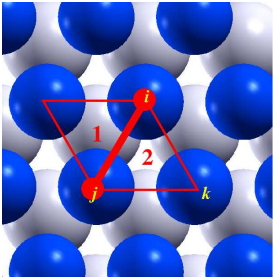}
\caption{\label{fig:THEESPIN-Co-Ir-a} Geometry of the smallest three-atom clusters 
  in the monolayer of 3$d$-atoms on $M(111)$ surface ($M$= Au, Ir):
  $M$-centered triangle $\Delta_1$ and hole-centered triangle
  $\Delta_2$.     }  
\end{figure}

\begin{figure}[h]
\includegraphics[width=0.35\textwidth,angle=0,clip]{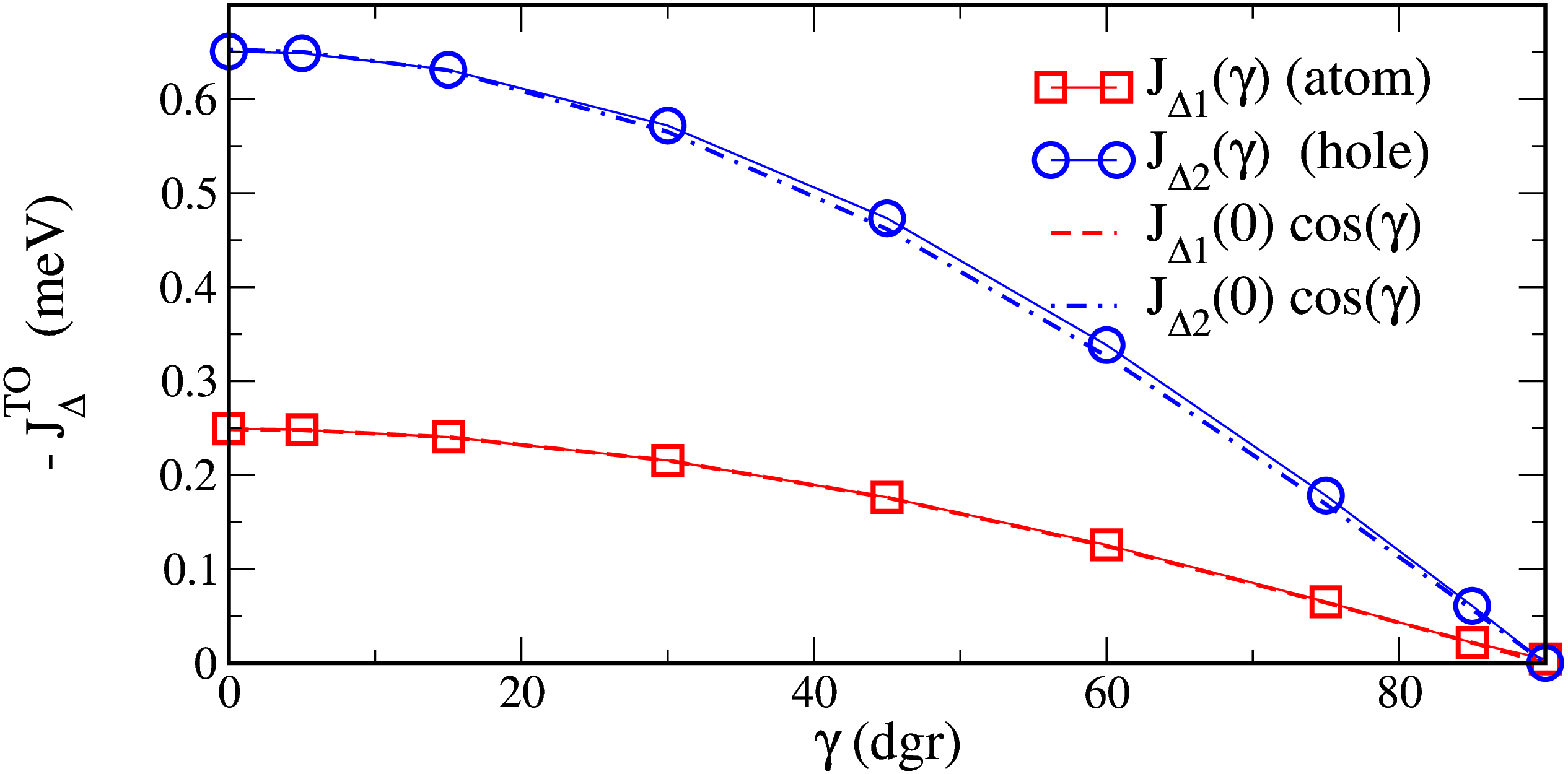}\;(a)\\
\includegraphics[width=0.35\textwidth,angle=0,clip]{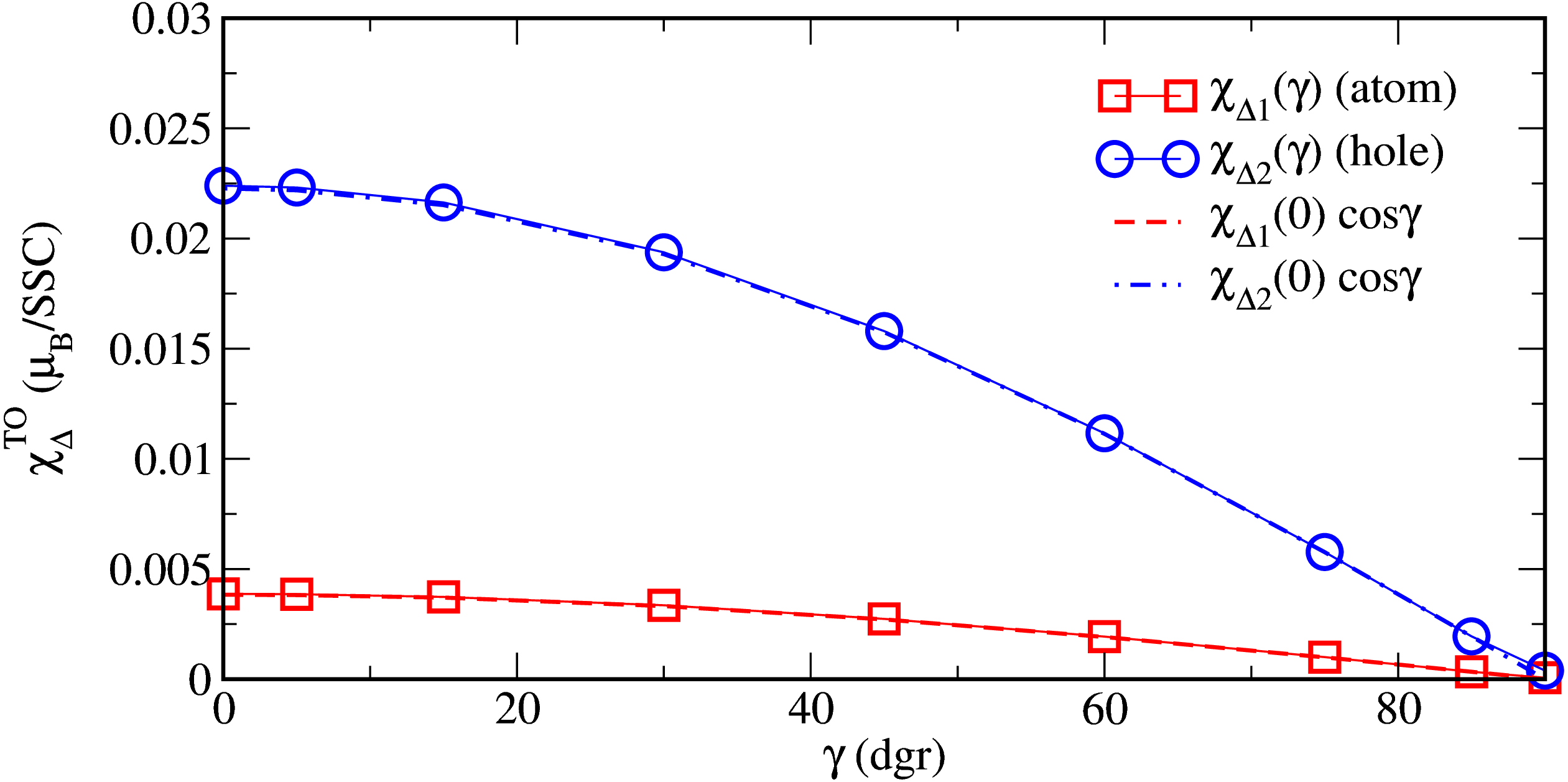}\;(b)
\caption{\label{fig:THEESPIN-Co-Ir} (a) Three-spin chiral exchange interaction
  parameters $-J_{\Delta}(\gamma)$  ('minus' is used to stress the
  relation between $J_{\Delta}$ and $\chi^{\rm{TO}}_{\Delta}$),  and (bc)
  topological orbital susceptibility (TOS, for SOC = 
  0), calculated for Fe on Ir (111), as a function of the angle between
  the magnetization and normal $\hat{n}$ to the surface, for  the smallest
  triangles $\Delta_1$ and $\Delta_2$.  The dashed lines represent
  $J_{\Delta}(0) \, \cos( \gamma )$ (a) and
  $\chi^{\rm{TO}}_{\Delta}(0) \, \cos( \gamma )$ (b), respectively.  All data are taken from Ref.\ \onlinecite{MPE21}.
    }  
\end{figure}

The microscopic energy term of the electron system, giving access to the
chiral three-spin interaction in the spin Hamiltonian is described by the
second-order term 
\begin{eqnarray}
\Delta {\cal E}^{(2)} &=& -\frac{1}{\pi} \mbox{Im}\,\mbox{Tr} \int^{E_F}
                    dE (E - E_F)\, \nonumber \\
  && G_0 \Delta V G_0 \Delta V G_0 \;
\label{Eq_Free_Energy-2}
\end{eqnarray}
of the free energy expansion.
Taking the first-order derivative with respect to $q_1$ and $q_2$ in
the limit $q_1 \to 0$,  $q_2 \to 0$, and equating the terms proportional
to $\big( \hat{z} \cdot [(\vec{R}_i - \vec{R}_j) \times(\vec{R}_k -
  \vec{R}_j) ] \big)$ with the corresponding terms in the spin Hamiltonian, one
 obtains the following expression for the three-spin interaction parameter
\begin{eqnarray}
  J_{ijk}  &=&   \frac {1}{8\pi} \mbox{Im}\, \mbox{Tr} \int^{E_F} dE (E - E_F)\,  \nonumber \\
 &&                    
\Big[ \underline{T}^{i, x}\, \underline{\tau}^{ij}
\underline{T}^{j, y}\, \underline{\tau}^{jk}
               \underline{O}^{k}\,  \underline{\tau}^{ki} 
- \underline{T}^{i, y}\, \underline{\tau}^{ij}
 \underline{T}^{j, x}\, \underline{\tau}^{jk}
     \underline{O}^{k}\,  \underline{\tau}^{ki}
               \nonumber \\       
&&     - \underline{T}^{i, x}\, \underline{\tau}^{ij}
\underline{O}^{j}\, \underline{\tau}^{jk}
               \underline{T}^{k, y}\, \underline{\tau}^{ki}\;
+ \underline{T}^{i, y}\, \underline{\tau}^{ij}
\underline{O}^{j}\, \underline{\tau}^{jk}
               \underline{T}^{k, x}\, \underline{\tau}^{ki} \;
               \nonumber \\       
&& +
   \underline{O}^{i} \, \underline{\tau}^{ij} 
\underline{T}^{i, x}\, \underline{\tau}^{jk}
               \underline{T}^{k, y}\, \underline{\tau}^{ki}\;
 - \underline{O}^{j}\,  \underline{\tau}^{ij}
\underline{T}^{i, y}\,\underline{\tau}^{jk}
               \underline{T}^{k, x}\, \underline{\tau}^{ki} \Big] \;,
\label{Eq:J_XYZ} 
\end{eqnarray}
giving access to the three-spin chiral interaction determined as $J_{\Delta} = J_{ijk} -
J_{ikj}$. Its interpretation was discussed in Ref.\ 
\onlinecite{MPE21}, where its dependence on the SOC as well
as on the topological orbital susceptibility $\chi_{\Delta}^{TO} = \chi_{ijk}^{TO} -
\chi_{ikj}^{TO}$ was demonstrated. In fact that the expression for  $\chi_{ijk}^{TO}$ worked
out in  Ref.\ 
\onlinecite{MPE21} has a rather similar form as $J_{ijk}$, as that can be
seen from the expression
\begin{eqnarray}
  \chi^{\rm{TO}}_{ijk}  &=& - \frac {1}{4\pi} \mbox{Im}\, \mbox{Tr} \int^{E_F} dE \nonumber \\
 &&                    
\times \Big[ \underline{T}^{i, x}\, \underline{\tau}^{ij}
\underline{T}^{j, y}\, \underline{\tau}^{jk}
               \underline{l}_z^{k}\,  \underline{\tau}^{ki} 
- \underline{T}^{i, y}\, \underline{\tau}^{ij}
 \underline{T}^{j, x}\, \underline{\tau}^{jk}
     \underline{l}_z^{k}\,  \underline{\tau}^{ki}
               \nonumber \\       
&&     - \underline{T}^{i, x}\, \underline{\tau}^{ij}
\underline{l}_z^{j}\, \underline{\tau}^{jk}
               \underline{T}^{k, y}\, \underline{\tau}^{ki}\;
+ \underline{T}^{i, y}\, \underline{\tau}^{ij}
\underline{l}_z^{j}\, \underline{\tau}^{jk}
               \underline{T}^{k, x}\, \underline{\tau}^{ki} \;
               \nonumber \\       
&& +
    \underline{l}_z^{i} \, \underline{\tau}^{ij} 
\underline{T}^{j, x}\, \underline{\tau}^{jk}
               \underline{T}^{k, y}\, \underline{\tau}^{ki}\;
 - \underline{l}_z^{i}\,  \underline{\tau}^{ij}
\underline{T}^{j, y}\,\underline{\tau}^{jk}
               \underline{T}^{k, x}\, \underline{\tau}^{ki} \Big] \; .
               \nonumber \\       
\label{Eq:TOS} 
\end{eqnarray}
For every trimer of atoms, both quantities, $\chi_{ijk}^{TO}$ and
$J_{ijk}$, are non-zero only in the 
case of non-zero scalar spin chirality $\hat{s}_i\cdot (\hat{s}_j \times
\hat{s}_k) $  and depend on the orientation of
the trimer magnetic moment with respect to the trimer plain.
This is shown in Fig. \ref{fig:THEESPIN-Co-Ir} \cite{MPE21} representing
$\Delta J$ and $\Delta \chi^{TO}$ as a function of the angle between
  the magnetization and normal $\hat{n}$ to the surface, which are calculated
for the two smallest trimers, $\Delta_1$ and $\Delta_2$, centered at the
Ir atom and the hole site in the Ir surface layer for 1ML Fe/Ir(111),
respectively (Fig.\ \ref{fig:THEESPIN-Co-Ir-a}). 

\begin{figure}[hbt]
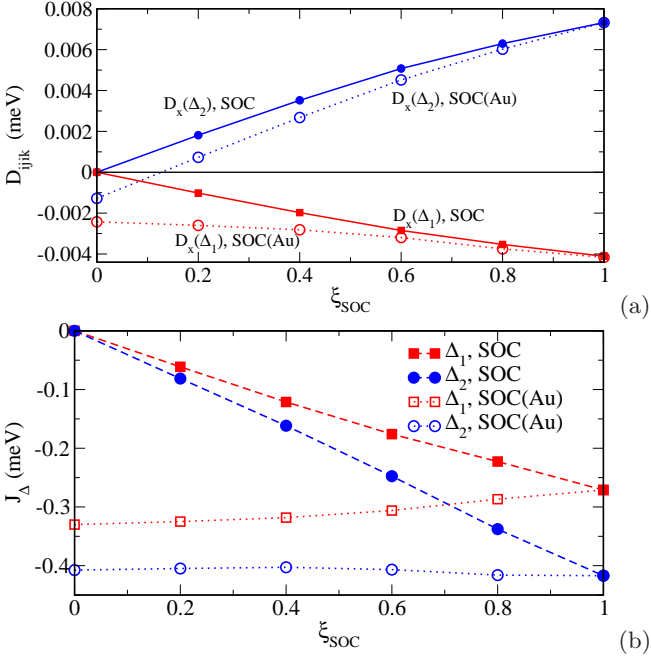

\includegraphics[width=0.45\textwidth,angle=0,clip]{FeAu_TDMI-SOC.eps} (a)
\includegraphics[width=0.45\textwidth,angle=0,clip]{FeAu_TCI-SOC.eps} (b)
\caption{\label{fig:THEESPIN-SOC}
  (a) Three-site 4-spin DMI-like interaction, ${\cal D}^z_{ijkj}  $
  and (c) three-spin chiral exchange interaction (TCI)
  parameters $J_{\Delta}$ calculated for Fe on Au (111) on the basis of
  Eq.\ (\ref{Eq:J_XYZ}) as a function of SOC scaling parameter
  $\xi_{SOC}$ for  the smallest triangles  $\Delta_1$ and $\Delta_2$.
  In figure (b), full   symbols represent the results obtained when 
 scaling the SOC for all elements in the system, while open symbols show
 the results when scaling  only the SOC for Au. All data are taken from
 Ref.\ \onlinecite{MPE21}. 
} 
\end{figure}

The role of the SOC for the three-site 4-spin DMI-like interaction, ${\cal
  D}^z_{ijik}  $, and the
three-spin chiral interaction, $J_{\Delta}$ is shown in Fig.\ \ref{fig:THEESPIN-SOC}.
These quantities are calculated for 1ML Fe on Au (111), 
for the two smallest triangles $\Delta_1$ 
and $\Delta_2$ centered at an Au atom or a hole site, respectively
(see Fig.\ \ref{fig:THEESPIN-Co-Ir-a}).
Here, setting the SOC scaling factor $\xi_{\rm{SOC}}=0$ implies a
suppression of the  SOC, while $\xi_{\rm{SOC}} = 1$ corresponds to  the
fully relativistic case.
 Fig.\ \ref{fig:THEESPIN-SOC} (a) shows the three-site 4-spin DMI-like
 interaction parameter, ${\cal D}^z_{ijik}(\xi_{\rm{SOC}})$ when the SOC scaling
 parameter $\xi_{\rm{SOC}}$ applied to all components in the system, shown
 by full symbols, and with the SOC scaling applied only to the Au substrate.
One can see a dominating role of the SOC of substrate atoms
for ${\cal D}^z_{ijik}$. Also in Fig.\
\ref{fig:THEESPIN-SOC} (b), a nearly linear variation can be seen for
$J_{\Delta}(\xi_{\rm{SOC}})$ when the SOC scaling parameter
$\xi_{\rm{SOC}}$ is applied to all components
in the system (full symbols). Similar to ${\cal D}^z_{ijik}$, this shows
that the SOC is an ultimate prerequisite for a non-vanishing TCI
$J_{\Delta}$. When scaling the SOC  only for Au (open symbols), Fig.\
\ref{fig:THEESPIN-SOC} (b) show only weak changes for the TCI parameters
$J_{\Delta}(\xi_{\rm{SOC}})$, demonstrating a minor impact of the SOC of the 
substrate on these interactions, in contrast to the DMI-like
interaction shown in  Fig.\ \ref{fig:THEESPIN-SOC} (a).
One can see also that  ${\cal D}^z_{ijik}$ is about two orders of magnitude 
smaller than $J_{\Delta}$ for this particular system.

The origin of the TCI parameters have been discussed in the
literature suggesting a different interpretation of the
corresponding terms derived also within the multiple-scattering theory
Green function formalism \cite{BSL19,GHH+20,SBL+21a}.
However, the expression worked out in Ref.\
\onlinecite{GHH+20} has obviously not been applied for calculations so far. 
As pointed out in Ref.\ \onlinecite{MPE21}, the different interpretation
of this type of interactions can be explained by their
different origin. In particular, one has to stress that the parameters in
Refs.\ \onlinecite{MPE21} and \onlinecite{GHH+20} were derived in
a different order of perturbation theory. 
On the other hand, the approach used for calculations of the multispin exchange
parameters reported in Ref.\ \onlinecite{BSL19,GHH+20,Lou20} is very
similar to the one used in Refs.\ \onlinecite{MPE20,MPE21}.
The corresponding expressions have been worked out within the
framework of multiple-scattering Green function formalism using
the magnetic force theorem. In particular, the Lloyd  formula has been used to
express the energy change due to the perturbation $\Delta V$ leading to the
expression 
\begin{eqnarray}
 \Delta {\cal E} &=&  -\frac{1}{\pi} \mbox{Im\, Tr\,} \int^{E_F}dE\, \sum_p
                     \frac{1}{p} \mbox{Tr}
                     \left[\underline{\underline{G}}(E)\,\Delta\underline{\underline{V}}
                     \right]^p \,.
\label{eq:Free_energy-Lounis1}
\end{eqnarray}

Using the off-site part of the GF in Eq.\ (\ref{Eq:GF_KKR_Julich}), as
defined by Dederichs et al.\ \cite{DDZ92}, Eq.\
(\ref{eq:Free_energy-Lounis1}) is transformed to the form
\begin{eqnarray}
 \Delta {\cal E} &=&  -\frac{1}{\pi} \mbox{Im\, Tr\,} \int^{E_F}dE\, \sum_p
                     \frac{1}{p} \mbox{Tr}
                     \left[\underline{\underline{G}}^{\mbox{str}}(E)\,\underline{\underline{\Delta t}}(E) \right]^p\,.
\label{eq:Free_energy-Lounis2}
\end{eqnarray}
 By splitting the structural Green
 function ${\underline{G}}_{ij}^{\mbox{str}}$  into a spin-dependent
($\vec{\underline{B}}_{ij}^{\mbox{str}}$) and 
 a spin-independent (${\underline{A}}_{ij}^{\mbox{str}}$) parts
 according to
\begin{eqnarray}
{\underline{G}}_{ij}^{\mbox{str}} &=& {\underline{A}}_{ij}^{\mbox{str}}
                                      \sigma_0 + \vec{\underline{B}}_{ij}^{\mbox{str}}
                                      \cdot \vec{\sigma}
\label{eq:Lounis3}
\end{eqnarray}
and expressing the change of the single-site scattering matrix
\begin{eqnarray}
\underline{\Delta t}_i (E) = (\underline{t}^{\uparrow}_i (E) -
  \underline{t}^{\downarrow}) \delta \hat{s}_i \times \vec{\sigma}\,,
\label{eq:Lounis4}
\end{eqnarray}
by means of the rigid spin approximation, the different terms in Eq.\
(\ref{eq:Free_energy-Lounis2}) corresponding to 
 different numbers $p$ give access to corresponding multispin terms,
 chiral and non-chiral, in the extended spin Hamiltonian.
 In particular, the isotropic six-spin interactions, that are
 responsible for the non-collinear magnetic structure of B20-MnGe according
 to Grytsiuk et al \cite{GHH+20}, is given by the expression
\begin{eqnarray}
  \kappa^{6-spin}_{ijklmn} &=& \frac{1}{3\pi} \mbox{Im Tr} \int^{E_F} dE
  \nonumber \\ 
    && \times    A_{ij}t^\sigma_jA_{jk} t^\sigma_k
                               A_{kl}t^\sigma_l A_{lm} t^\sigma_mA_{mn}
                               t^\sigma_n A_{ni} t^\sigma_i \,.
\label{eq:Lounis5}
\end{eqnarray}


A rather different point of view concerning the multispin extension of the spin
Hamiltonian was adopted by Streib et al. \cite{SSB+21,SCP+22},
who suggested to distinguish so-called local and global Hamiltonians.
According to that classification, a global Hamiltonian implies 
to include in principle all possible spin configurations for the energy
mapping in order to calculate exchange parameters that characterize in
turn the energy of any spin configuration. On the other hand, a local 
Hamiltonian is {\it 'designed to describe the energetics of spin configurations in the
vicinity of the ground state  or, more generally, in the vicinity of a
predefined spin configuration'} \cite{SSB+21}. This implies that taking
the ground state as a reference state, it has to be determined first
before the calculating the exchange parameters which are in principle
applicable only 
for small spin tiltings around the reference state and can be
used e.g. to investigate spin fluctuations around the ground state spin
configuration. 
In Ref.\ \onlinecite{SSB+21}, the authors used a constraining field to
stabilize the non-collinear   
  magnetic configuration. This leads to the
 effective two-spin exchange interactions corresponding to a non-collinear magnetic
 spin configuration \cite{SSB+21,SCP+22}. According to the authors, {\it 'local
 spin Hamiltonians do not require any spin interactions beyond the
 bilinear order (for Heisenberg exchange as well as
 Dzyaloshinskii-Moriya interactions)'}.
On the other hand, they point out the limitations for these exchange
interactions in the case of non-collinear system in the regime when the
standard Heisenberg model is not valid\cite{SCP+22}, and multi-spin
interactions get more important.

\section{Gilbert damping} \label{SEC:TOM}

Another parameter entering the Landau-Lifshitz-Gilbert (LLG) equation in
Eq.\ (\ref{eq:LLG_0}) 
is the Gilbert damping parameter $\tilde{G}$ characterizing energy
dissipation associated with 
the magnetization dynamics.

Theoretical investigations on the Gilbert damping parameter have been
performed by various groups and accordingly the properties of GD is
discussed in detail in the literature. Many of these investigations are
performed assuming a certain 
dissipation mechanism, like Kambersky's breathing Fermi surface  (BFS)
\cite{Kam70,FS06}, or more general torque-correlation models (TCM) 
\cite{Kam76,GIS07} to be evaluated on the basis of electronic structure
calculations. 
The earlier works in the field relied on the relaxation time parameter
that represents scattering processes responsible for the energy
dissipation. Only few computational schemes for Gilbert damping
parameter account explicitly for disorder in the systems,
which is responsible for the spin-flip scattering
process.
This issue was addressed in particular by
Brataas \emph{et al.}\ \cite{BTB08} who described the Gilbert
 damping mechanism by means of scattering theory. This development 
supplied the formal basis for the first parameter-free 
investigations on disordered alloys \cite{SKB+10,EMKK11,MKWE13}. 

A formalism for the calculation of the Gilbert damping parameter 
based on linear response theory has been reported in Ref.\
\onlinecite{MKWE13} and implemented using fully relativistic multiple
scattering or Korringa-Kohn-Rostoker (KKR) formalism. Considering the FM
state as a reference state of the   
system, the energy dissipation can be expressed in terms of
the GD parameter by: 
%
\begin{eqnarray}
\dot{E}_{\rm mag} = \vec{H}_{\rm eff}\cdot\frac{d\vec{M}}{d\tau} 
= \frac{1}{\gamma^2}\dot{\vec{m}}[\tilde{G}(\vec{m})\dot{\vec{m}}]\,.
\label{MagnE}
\end{eqnarray}
%
On the other hand, the energy dissipation in the electronic system is
determined by the underlying Hamiltonian $\hat{H}(\tau)$ as follows
$\dot{E}_{\rm dis} = \left\langle \frac{d\hat{H}}{d\tau}\right\rangle$.
Assuming a small deviation of the magnetic moment from the equilibrium
$\vec{u}(\tau)$, the normalized magnetization $\vec{m}(\tau)$ can be
written in a linearized form $\vec{m}(\tau) = \vec{m}_0 +
\vec{u}(\tau)$, that in turn leads to the linearized time dependent
electronic Hamiltonian $\hat{H}(\tau)$
%
\begin{eqnarray}
\hat{H} = \hat{H}_{0}(\vec{m}_0) + \sum_\mu \vec{u}_\mu\frac{\partial}{\partial \vec{u}_\mu} \hat{H}(\vec{m}_0) 
\; .
\label{E_expansion}
\end{eqnarray}
%
As shown in Ref.\ \onlinecite{EMKK11}, the energy dissipation within the linear
response formalism is given by:
\begin{eqnarray}
\dot{\cal E}_{\rm dis} &=& -\pi\hbar \sum_{ij}\sum_{\mu\nu}
\dot{u}_\mu \dot{u}_\nu
\left\langle  \psi_{i}\bigg|  \frac{\partial \hat{H}}{\partial u_{\mu}} \bigg| \psi_{j}\right\rangle
\left\langle  \psi_{j}\bigg| \frac{\partial \hat{H}}{\partial u_{\nu}}
  \bigg| \psi_i\right\rangle 
\nonumber \\
&&\;\;\;\;\;\;\;\;\;\;\;\;\;\;\;\;\;\;\times \delta(E_F - E_i)\delta(E_F - E_{j})
\; .
\label{E_dot}
\end{eqnarray}
Identifying it with the  corresponding phenomenological
quantity in Eq.\ (\ref{MagnE}), $\dot{E}_{\rm mag} = \dot{\cal E}_{\rm dis}$
one obtains for the GD parameter $\alpha $ a Kubo-Greenwood-like expression:
%
\begin{eqnarray}
\alpha_{\mu\nu} &=& -\frac{\hbar \gamma}{\pi M_s} 
\mbox{Trace}
\left\langle  \frac{\partial \hat{H}}{\partial u_{\mu}} \mbox{Im}\; G^{+}(E_F)
\frac{\partial \hat{H}}{\partial u_{\nu}}  \mbox{Im}\; G^{+}(E_F)
\right\rangle_{c} 
\;, \nonumber \\ 
\label{alpha2}
\end{eqnarray}
%
where  $\alpha = \tilde G /(\gamma M_s)$, and $\langle ... \rangle_{c}$
indicates a configurational average required in the presence of
 chemical or thermally induced  disorder responsible for the dissipation
 processes.  
Within the multiple scattering formalism with the 
representation of the Green function given by
Eq.\ (\ref{Eq:GF_KKR}), Eq.\ (\ref{alpha2}) leads to
\begin{eqnarray}
\alpha_{\mu\mu} =   \frac{g}{\pi\mu_{tot}} \sum_{n } \mbox{ Trace}
\left\langle \underline{T}^{0\mu} \,
 \tilde{\underline{\tau}}^{0n}\,
\underline{T}^{n\mu} \,
 \tilde{\underline{\tau}}^{n0} \right\rangle_{c}
\label{alpha_MST}
\end{eqnarray}
%
with the g-factor $2(1 + {\mu_{orb}}/{\mu_{spin}})$ in terms of
the spin and orbital moments, $\mu_{spin}$ and $\mu_{orb}$,
respectively, the total magnetic moment $\mu_{tot} =
\mu_{spin}+\mu_{orb}$, and
$\tilde{\tau}_{\Lambda\Lambda'}^{0n} =
\frac{1}{2i}(\tau_{\Lambda\Lambda'}^{0n} -
  \tau_{\Lambda'\Lambda}^{0n})$ and with the energy argument $E_F$
omitted.
 The matrix elements  $\underline{T}^{n\mu}$  are identical to
those occurring in the context of exchange coupling  \cite{EM09a} and can be expressed in
 terms of the spin-dependent part $B$ of the electronic potential with matrix elements:
%

\begin{eqnarray}
T_{\Lambda'\Lambda}^{n\mu} & = & \int d^3r\; Z^{n\times}_{\Lambda'}(\vec{r})\;\left[\beta
\sigma_{\mu}B_{xc}(\vec{r})\right] Z^{n}_{\Lambda}(\vec{r}) 
\label{matrix-element}
\; .
\end{eqnarray}
%

As is discussed in Ref.\ \onlinecite{MKWE13}, for a system having
chemical disorder, the configurational average is
performed using the scattering path operators evaluated on the basis of
the coherent potential approximation (CPA) alloy theory.
In the case of thermally induced disorder, the alloy analogy model is
used, which was discussed already above.
When evaluating Eq.\ (\ref{alpha_MST}), the so-called vertex corrections
have to be included\cite{But85} that accounts for the difference between the
averages $\left\langle  T_{\mu} \mbox{Im} G^{+} T_{\nu} \mbox{Im}
  G^{+} \right\rangle_{c}$ and 
$\left\langle  T_{\mu} \mbox{Im} G^{+}\right\rangle_{c}
\left\langle  T_{\nu} \mbox{Im} G^{+}\right\rangle_{c}$.
Within the Boltzmann formalism these corrections 
account for scattering-in processes.

\begin{figure}
  \begin{center}
 \includegraphics[scale=0.35]{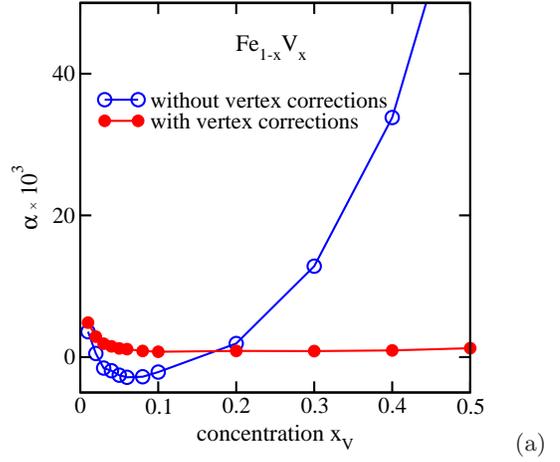} \;\; (a) \\
  \end{center}
  \vskip-5mm
  \caption{\label{Fig-vertex}
The Gilbert damping parameter for (a) bcc Fe$_{1-x}$V$_x$  ($T = 0$~K) as a function
    of V concentration. Full (open) symbols give results with (without) the
    vertex corrections. All data are taken from Ref.\ \onlinecite{MKWE13}.} 
  \vskip-6mm  
\end{figure}

The crucial role of these corrections is demonstrated \cite{MKWE13} in
Fig. \ref{Fig-vertex} representing the Gilbert damping
parameter for an Fe$_{1-x}$V$_x$ disordered alloy as a function of the  
concentration $x$, calculated  with and without vertex corrections.
As one can see, neglect of the vertex corrections may lead to the
nonphysical result $\alpha < 0$.
This wrong behavior does not occur when the vertex corrections are
included, that obviously account for energy transfer processes connected
with scattering-in processes. 

\begin{figure}
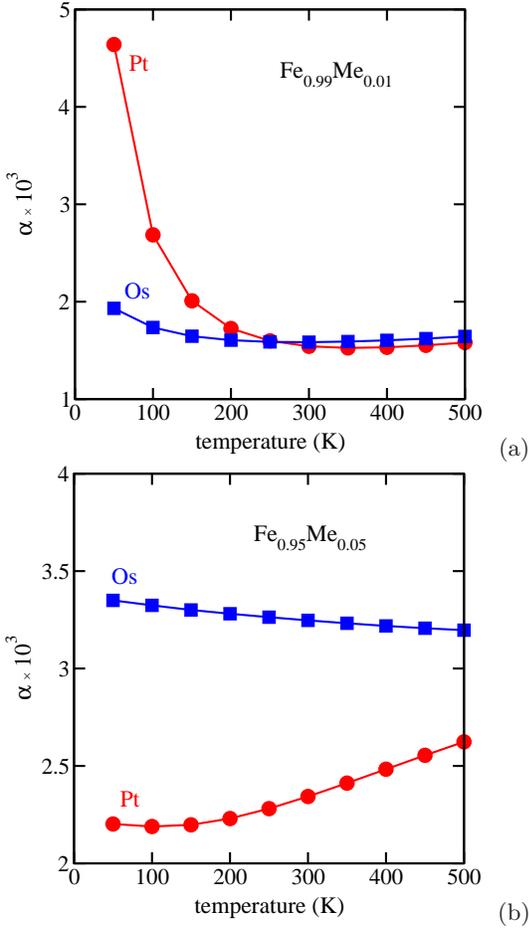

  \begin{center}
 \includegraphics[scale=0.35]{gd_prb_fig_8a.eps} \;(a)
 \includegraphics[scale=0.35]{gd_prb_fig_8b.eps} \;(b)
  \end{center}
  \caption{\label{Fig-FePtOs}
Gilbert damping parameter for bcc Fe$_{1-x}$$M_{x}$ with $M =$ Pt (circles) and $M =$ Os (squares)
impurities as a function of temperature for  1\% (a)  and 5\% (b) of the
impurities. All data are taken from Ref.\ \onlinecite{MKWE13}.} 
\end{figure}

The impact of thermal vibrations onto the Gilbert damping can be taken
into account within the alloy-analogy model (see above)
by averaging over a discrete set of thermal atom displacements
for a given temperature $T $. 
Fig. \ref{Fig-FePtOs} represents the temperature dependent behavior of the
Gilbert damping parameter $\alpha$ for bcc Fe with $1\%$ and $5\%$ of impurities
of Os and Pt \cite{EMKK11,MKWE13}. One can see  a strong impact of impurities
on GD. In the case of $1\%$ of Pt in Fig. \ref{Fig-FePtOs} (a), $\alpha$
decreases in the low-temperature regime much steeper upon increasing the
temperature, indicating that the breathing Fermi surface mechanism dominates.
When the concentration of the impurities increases up to $5\%$
(Fig. \ref{Fig-FePtOs} (a)), the spin-flip 
scattering mechanism takes the leading role for the magnetization dissipation
practically for the whole region of temperatures under consideration. 
The different behavior of GD for Fe with Os and Pt is a result of the
different density of states (DOS) of the impurities at the Fermi energy
(see Ref.\ \onlinecite{MKWE13} for a discussion). 

The role of the electron-phonon scattering for the ultrafast
laser-induced demagnetization was investigated by Carva et
al. \cite{CBLO13} based on the Elliott-Yafet theory of spin relaxation
in metals, that puts the focus on spin-flip (SF) transitions upon the
electron-phonon scattering.
As the evaluation of the spin-dependent
electron-phonon matrix elements entering the expression for the rate
of the spin-flip transition is a demanding problem,
various approximations are used for this. In
particular, Carva et al. \cite{CBO11,CBLO13} use the so-called Elliott
approximation to evaluate a SF probability $P_S^b =
\frac{\tau}{\tau_{sf}}$ with the spin lifetime $\tau_{sf}$ and a
spin-diagonal lifetime $\tau$: 
\begin{eqnarray}
 P_S^b &=& \frac{\tau}{\tau_{sf}} = 4\langle b^2 \rangle
\label{eq_SF}
\end {eqnarray}
with the Fermi-surface averaged spin mixing of Bloch wave eigenstates  
\begin{eqnarray}
\langle b^2 \rangle &=& \sum_{\sigma,n} \int d^3k |b^\sigma_{\vec{k} n}|
                        \delta (E^\sigma_{\vec{k} n} - E_F)\,.
\label{eq_SF2}
\end {eqnarray}

In the case of a non-collinear magnetic structure, the description of
the Gilbert damping 
can be extended by adding higher-order
non-local contributions.
The role of non-local damping contributions has been investigated by
calculating the precession damping $\alpha(\vec{q})$ for magnons in FM
metals, characterized by a wave vector  $\vec{q}$. 
Following the same idea, Thonig et al.\ \cite{TKEP18} used a
torque-torque correlation model based on a tight binding approach,
and calculated the Gilbert damping for the itinerant-electron
ferromagnets Fe, Co 
and Ni, both  in the reciprocal, $\alpha(\vec{q})$, and real
$\alpha_{ij}$  space representations. The important role of non-local
contributions to the GD for spin dynamics has been demonstrated
using  atomistic magnetization dynamics simulations. 

A formalism for calculating the non-local contributions to the GD
has been recently worked out within the KKR Green function formalism \cite{MWE18}.
Using linear response theory for weakly-noncollinear magnetic
systems it gives access to the GD parameters represented as a function
of a wave vector $\vec{q}$.
Using the definition for the spin susceptibility tensor
$\chi_{\alpha\beta}(\vec{q}, \omega)$, the Fourier transformation of the
real-space Gilbert damping can be represented by the expression \cite{QV02, HVT07}
\begin{eqnarray}
  \alpha_{\alpha\beta}(\vec{q}) &=&  
  \frac{\gamma}{M_0V} \lim_{\omega \to 0} \frac{\partial  \Im
[\chi^{-1}]_{\alpha\beta}(\vec{q}, \omega)}{\partial \omega} \;.
\label{GD_HTV08}
\end{eqnarray}
%
Here $\gamma = g \mu_B$ is the gyromagnetic ratio, $M_0 =
\mu_{tot} \mu_B / V$ is the equilibrium
magnetization and $V$ is the volume of the system.
As is shown in Ref.\ \onlinecite{MWE18}, this expression can be
transformed to the form which allows an
expansion of GD in powers of wave vector $\vec{q}$: 
\begin{eqnarray}
{\underline \alpha}(\vec{q})  & = &  {\underline \alpha} 
+ \sum_\mu  \underline {\alpha}^\mu q_\mu 
+ \frac{1}{2} \sum_{\mu\nu} \underline {\alpha}^{\mu\nu}   q_\mu q_\nu
 + ... \,.
\label{GD_CHI_Q}
\end{eqnarray}
%
with the following expansion coefficients:
%
\begin{eqnarray}
\label{GD_CHI0}
\alpha^{0 \pm\pm}_{\alpha\alpha} & =  &  \frac{g}{\pi \mu_{tot}}
\frac{1}{\Omega_{BZ}} {\mbox{Tr}} \int d^3k  \bigg\langle
 \underline{T}_{\beta}   \underline{\tau}(\vec{k}, E_F^{\pm})
 \underline{T}_{\beta}   
 \underline{\tau}(\vec{k}, E_F^{\pm})  \bigg\rangle_c \nonumber
\\
%
\alpha_{\alpha\alpha}^{\mu \pm\pm} & = &  \frac{g}{\pi \mu_{tot}} 
\label{GD_CHI1}
\frac{1}{\Omega_{BZ}} {\mbox{Tr}} \int d^3k  \bigg\langle
 \underline{T}^{\beta}    \frac{\partial
  \underline{\tau}(\vec{k}, E_F^{\pm})}{\partial k_\mu} 
  \underline{T}^{\beta}    
  \underline{\tau}(\vec{k}, E_F^{\pm}) \bigg\rangle_c
  \nonumber \\
\alpha_{\alpha\alpha}^{\mu\nu \pm\pm} & = &
                                            - \frac{g}{2\pi\mu_{tot}}
                                            \frac{1}{\Omega_{BZ}}
                                            \nonumber \\
                                         && \times   {\mbox{Tr}
\int d^3k  } \bigg\langle \underline{T}^{\beta}   \frac{\partial
  \underline{\tau}(\vec{k}, E_F^\pm)}{\partial k_\mu} 
  \underline{T}^{\beta}  \frac{\partial
   \underline{\tau}(\vec{k}, E_F^\pm)}{\partial k_\nu} \bigg\rangle_c
   \; .
\end{eqnarray}
%

\begin{figure}[t]
\includegraphics[width=0.45\textwidth,angle=0,clip]{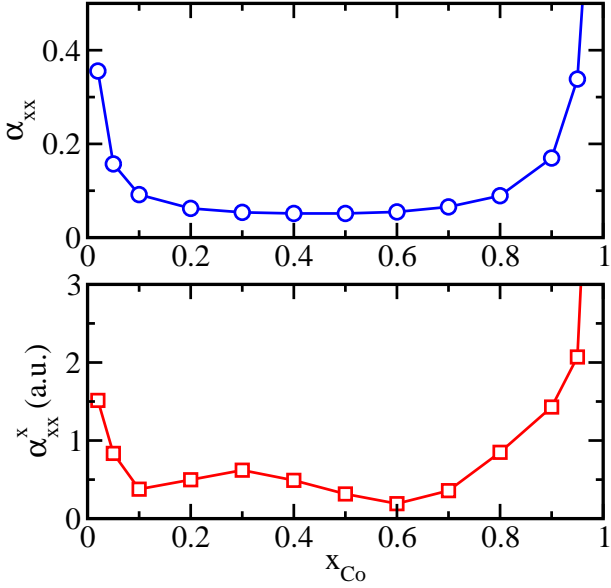}\;
\caption{\label{fig:GD1} The  Gilbert damping parameters 
$\alpha_{xx}$ (top) and $\alpha^x_{xx}$  (bottom) calculated 
for the model multilayer system (Cu/Fe$_{1-x}$Co$_x$/Pt)$_n$  using
first and second expressions in Eq.\ (\ref{GD_CHI1}), respectively. All
data are taken from Ref.\ \onlinecite{MWE18}. }     
\end{figure}

 For the prototype multilayer system (Cu/Fe$_{1-x}$Co$_x$/Pt)$_n$ 
the calculated  zero-order (uniform) GD parameter
 $\alpha_{xx}$ and the corresponding 
first-order (chiral) $\alpha^{x}_{xx}$ correction
 term  for $\vec{q}\, \| \hat{x}$ are plotted in
Fig.\ \ref{fig:GD1} top  and bottom, respectively,
as a function of the Co concentration $x$.
%
Both terms,
$\alpha_{xx}$  and $\alpha^x_{xx}$, increase approaching
the pure limits w.r.t.\ the  Fe$_{1-x}$Co$_x$ alloy subsystem. 
As is pointed out in Ref. \onlinecite{MWE18}, 
this increase is associated with the dominating
so-called breathing Fermi-surface damping mechanism
due to the modification of the Fermi surface (FS) 
 induced by the SOC, which 
 follows the magnetization direction that slowly varies with time. 
As $\underline\alpha$ is caused for a ferromagnet exclusively by the SOC
one can expect that it vanishes 
for vanishing SOC. This was indeed demonstrated before \cite{MKWE13}.
The same holds also for  $\underline\alpha^x$ that is caused by SOC as well.

Alternatively, a real-space extension for classical Gilbert damping tensor
was proposed recently by Brinker et al. \cite{BSL22}, by introducing 
two-site Gilbert damping tensor $\underline{\cal G}_{ij}$ entering the site-resolved LLG equation 
\begin{eqnarray}
\frac{1}{\gamma}\frac{d\vec{M}_i}{d\tau}  &= & 
-\gamma \vec{M}_i \times \bigg( \vec{H}_{i,\rm eff}
+ \sum_j \left[ \underline{\cal G}_{ij}(\vec{M})\cdot
\frac{d\vec{M}_i}{d\tau} \right] \bigg) \;,
\label{eq:LLG-Brinker}
\end{eqnarray}
which is related to the inverse dynamical susceptibility
$\underline{\chi}_{ij}$ via the expression 
\begin{eqnarray}
  \frac{d}{d \omega} \mbox{Im} [\underline{\chi}]^{\alpha\beta}_{ij} &=
  &
    \delta_{ij}\bigg(\frac{1}{\gamma
    M_i}\epsilon_{\alpha\beta\gamma}\bigg) + \bigg({\cal R}_i
    \underline{\cal G}_{ij} {\cal R}_j^T \bigg)_{\alpha\beta} \;,
\label{eq:GD-Brinker}
\end{eqnarray}
where ${\cal R}_i$ and ${\cal R}_j$ are the rotation matrices to go from
the global to the local frames of reference for atoms $i$ and $j$, respectively,
assuming a non-collinear magnetic ground state in the system.
Thus, an expression for the GD can be directly obtained using the adiabatic
approximation for the slow spin-dynamics processes.
This justifies the approximation $([\underline{\chi}]^{-1}(\omega))_\omega' \approx
([\underline{\chi}_0]^{-1}(\omega))_\omega'$, with the un-enhanced
dynamical susceptibility given in  terms of electronic Green function $G_{ij}$
\begin{eqnarray}
  \chi^{\alpha\beta}_{ij} (\omega + i\eta) &= & -\frac{1}{\pi} \mbox{Tr}
                                                \int^{E_F} dE   \nonumber \\
 && \bigg[\sigma^\alpha
                                                {G}_{ij}(E+ \omega +
                                                i\eta) \sigma^\beta
    \mbox{Im}{G}_{ij}(E)  \nonumber \\ 
 &&                                             + \sigma^\alpha
                                                {G}_{ij}(E)\sigma^\beta
                                                \mbox{Im}{G}_{ij}(E - \omega -
                                                i\eta)  \bigg]\,,
\label{eq:chi_ij}
\end{eqnarray}
 with the Green function $G(E \pm i\eta) = (E - {\cal H}  \pm
  i\eta)^{-1}$  corresponding to the Hamiltonian $\cal {H}$. 

  Moreover, this approach allows a multisite expansion of the GD
  accounting for higher-order non-local contributions for non-collinear
  structures \cite{BSL22}. 
  For this purpose, the Hamiltonian $\cal {H}$ is split into the
  on-site contribution $\cal {H}_0$ and the intersite hopping term
  $t_{ij}$, which is spin dependent in the general case. The GF
  can then be expanded in a perturbative way using the Dyson equation
\begin{eqnarray}
 {G}_{ij} &= &  {G}_{i}^0 \delta_{ij}  + {G}_{i}^0
               t_{ij}{G}_{j}^0  + {G}_{i}^0
               t_{ik}{G}_{k}^0 t_{kj}{G}_{j}^0  + ... \,.
\label{eq:chi-G_ij}
\end{eqnarray}
As a result, the authors generalize the LLG equation by splitting the
Gilbert damping tensor in terms proportional to scalar,
anisotropic, vector-chiral and scalar-chiral products of the magnetic
moments, i.e.\ terms like $\hat{e}_i \cdot \hat{e}_j$, $(\hat{n}_{ij}
\cdot \hat{e}_i)(\hat{n}_{ij} \cdot \hat{e}_j)$, 
$\hat{n}_{ij} \cdot (\hat{e}_i \times \hat{e}_j)$, etc.

 It should be stressed that the Gilbert damping parameter accounts for
the energy transfer connected with the magnetization dynamics but gives
no information on the angular momentum transfer that plays an important
role e.g. for ultrafast demagnetization processes. The formal basis to
account simultaneously for the spin and lattice degrees of freedom was
considered recently by A{\ss}mann and Nowak \cite{AN19} and Hellsvik
et al. \cite{HTM+19}. 
Hellsvik et al. \cite{HTM+19,SBK+22} report on an approach
solving simultaneously the equations for spin and lattice
 dynamics, accounting for spin-lattice interactions in the
 Hamiltonian, calculated on a first-principles level.  These interactions
 appear as a correction  
 to the exchange coupling parameters due to atomic
 displacements. As a  result, this leads to the three-body spin-lattice 
 coupling  parameters $\Gamma_{ijk}^{\alpha\beta\mu} =
 \frac{\partial J^{\alpha\beta}_{ij}}{\partial u^\mu_k}$ and four-body
 parameters $\Lambda_{ijkl}^{\alpha\beta\mu\nu} =
 \frac{\partial J^{\alpha\beta}_{ij}}{\partial u^\mu_k \partial
   u^\nu_l}$ represented by rank 3 and rank 4 tensors, respectively,
 entering the spin-lattice Hamiltonian 
\begin{eqnarray}
{\cal H}_{sl} &=&  - \frac{1}{2}\sum_{i,j,k,\alpha \beta,\mu}
                  \Gamma_{ijk}^{\alpha\beta\mu} e_i^{\alpha}e_j^{\beta}
                  u^{\mu}_k \nonumber \\
 &&  - \frac{1}{4}\sum_{i,j,k,l,\alpha \beta,\mu,\nu}
                  \Lambda_{ijkl}^{\alpha\beta\mu\nu} e_i^{\alpha}e_j^{\beta}
                  u^{\mu}_k u^{\nu}_l               \;.
\label{eq:Hamilt_extended_magneto-elastic-Hellsvik}
\end{eqnarray}
The parameters $\Gamma_{ijk}^{\alpha\beta\mu}$  in Ref.\
\onlinecite{HTM+19} are calculated using a finite difference  method,
using the exchange coupling parameters $\underline{J}_{ij}$ for the
system without displacements ($\underline{J}^0_{ij}$)  and with a 
displaced atom $k$  ($\underline{J}^{\Delta}_{ij}(\vec{u}_k)$), used to
estimate the coefficient $\Gamma_{ijk}^{\alpha\beta\mu} \approx
\frac{(\underline{J}^{\Delta}_{ij}(\vec{u}_k) - \underline{J}^0_{ij})}{u_\mu}$.

Alternatively, to describe the coupling of spin and spatial degrees of
freedom the present authors (see Ref.\ \onlinecite{MPL+22}) adopt an atomistic
approach and start with the expansion of a 
phenomenological spin-lattice Hamiltonian  
\begin{eqnarray}
{\cal H}_{sl} &=&   -  \sum_{i,j,\alpha,\beta} \sum_{k,\mu} J_{ij,k}^{\alpha
     \beta,\mu} 
                   e_i^{\alpha}e_j^{\beta}  
                   u^{\mu}_k
\nonumber \\ 
  &&  
\qquad \qquad \quad  
  -  \sum_{i,j}\sum_{k,l}  J_{ij,kl}^{\alpha
     \beta,\mu\nu}  
                   e_i^{\alpha}e_j^{\beta}  
                    u^{\mu}_k u^{\nu}_l
                   \;,
\label{eq:Hamilt_extended_magneto-elastic}
\end{eqnarray}
%
that can be seen as a lattice extension of a Heisenberg model.
Accordingly, the spin and lattice degrees of freedom
are represented by the orientation vectors $\hat{e}_i$ of the
magnetic moments $\vec m_i$ and displacement  vectors $\vec u_i$ for each 
atomic site $i$. 
The spin-lattice Hamiltonian in Eq.\
(\ref{eq:Hamilt_extended_magneto-elastic}) is restricted to three and
four-site terms.
As  relativistic effects are taken into account, the SLC is described
in  tensorial form with $J_{ij,k}^{\alpha \beta,\mu}$ and
$J_{ij,kl}^{\alpha \beta,\mu\nu}$ represented by rank 3 and rank 4
tensors, similar to those discussed by Hellsvik et al. \cite{HTM+19}.

The same strategy as for the exchange coupling parameters $J_{ij}$
\cite{LKAG87} or $J_{ij}^{\alpha\beta}$ \cite{USPW03,EM09a}, is used to
map the free energy landscape ${\cal F}(\{\hat{e}_i\},\{\vec u_i\})$
accounting for its dependence on the spin configuration $\{\hat{e}_i\}$
as well as  atomic displacements $\{\vec u_i\}$, making use of the
magnetic force theorem and the Lloyd formula to evaluate integrated DOS
$\Delta N(E)$. 
With this, the free energy change due to any perturbation in the system is given by
 Eq.\ (\ref{eq:Free_energy-22}).

Using as a reference the ferromagnetically ordered state of the system with
a non-distorted lattice, 
and the perturbed state characterized by finite  spin tiltings $\delta \hat{e}_i$
and finite  atomic displacements $\vec{u}_i$ at site $i$, one can write
the corresponding changes
of the inverse $t$-matrix as $\Delta^s_{\mu} \underline{m}_i =
\underline{m}_i(\delta \hat{e}_i^\mu) - \underline{m}^0_i $ 
and  $\Delta_{\nu}^u \underline{m}_i   = \underline{m}_i({u}_i^\nu) -
\underline{m}^0_i $. This allows to replace the integrand in Eq.\
(\ref{eq:Free_energy-2}) by 
%
\begin{eqnarray}
  \mbox{ln} \,\underline{\underline{\tau}}
- \mbox{ln} \,\underline{\underline{\tau}}^0
  &=& - \ln \Big(1 +
       \underline{\underline{\tau}}\,[\Delta^s_{\mu} \underline{\underline{m}}_i + \Delta_{\nu}^u
                        \underline{\underline{m}}_j +  ... ] \Big) \; ,
\label{eq:tau2}
\end{eqnarray}
where all site-dependent changes in the spin configuration
$\{\hat{e}_i\}$ and atomic positions  $\{\vec u_i\}$ 
are accounted for in a one-to-one manner by the various terms on the
right hand side.
Due to the use of the magnetic force theorem these blocks may be written 
in terms of the  spin tiltings $\delta \hat{e}_i^\mu$ and  atomic displacements of the atoms ${u}_i^\nu$ 
together with the corresponding auxiliary matrices
$\underline{T}^{\mu}_i $ and ${\cal U}_{i}^{\nu}$, respectively, 
as
\begin{eqnarray}
  \Delta^s_{\mu} {\underline{m}}_i & = &  \delta \hat{e}^\mu_i\,
       \underline{T}^{\mu}_i \,, \\
  \Delta^{u}_{\nu} \underline{m}_i & = & 
  u^\nu_i  \underline{\cal U}_{i}^{\nu}  \,.
                                         \label{eq:linear_distor}
\end{eqnarray}
%
Inserting these expressions into Eq.\ (\ref{eq:tau2}) and the result in
turn into Eq.\ (\ref{eq:Free_energy-22}) 
allows us to calculate the parameters of the
spin-lattice Hamiltonian as the derivatives of the free energy with
respect to tilting angles and displacements.
This way one gets for example for the three-site term:
%
\begin{eqnarray}
  {\cal J}^{\alpha\beta,\mu}_{ij,k} &=& \frac{\partial^3 {\cal F}}{\partial e^\alpha_i \,\partial e^\beta_j
                                        \, \partial u^{\mu}_k} = \frac{1}{2\pi} \mbox{Im\, Tr\,}
                                  \int^{E_F}dE \,\, \nonumber \\
   &
                          \times  & \Big[   \underline{T}^{\alpha}_i \,\underline{\tau}_{ij}
                                  \underline{T}^{\beta}_j 
                                 \, \underline{\tau}_{jk}
                                \underline{\cal U}^{\mu}_k \,\underline{\tau}_{ki}
                             +  \underline{T}^{\alpha}_i
                                   \,\underline{\tau}_{ik}
       \underline{\cal U}^{\mu}_k \,\underline{\tau}_{kj}
                                  \underline{T}^{\beta}_j 
                                 \, \underline{\tau}_{ji}\Big] \;\;
       \label{eq:Parametes_linear} 
\end{eqnarray}
and for the four-site term:
\begin{eqnarray}
  {\cal J}^{\alpha\beta,\mu\nu}_{ij,kl} &=& \frac{\partial^4 {\cal
                                F}}{\partial e^\alpha_i \,\partial e^\beta_j
                                \, \partial u^{\mu}_k \partial
                                            u^{\nu}_l} = \frac{1}{4\pi} \mbox{Im\, Tr\,}
      \int^{E_F}dE \, \,  \nonumber \\
  & &  \times \Bigg[    \underline{\cal U}^{\mu}_k\, \underline{\tau}_{ki} \,
                                  \underline{T}^{\alpha}_i \,\underline{\tau}_{ij}\,
                                  \underline{T}^{\beta}_j \,
                                 \, \underline{\tau}_{jl}\,
      \underline{\cal U}^{\nu}_l \,\underline{\tau}_{lk} \,\nonumber \\
  & &+            \underline{T}^{\alpha}_i \,\underline{\tau}_{ik}\,
                        \underline{\cal U}^{\mu}_k\, \underline{\tau}_{kj} \,
                                  \underline{T}^{\beta}_j \,
                                 \, \underline{\tau}_{jl}\,
      \underline{\cal U}^{\nu}_l \,\underline{\tau}_{li} \,\nonumber \\
  & &+           \underline{\cal U}^{\mu}_k\, \underline{\tau}_{ki} \,
                                  \underline{T}^{\alpha}_i
                                 \,\underline{\tau}_{il}\,      
      \underline{\cal U}^{\nu}_l \,\underline{\tau}_{lj} \,
                                  \underline{T}^{\beta}_j \,
                                 \, \underline{\tau}_{jk}\, \nonumber \\                                        
  & &+            \underline{T}^{\alpha}_i \,\underline{\tau}_{ik}\,
                        \underline{\cal U}^{\mu}_k\,
      \underline{\tau}_{kl} \,
      \underline{\cal U}^{\nu}_l \,\underline{\tau}_{lj} \,
                                  \underline{T}^{\beta}_j \,
                                 \, \underline{\tau}_{ji} \Bigg] \,.
\label{eq:Parametes_quadratic}
\end{eqnarray}

\begin{figure}[t]
\includegraphics[width=0.44\textwidth,angle=0]{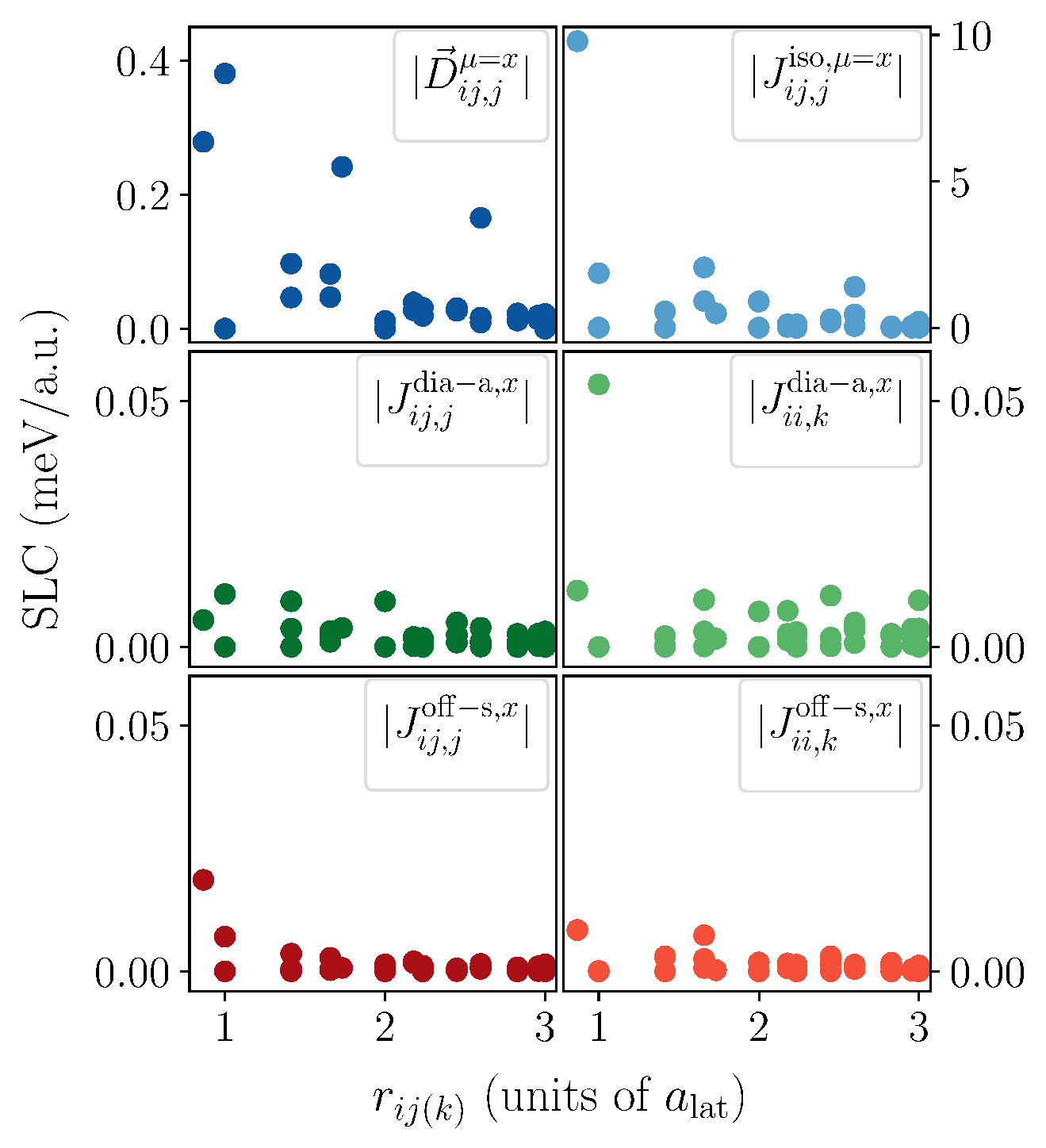}\,
\caption{\label{fig:DSLC_vs_Rij}  
  The absolute values of site-off-diagonal and site-diagonal SLC
  parameters: DMI $|\vec{\cal D}^x_{ij,j}|$ and isotropic SLC ${\cal
    J}^{\mathrm{iso},x}_{ij,j}$ (top), anti-symmetric diagonal
  components ${\cal J}^{\mathrm{dia-a},x}_{ij,j}$ and ${\cal
    J}^{\mathrm{dia-a},x}_{ii,k}$ (middle), and symmetric off-diagonal
  components ${\cal J}^{\mathrm{off-s},x}_{ij,j}$ and ${\cal
    J}^{\mathrm{off-s},x}_{ii,k}$ (bottom) for bcc Fe, as a function of
  the interatomic distance $r_{ij}$
}     
\end{figure}

Fig. \ref{fig:DSLC_vs_Rij} shows corresponding results for the SLC parameters of
bcc Fe, plotted as a function of the 
distance $r_{ij}$ for $i = k$ which implies that a displacement along the
$x$ direction is applied for one of the interacting atoms. 
The absolute values of the DMI-like SLC parameters (DSLC) $|\vec{\cal
  D}|^{\mu=x}_{ij,k}$ (note that ${\cal D}^{z,\mu}_{ij,k} =
\frac{1}{2}({\cal J}^{xy,\mu}_{ij,k} - {\cal J}^{yx,\mu}_{ij,k})$ )
show a rather slow decay with the distance $r_{ij}$. The isotropic SLC parameters  ${\cal
  J}^{\mathrm{iso},\mu=x}_{ij,j}$, which have only a weak dependence on the SOC, are about one order of magnitude larger
than the DSLC. All other SOC-driven parameters shown in Fig. \ref{fig:DSLC_vs_Rij}, characterizing the
displacement-induced contributions to MCA, are much smaller than the DSLC.

\section{Summary}

To summarize, we have considered a multi-level atomistic approach
commonly used to simulate finite temperature and dynamical magnetic
properties of  
solids, avoiding in particular time-consuming TD-SDFT calculations.
The approach is based on a phenomenological parameterized spin Hamiltonian
which allows to separate the spin and orbital degrees of
freedom and that way to avoid the demanding treatment of complex
spin-dependent many-body effects. As these parameters are fully determined by the
electronic structure of a system, they can be deduced from the
information provided by relativistic band structure calculations
based on SDFT.
We gave a short overview of the various methods
to calculate these parameters  entering for example the LLG
equation. It is shown that the KKR Green function formalism
is one of the most powerful band structure methods as it gives
straightforward access to
practically all parameters of the phenomenological models.
It allows in particular to add in a very simple way further extensions
to the model Hamiltonians,
 accounting for example for multi-site interaction terms.
 Another important issue  are spin-lattice interactions,
 that couple the degrees of freedom of the spin and
 lattice subsystems.
 The key role of the SOC for the interaction parameters
 is pointed out as it gives not only rise to the MCA but also
 to the Gilbert damping as well as  the anisotropy of the
exchange coupling  and spin-lattice interaction
with many important physical phenomena connected
to these.


\begin{thebibliography}{90}%
\makeatletter
\providecommand \@ifxundefined [1]{%
 \@ifx{#1\undefined}
}%
\providecommand \@ifnum [1]{%
 \ifnum #1\expandafter \@firstoftwo
 \else \expandafter \@secondoftwo
 \fi
}%
\providecommand \@ifx [1]{%
 \ifx #1\expandafter \@firstoftwo
 \else \expandafter \@secondoftwo
 \fi
}%
\providecommand \natexlab [1]{#1}%
\providecommand \enquote  [1]{``#1''}%
\providecommand \bibnamefont  [1]{#1}%
\providecommand \bibfnamefont [1]{#1}%
\providecommand \citenamefont [1]{#1}%
\providecommand \href@noop [0]{\@secondoftwo}%
\providecommand \href [0]{\begingroup \@sanitize@url \@href}%
\providecommand \@href[1]{\@@startlink{#1}\@@href}%
\providecommand \@@href[1]{\endgroup#1\@@endlink}%
\providecommand \@sanitize@url [0]{\catcode `\\12\catcode `\$12\catcode
  `\&12\catcode `\#12\catcode `\^12\catcode `\_12\catcode `\%12\relax}%
\providecommand \@@startlink[1]{}%
\providecommand \@@endlink[0]{}%
\providecommand \url  [0]{\begingroup\@sanitize@url \@url }%
\providecommand \@url [1]{\endgroup\@href {#1}{\urlprefix }}%
\providecommand \urlprefix  [0]{URL }%
\providecommand \Eprint [0]{\href }%
\providecommand \doibase [0]{http://dx.doi.org/}%
\providecommand \selectlanguage [0]{\@gobble}%
\providecommand \bibinfo  [0]{\@secondoftwo}%
\providecommand \bibfield  [0]{\@secondoftwo}%
\providecommand \translation [1]{[#1]}%
\providecommand \BibitemOpen [0]{}%
\providecommand \bibitemStop [0]{}%
\providecommand \bibitemNoStop [0]{.\EOS\space}%
\providecommand \EOS [0]{\spacefactor3000\relax}%
\providecommand \BibitemShut  [1]{\csname bibitem#1\endcsname}%
\let\auto@bib@innerbib\@empty
\bibitem [{\citenamefont {Engel}\ and\ \citenamefont {Dreizler}(2011)}]{ED11}%
  \BibitemOpen
  \bibfield  {author} {\bibinfo {author} {\bibfnamefont {E.}~\bibnamefont
  {Engel}}\ and\ \bibinfo {author} {\bibfnamefont {R.~M.}\ \bibnamefont
  {Dreizler}},\ }\href {\doibase 10.1007/978-3-642-14090-7}  {\bibinfo
  {title} {Density Functional Theory -- An advanced course}}\ (\bibinfo
  {publisher} {Springer},\ \bibinfo {address} {Berlin},\ \bibinfo {year}
  {2011})\BibitemShut {NoStop}%
\bibitem [{\citenamefont {Krieger}\ \emph {et~al.}(2015)\citenamefont
  {Krieger}, \citenamefont {Dewhurst}, \citenamefont {Elliott}, \citenamefont
  {Sharma},\ and\ \citenamefont {Gross}}]{KDE+15a}%
  \BibitemOpen
  \bibfield  {author} {\bibinfo {author} {\bibfnamefont {K.}~\bibnamefont
  {Krieger}}, \bibinfo {author} {\bibfnamefont {J.~K.}\ \bibnamefont
  {Dewhurst}}, \bibinfo {author} {\bibfnamefont {P.}~\bibnamefont {Elliott}},
  \bibinfo {author} {\bibfnamefont {S.}~\bibnamefont {Sharma}}, \ and\ \bibinfo
  {author} {\bibfnamefont {E.~K.~U.}\ \bibnamefont {Gross}},\ }\href {\doibase
  10.1021/acs.jctc.5b00621} {\bibfield  {journal} {\bibinfo  {journal} {Journal
  of Chemical Theory and Computation}\ }\textbf {\bibinfo {volume} {11}},\
  \bibinfo {pages} {4870} (\bibinfo {year} {2015})}\BibitemShut {NoStop}%
\bibitem [{\citenamefont {Liechtenstein}\ \emph {et~al.}(1984)\citenamefont
  {Liechtenstein}, \citenamefont {Katsnelson},\ and\ \citenamefont
  {Gubanov}}]{LKG84}%
  \BibitemOpen
  \bibfield  {author} {\bibinfo {author} {\bibfnamefont {A.~I.}\ \bibnamefont
  {Liechtenstein}}, \bibinfo {author} {\bibfnamefont {M.~I.}\ \bibnamefont
  {Katsnelson}}, \ and\ \bibinfo {author} {\bibfnamefont {V.~A.}\ \bibnamefont
  {Gubanov}},\ }\href {\doibase 10.1088/0305-4608/14/7/007} {\bibfield
  {journal} {\bibinfo  {journal} {J. Phys. F: Met. Phys.}\ }\textbf {\bibinfo
  {volume} {14}},\ \bibinfo {pages} {L125} (\bibinfo {year}
  {1984})}\BibitemShut {NoStop}%
\bibitem [{\citenamefont {Liechtenstein}\ \emph {et~al.}(1987)\citenamefont
  {Liechtenstein}, \citenamefont {Katsnelson}, \citenamefont {Antropov},\ and\
  \citenamefont {Gubanov}}]{LKAG87}%
  \BibitemOpen
  \bibfield  {author} {\bibinfo {author} {\bibfnamefont {A.~I.}\ \bibnamefont
  {Liechtenstein}}, \bibinfo {author} {\bibfnamefont {M.~I.}\ \bibnamefont
  {Katsnelson}}, \bibinfo {author} {\bibfnamefont {V.~P.}\ \bibnamefont
  {Antropov}}, \ and\ \bibinfo {author} {\bibfnamefont {V.~A.}\ \bibnamefont
  {Gubanov}},\ }\href {\doibase 10.1016/0304-8853(87)90721-9} {\bibfield
  {journal} {\bibinfo  {journal} {J. Magn. Magn. Materials}\ }\textbf {\bibinfo
  {volume} {67}},\ \bibinfo {pages} {65} (\bibinfo {year} {1987})}\BibitemShut
  {NoStop}%
\bibitem [{\citenamefont {Udvardi}\ \emph {et~al.}(2003)\citenamefont
  {Udvardi}, \citenamefont {Szunyogh}, \citenamefont {Palot\'as},\ and\
  \citenamefont {Weinberger}}]{USPW03}%
  \BibitemOpen
  \bibfield  {author} {\bibinfo {author} {\bibfnamefont {L.}~\bibnamefont
  {Udvardi}}, \bibinfo {author} {\bibfnamefont {L.}~\bibnamefont {Szunyogh}},
  \bibinfo {author} {\bibfnamefont {K.}~\bibnamefont {Palot\'as}}, \ and\
  \bibinfo {author} {\bibfnamefont {P.}~\bibnamefont {Weinberger}},\ }\href
  {\doibase 10.1103/PhysRevB.68.104436} {\bibfield  {journal} {\bibinfo
  {journal} {Phys. Rev. B}\ }\textbf {\bibinfo {volume} {68}},\ \bibinfo
  {pages} {104436} (\bibinfo {year} {2003})}\BibitemShut {NoStop}%
\bibitem [{\citenamefont {Ebert}\ and\ \citenamefont
  {Mankovsky}(2009)}]{EM09a}%
  \BibitemOpen
  \bibfield  {author} {\bibinfo {author} {\bibfnamefont {H.}~\bibnamefont
  {Ebert}}\ and\ \bibinfo {author} {\bibfnamefont {S.}~\bibnamefont
  {Mankovsky}},\ }\href {\doibase 10.1103/PhysRevB.79.045209} {\bibfield
  {journal} {\bibinfo  {journal} {Phys. Rev. B}\ }\textbf {\bibinfo {volume}
  {79}},\ \bibinfo {pages} {045209} (\bibinfo {year} {2009})}\BibitemShut
  {NoStop}%
\bibitem [{\citenamefont {Heide}\ \emph {et~al.}(2008)\citenamefont {Heide},
  \citenamefont {Bihlmayer},\ and\ \citenamefont {Bl\"ugel}}]{HBB08}%
  \BibitemOpen
  \bibfield  {author} {\bibinfo {author} {\bibfnamefont {M.}~\bibnamefont
  {Heide}}, \bibinfo {author} {\bibfnamefont {G.}~\bibnamefont {Bihlmayer}}, \
  and\ \bibinfo {author} {\bibfnamefont {S.}~\bibnamefont {Bl\"ugel}},\ }\href
  {\doibase 10.1103/PhysRevB.78.140403} {\bibfield  {journal} {\bibinfo
  {journal} {Phys. Rev. B}\ }\textbf {\bibinfo {volume} {78}},\ \bibinfo
  {pages} {140403} (\bibinfo {year} {2008})}\BibitemShut {NoStop}%
\bibitem [{\citenamefont {Heide}\ \emph {et~al.}(2009)\citenamefont {Heide},
  \citenamefont {Bihlmayer},\ and\ \citenamefont {Blugel}}]{HBB09}%
  \BibitemOpen
  \bibfield  {author} {\bibinfo {author} {\bibfnamefont {M.}~\bibnamefont
  {Heide}}, \bibinfo {author} {\bibfnamefont {G.}~\bibnamefont {Bihlmayer}}, \
  and\ \bibinfo {author} {\bibfnamefont {S.}~\bibnamefont {Blugel}},\ }\href
  {\doibase http://dx.doi.org/10.1016/j.physb.2009.06.070} {\bibfield
  {journal} {\bibinfo  {journal} {Physica B: Condensed Matter}\ }\textbf
  {\bibinfo {volume} {404}},\ \bibinfo {pages} {2678 } (\bibinfo {year}
  {2009})},\ \bibinfo {note} {proceedings of the Workshop - Current Trends and
  Novel Materials}\BibitemShut {NoStop}%
\bibitem [{\citenamefont {Rusz}\ \emph {et~al.}(2006)\citenamefont {Rusz},
  \citenamefont {Bergqvist}, \citenamefont {Kudrnovsk\'y},\ and\ \citenamefont
  {Turek}}]{RBKT06}%
  \BibitemOpen
  \bibfield  {author} {\bibinfo {author} {\bibfnamefont {J.}~\bibnamefont
  {Rusz}}, \bibinfo {author} {\bibfnamefont {L.}~\bibnamefont {Bergqvist}},
  \bibinfo {author} {\bibfnamefont {J.}~\bibnamefont {Kudrnovsk\'y}}, \ and\
  \bibinfo {author} {\bibfnamefont {I.}~\bibnamefont {Turek}},\ }\href
  {\doibase 10.1103/PhysRevB.73.214412} {\bibfield  {journal} {\bibinfo
  {journal} {Phys. Rev. B}\ }\textbf {\bibinfo {volume} {73}},\ \bibinfo
  {pages} {214412} (\bibinfo {year} {2006})}\BibitemShut {NoStop}%
\bibitem [{\citenamefont {Antropov}\ \emph {et~al.}(1996)\citenamefont
  {Antropov}, \citenamefont {Katsnelson}, \citenamefont {Harmon}, \citenamefont
  {van Schilfgaarde},\ and\ \citenamefont {Kusnezov}}]{AKH+96}%
  \BibitemOpen
  \bibfield  {author} {\bibinfo {author} {\bibfnamefont {V.~P.}\ \bibnamefont
  {Antropov}}, \bibinfo {author} {\bibfnamefont {M.~I.}\ \bibnamefont
  {Katsnelson}}, \bibinfo {author} {\bibfnamefont {B.~N.}\ \bibnamefont
  {Harmon}}, \bibinfo {author} {\bibfnamefont {M.}~\bibnamefont {van
  Schilfgaarde}}, \ and\ \bibinfo {author} {\bibfnamefont {D.}~\bibnamefont
  {Kusnezov}},\ }\href {\doibase 10.1103/PhysRevB.54.1019} {\bibfield
  {journal} {\bibinfo  {journal} {Phys. Rev. B}\ }\textbf {\bibinfo {volume}
  {54}},\ \bibinfo {pages} {1019} (\bibinfo {year} {1996})}\BibitemShut
  {NoStop}%
\bibitem [{\citenamefont {Eriksson}\ \emph {et~al.}(2022)\citenamefont
  {Eriksson}, \citenamefont {Bergman}, \citenamefont {Bergqvist},\ and\
  \citenamefont {Hellsvik}}]{EBBH22}%
  \BibitemOpen
  \bibfield  {author} {\bibinfo {author} {\bibfnamefont {O.}~\bibnamefont
  {Eriksson}}, \bibinfo {author} {\bibfnamefont {A.}~\bibnamefont {Bergman}},
  \bibinfo {author} {\bibfnamefont {L.}~\bibnamefont {Bergqvist}}, \ and\
  \bibinfo {author} {\bibfnamefont {J.}~\bibnamefont {Hellsvik}},\ }
  {}  {\bibinfo {title} {Atomistic Spin Dynamics: Foundations and
  Applications.}}\ (\bibinfo  {publisher} {Oxford University Press},\ \bibinfo
  {year} {2022})\BibitemShut {NoStop}%
\bibitem [{\citenamefont {Bornemann}\ \emph {et~al.}(2012)\citenamefont
  {Bornemann}, \citenamefont {Min\'{a}r}, \citenamefont {Braun}, \citenamefont
  {K\"odderitzsch},\ and\ \citenamefont {Ebert}}]{BMB+12}%
  \BibitemOpen
  \bibfield  {author} {\bibinfo {author} {\bibfnamefont {S.}~\bibnamefont
  {Bornemann}}, \bibinfo {author} {\bibfnamefont {J.}~\bibnamefont
  {Min\'{a}r}}, \bibinfo {author} {\bibfnamefont {J.}~\bibnamefont {Braun}},
  \bibinfo {author} {\bibfnamefont {D.}~\bibnamefont {K\"odderitzsch}}, \ and\
  \bibinfo {author} {\bibfnamefont {H.}~\bibnamefont {Ebert}},\ }\href
  {\doibase 10.1016/j.ssc.2011.11.001} {\bibfield  {journal} {\bibinfo
  {journal} {Solid State Commun.}\ }\textbf {\bibinfo {volume} {152}},\
  \bibinfo {pages} {85} (\bibinfo {year} {2012})}\BibitemShut {NoStop}%
\bibitem [{\citenamefont {Bl\"ugel}(1999)}]{Blu99}%
  \BibitemOpen
  \bibfield  {author} {\bibinfo {author} {\bibfnamefont {S.}~\bibnamefont
  {Bl\"ugel}},\ }in\ \href@noop {}  {\bibinfo {booktitle} {30.\
  Ferienkurs des Instituts f\"ur Festk\"orperforschung 1999 ''Magnetische
  Schichtsysteme''}},\ \bibinfo {editor} {edited by\ \bibinfo {editor}
  {\bibnamefont {{\protect Institut f\"ur Festk\"orperforschung}}}}\ (\bibinfo
  {publisher} {Forschungszentrum J\"ulich GmbH},\ \bibinfo {address}
  {J\"ulich},\ \bibinfo {year} {1999})\ p.\ \bibinfo {pages} {C1.1}\BibitemShut
  {NoStop}%
\bibitem [{\citenamefont {Razee}\ \emph {et~al.}(1997)\citenamefont {Razee},
  \citenamefont {Staunton},\ and\ \citenamefont {Pinski}}]{RSP97}%
  \BibitemOpen
  \bibfield  {author} {\bibinfo {author} {\bibfnamefont {S.~S.~A.}\
  \bibnamefont {Razee}}, \bibinfo {author} {\bibfnamefont {J.~B.}\ \bibnamefont
  {Staunton}}, \ and\ \bibinfo {author} {\bibfnamefont {F.~J.}\ \bibnamefont
  {Pinski}},\ }\href {\doibase 10.1103/PhysRevB.56.8082} {\bibfield  {journal}
  {\bibinfo  {journal} {Phys. Rev. B}\ }\textbf {\bibinfo {volume} {56}},\
  \bibinfo {pages} {8082} (\bibinfo {year} {1997})}\BibitemShut {NoStop}%
\bibitem [{\citenamefont {Rose}(1961)}]{Ros61}%
  \BibitemOpen
  \bibfield  {author} {\bibinfo {author} {\bibfnamefont {M.~E.}\ \bibnamefont
  {Rose}},\ }\href
  {http://openlibrary.org/works/OL3517103W/Relativistic_electron_theory} 
  {\bibinfo {title} {Relativistic Electron Theory}}\ (\bibinfo  {publisher}
  {Wiley},\ \bibinfo {address} {New York},\ \bibinfo {year} {1961})\BibitemShut
  {NoStop}%
\bibitem [{\citenamefont {{H.\ Ebert et al.}}(2020)}]{SPR-KKR8.5}%
  \BibitemOpen
  \bibfield  {author} {\bibinfo {author} {\bibnamefont {{H.\ Ebert et al.}}},\
  }\href {https://www.ebert.cup.uni-muenchen.de/en/software-en/13-sprkkr}
  {}\bibinfo {howpublished} {{ The Munich SPR-KKR package}, version 8.5,
  https://www.ebert.cup.uni-muenchen.de/en/software-en/13-sprkkr} (\bibinfo
  {year} {2020})\BibitemShut {NoStop}%
\bibitem [{\citenamefont {Ebert}\ \emph
  {et~al.}(2011{\natexlab{a}})\citenamefont {Ebert}, \citenamefont
  {K\"odderitzsch},\ and\ \citenamefont {Min\'{a}r}}]{EKM11}%
  \BibitemOpen
  \bibfield  {author} {\bibinfo {author} {\bibfnamefont {H.}~\bibnamefont
  {Ebert}}, \bibinfo {author} {\bibfnamefont {D.}~\bibnamefont
  {K\"odderitzsch}}, \ and\ \bibinfo {author} {\bibfnamefont {J.}~\bibnamefont
  {Min\'{a}r}},\ }\href {\doibase 10.1088/0034-4885/74/9/096501} {\bibfield
  {journal} {\bibinfo  {journal} {Rep. Prog. Phys.}\ }\textbf {\bibinfo
  {volume} {74}},\ \bibinfo {pages} {096501} (\bibinfo {year}
  {2011}{\natexlab{a}})}\BibitemShut {NoStop}%
\bibitem [{\citenamefont {Ebert}\ \emph {et~al.}(2016)\citenamefont {Ebert},
  \citenamefont {Braun}, \citenamefont {K\"odderitzsch},\ and\ \citenamefont
  {Mankovsky}}]{EBKM16}%
  \BibitemOpen
  \bibfield  {author} {\bibinfo {author} {\bibfnamefont {H.}~\bibnamefont
  {Ebert}}, \bibinfo {author} {\bibfnamefont {J.}~\bibnamefont {Braun}},
  \bibinfo {author} {\bibfnamefont {D.}~\bibnamefont {K\"odderitzsch}}, \ and\
  \bibinfo {author} {\bibfnamefont {S.}~\bibnamefont {Mankovsky}},\ }\href
  {\doibase 10.1103/PhysRevB.93.075145} {\bibfield  {journal} {\bibinfo
  {journal} {Phys. Rev. B}\ }\textbf {\bibinfo {volume} {93}},\ \bibinfo
  {pages} {075145} (\bibinfo {year} {2016})}\BibitemShut {NoStop}%
\bibitem [{\citenamefont {Staunton}\ \emph {et~al.}(2006)\citenamefont
  {Staunton}, \citenamefont {Szunyogh}, \citenamefont {Buruzs}, \citenamefont
  {Gyorffy}, \citenamefont {Ostanin},\ and\ \citenamefont {Udvardi}}]{SSB+06}%
  \BibitemOpen
  \bibfield  {author} {\bibinfo {author} {\bibfnamefont {J.~B.}\ \bibnamefont
  {Staunton}}, \bibinfo {author} {\bibfnamefont {L.}~\bibnamefont {Szunyogh}},
  \bibinfo {author} {\bibfnamefont {A.}~\bibnamefont {Buruzs}}, \bibinfo
  {author} {\bibfnamefont {B.~L.}\ \bibnamefont {Gyorffy}}, \bibinfo {author}
  {\bibfnamefont {S.}~\bibnamefont {Ostanin}}, \ and\ \bibinfo {author}
  {\bibfnamefont {L.}~\bibnamefont {Udvardi}},\ }\href {\doibase
  10.1103/PhysRevB.74.144411} {\bibfield  {journal} {\bibinfo  {journal} {Phys.
  Rev. B}\ }\textbf {\bibinfo {volume} {74}},\ \bibinfo {pages} {144411}
  (\bibinfo {year} {2006})}\BibitemShut {NoStop}%
\bibitem [{\citenamefont {Gyorffy}\ \emph {et~al.}(1985)\citenamefont
  {Gyorffy}, \citenamefont {Pindor}, \citenamefont {Staunton}, \citenamefont
  {Stocks},\ and\ \citenamefont {Winter}}]{GPS+85}%
  \BibitemOpen
  \bibfield  {author} {\bibinfo {author} {\bibfnamefont {B.~L.}\ \bibnamefont
  {Gyorffy}}, \bibinfo {author} {\bibfnamefont {A.~J.}\ \bibnamefont {Pindor}},
  \bibinfo {author} {\bibfnamefont {J.}~\bibnamefont {Staunton}}, \bibinfo
  {author} {\bibfnamefont {G.~M.}\ \bibnamefont {Stocks}}, \ and\ \bibinfo
  {author} {\bibfnamefont {H.}~\bibnamefont {Winter}},\ }\href {\doibase
  10.1088/0305-4608/15/6/018} {\bibfield  {journal} {\bibinfo  {journal} {J.
  Phys. F: Met. Phys.}\ }\textbf {\bibinfo {volume} {15}},\ \bibinfo {pages}
  {1337} (\bibinfo {year} {1985})}\BibitemShut {NoStop}%
\bibitem [{\citenamefont {Soven}(1967)}]{Sov67}%
  \BibitemOpen
  \bibfield  {author} {\bibinfo {author} {\bibfnamefont {P.}~\bibnamefont
  {Soven}},\ }\href {\doibase 10.1103/PhysRev.156.809} {\bibfield  {journal}
  {\bibinfo  {journal} {Phys. Rev.}\ }\textbf {\bibinfo {volume} {156}},\
  \bibinfo {pages} {809} (\bibinfo {year} {1967})}\BibitemShut {NoStop}%
\bibitem [{\citenamefont {Staunton}\ \emph {et~al.}(2000)\citenamefont
  {Staunton}, \citenamefont {Poulter}, \citenamefont {Ginatempo}, \citenamefont
  {Bruno},\ and\ \citenamefont {Johnson}}]{SPG+00}%
  \BibitemOpen
  \bibfield  {author} {\bibinfo {author} {\bibfnamefont {J.~B.}\ \bibnamefont
  {Staunton}}, \bibinfo {author} {\bibfnamefont {J.}~\bibnamefont {Poulter}},
  \bibinfo {author} {\bibfnamefont {B.}~\bibnamefont {Ginatempo}}, \bibinfo
  {author} {\bibfnamefont {E.}~\bibnamefont {Bruno}}, \ and\ \bibinfo {author}
  {\bibfnamefont {D.~D.}\ \bibnamefont {Johnson}},\ }\href
  {http://dx.doi.org/10.1103/PhysRevB.62.1075} {\bibfield  {journal} {\bibinfo
  {journal} {Phys. Rev. B}\ }\textbf {\bibinfo {volume} {62}},\ \bibinfo
  {pages} {1075} (\bibinfo {year} {2000})}\BibitemShut {NoStop}%
\bibitem [{\citenamefont {Faulkner}\ and\ \citenamefont {Stocks}(1980)}]{FS80}%
  \BibitemOpen
  \bibfield  {author} {\bibinfo {author} {\bibfnamefont {J.~S.}\ \bibnamefont
  {Faulkner}}\ and\ \bibinfo {author} {\bibfnamefont {G.~M.}\ \bibnamefont
  {Stocks}},\ }\href {\doibase 10.1103/PhysRevB.21.3222} {\bibfield  {journal}
  {\bibinfo  {journal} {Phys. Rev. B}\ }\textbf {\bibinfo {volume} {21}},\
  \bibinfo {pages} {3222} (\bibinfo {year} {1980})}\BibitemShut {NoStop}%
\bibitem [{\citenamefont {Staunton}\ \emph {et~al.}(2004)\citenamefont
  {Staunton}, \citenamefont {Ostanin}, \citenamefont {Razee}, \citenamefont
  {Gyorffy}, \citenamefont {Szunyogh}, \citenamefont {Ginatempo},\ and\
  \citenamefont {Bruno}}]{SOR+04}%
  \BibitemOpen
  \bibfield  {author} {\bibinfo {author} {\bibfnamefont {J.~B.}\ \bibnamefont
  {Staunton}}, \bibinfo {author} {\bibfnamefont {S.}~\bibnamefont {Ostanin}},
  \bibinfo {author} {\bibfnamefont {S.~S.~A.}\ \bibnamefont {Razee}}, \bibinfo
  {author} {\bibfnamefont {B.~L.}\ \bibnamefont {Gyorffy}}, \bibinfo {author}
  {\bibfnamefont {L.}~\bibnamefont {Szunyogh}}, \bibinfo {author}
  {\bibfnamefont {B.}~\bibnamefont {Ginatempo}}, \ and\ \bibinfo {author}
  {\bibfnamefont {E.}~\bibnamefont {Bruno}},\ }\href {\doibase
  10.1103/PhysRevLett.93.257204} {\bibfield  {journal} {\bibinfo  {journal}
  {Phys. Rev. Lett.}\ }\textbf {\bibinfo {volume} {93}},\ \bibinfo {pages}
  {257204} (\bibinfo {year} {2004})}\BibitemShut {NoStop}%
\bibitem [{\citenamefont {Szilva}\ \emph {et~al.}(2022)\citenamefont {Szilva},
  \citenamefont {Kvashnin}, \citenamefont {Stepanov}, \citenamefont
  {Nordström}, \citenamefont {Eriksson}, \citenamefont {Lichtenstein},\ and\
  \citenamefont {Katsnelson}}]{SKS+22}%
  \BibitemOpen
  \bibfield  {author} {\bibinfo {author} {\bibfnamefont {A.}~\bibnamefont
  {Szilva}}, \bibinfo {author} {\bibfnamefont {Y.}~\bibnamefont {Kvashnin}},
  \bibinfo {author} {\bibfnamefont {E.~A.}\ \bibnamefont {Stepanov}}, \bibinfo
  {author} {\bibfnamefont {L.}~\bibnamefont {Nordström}}, \bibinfo {author}
  {\bibfnamefont {O.}~\bibnamefont {Eriksson}}, \bibinfo {author}
  {\bibfnamefont {A.~I.}\ \bibnamefont {Lichtenstein}}, \ and\ \bibinfo
  {author} {\bibfnamefont {M.~I.}\ \bibnamefont {Katsnelson}},\ }\href
  {\doibase 10.48550/ARXIV.2206.02415} {\bibfield  {journal} {\bibinfo
  {journal} {arXiv:2206.02415}\ } (\bibinfo {year} {2022}),\
  10.48550/ARXIV.2206.02415}\BibitemShut {NoStop}%
\bibitem [{\citenamefont {Uhl}\ \emph {et~al.}(1994)\citenamefont {Uhl},
  \citenamefont {Sandratskii},\ and\ \citenamefont {K\"ubler}}]{USK94}%
  \BibitemOpen
  \bibfield  {author} {\bibinfo {author} {\bibfnamefont {M.}~\bibnamefont
  {Uhl}}, \bibinfo {author} {\bibfnamefont {L.~M.}\ \bibnamefont
  {Sandratskii}}, \ and\ \bibinfo {author} {\bibfnamefont {J.}~\bibnamefont
  {K\"ubler}},\ }\href {\doibase 10.1103/PhysRevB.50.291} {\bibfield  {journal}
  {\bibinfo  {journal} {Phys. Rev. B}\ }\textbf {\bibinfo {volume} {50}},\
  \bibinfo {pages} {291} (\bibinfo {year} {1994})}\BibitemShut {NoStop}%
\bibitem [{\citenamefont {Halilov}\ \emph {et~al.}(1998)\citenamefont
  {Halilov}, \citenamefont {Eschrig}, \citenamefont {Perlov},\ and\
  \citenamefont {Oppeneer}}]{HEPO98}%
  \BibitemOpen
  \bibfield  {author} {\bibinfo {author} {\bibfnamefont {S.~V.}\ \bibnamefont
  {Halilov}}, \bibinfo {author} {\bibfnamefont {H.}~\bibnamefont {Eschrig}},
  \bibinfo {author} {\bibfnamefont {A.~Y.}\ \bibnamefont {Perlov}}, \ and\
  \bibinfo {author} {\bibfnamefont {P.~M.}\ \bibnamefont {Oppeneer}},\ }\href
  {\doibase 10.1103/PhysRevB.58.293} {\bibfield  {journal} {\bibinfo  {journal}
  {Phys. Rev. B}\ }\textbf {\bibinfo {volume} {58}},\ \bibinfo {pages} {293}
  (\bibinfo {year} {1998})}\BibitemShut {NoStop}%
\bibitem [{\citenamefont {Sandratskii}\ and\ \citenamefont
  {Bruno}(2002)}]{SB02}%
  \BibitemOpen
  \bibfield  {author} {\bibinfo {author} {\bibfnamefont {L.~M.}\ \bibnamefont
  {Sandratskii}}\ and\ \bibinfo {author} {\bibfnamefont {P.}~\bibnamefont
  {Bruno}},\ }\href {\doibase 10.1103/PhysRevB.66.134435} {\bibfield  {journal}
  {\bibinfo  {journal} {Phys. Rev. B}\ }\textbf {\bibinfo {volume} {66}},\
  \bibinfo {pages} {134435} (\bibinfo {year} {2002})}\BibitemShut {NoStop}%
\bibitem [{\citenamefont {Pajda}\ \emph {et~al.}(2000)\citenamefont {Pajda},
  \citenamefont {Kudrnovsk\'y}, \citenamefont {Turek}, \citenamefont {Drchal},\
  and\ \citenamefont {Bruno}}]{PKT+00}%
  \BibitemOpen
  \bibfield  {author} {\bibinfo {author} {\bibfnamefont {M.}~\bibnamefont
  {Pajda}}, \bibinfo {author} {\bibfnamefont {J.}~\bibnamefont {Kudrnovsk\'y}},
  \bibinfo {author} {\bibfnamefont {I.}~\bibnamefont {Turek}}, \bibinfo
  {author} {\bibfnamefont {V.}~\bibnamefont {Drchal}}, \ and\ \bibinfo {author}
  {\bibfnamefont {P.}~\bibnamefont {Bruno}},\ }\href {\doibase
  10.1103/PhysRevLett.85.5424} {\bibfield  {journal} {\bibinfo  {journal}
  {Phys. Rev. Lett.}\ }\textbf {\bibinfo {volume} {85}},\ \bibinfo {pages}
  {5424} (\bibinfo {year} {2000})}\BibitemShut {NoStop}%
\bibitem [{\citenamefont {Solovyev}(2021)}]{Sol21}%
  \BibitemOpen
  \bibfield  {author} {\bibinfo {author} {\bibfnamefont {I.~V.}\ \bibnamefont
  {Solovyev}},\ }\href {\doibase 10.1103/PhysRevB.103.104428} {\bibfield
  {journal} {\bibinfo  {journal} {Phys. Rev. B}\ }\textbf {\bibinfo {volume}
  {103}},\ \bibinfo {pages} {104428} (\bibinfo {year} {2021})}\BibitemShut
  {NoStop}%
\bibitem [{\citenamefont {Grotheer}\ \emph {et~al.}(2001)\citenamefont
  {Grotheer}, \citenamefont {Ederer},\ and\ \citenamefont {F\"ahnle}}]{GEF01}%
  \BibitemOpen
  \bibfield  {author} {\bibinfo {author} {\bibfnamefont {O.}~\bibnamefont
  {Grotheer}}, \bibinfo {author} {\bibfnamefont {C.}~\bibnamefont {Ederer}}, \
  and\ \bibinfo {author} {\bibfnamefont {M.}~\bibnamefont {F\"ahnle}},\ }\href
  {\doibase 10.1103/PhysRevB.63.100401} {\bibfield  {journal} {\bibinfo
  {journal} {Phys. Rev. B}\ }\textbf {\bibinfo {volume} {63}},\ \bibinfo
  {pages} {100401} (\bibinfo {year} {2001})}\BibitemShut {NoStop}%
\bibitem [{\citenamefont {Antropov}(2003)}]{Ant03}%
  \BibitemOpen
  \bibfield  {author} {\bibinfo {author} {\bibfnamefont {V.}~\bibnamefont
  {Antropov}},\ }\href {\doibase https://doi.org/10.1016/S0304-8853(03)00206-3}
  {\bibfield  {journal} {\bibinfo  {journal} {Journal of Magnetism and Magnetic
  Materials}\ }\textbf {\bibinfo {volume} {262}},\ \bibinfo {pages} {L192}
  (\bibinfo {year} {2003})}\BibitemShut {NoStop}%
\bibitem [{\citenamefont {Bruno}(2003)}]{Bru03}%
  \BibitemOpen
  \bibfield  {author} {\bibinfo {author} {\bibfnamefont {P.}~\bibnamefont
  {Bruno}},\ }\href {\doibase 10.1103/PhysRevLett.90.087205} {\bibfield
  {journal} {\bibinfo  {journal} {Phys. Rev. Lett.}\ }\textbf {\bibinfo
  {volume} {90}},\ \bibinfo {pages} {087205} (\bibinfo {year}
  {2003})}\BibitemShut {NoStop}%
\bibitem [{\citenamefont {Dederichs}\ \emph {et~al.}(1992)\citenamefont
  {Dederichs}, \citenamefont {Drittler},\ and\ \citenamefont {Zeller}}]{DDZ92}%
  \BibitemOpen
  \bibfield  {author} {\bibinfo {author} {\bibfnamefont {P.~H.}\ \bibnamefont
  {Dederichs}}, \bibinfo {author} {\bibfnamefont {B.}~\bibnamefont {Drittler}},
  \ and\ \bibinfo {author} {\bibfnamefont {R.}~\bibnamefont {Zeller}},\
  }\href@noop {} {\bibfield  {journal} {\bibinfo  {journal} {Mat. Res. Soc.
  Symp. Proc.}\ }\textbf {\bibinfo {volume} {253}},\ \bibinfo {pages} {185}
  (\bibinfo {year} {1992})}\BibitemShut {NoStop}%
\bibitem [{\citenamefont {Weinberger}(1990)}]{Wei90a}%
  \BibitemOpen
  \bibfield  {author} {\bibinfo {author} {\bibfnamefont {P.}~\bibnamefont
  {Weinberger}},\ }\href@noop {} { {\bibinfo {title} {Electron Scattering
  Theory for Ordered and Disordered Matter}}}\ (\bibinfo  {publisher} {Oxford
  University Press},\ \bibinfo {address} {Oxford},\ \bibinfo {year}
  {1990})\BibitemShut {NoStop}%
\bibitem [{\citenamefont {Mankovsky}\ \emph {et~al.}(2019)\citenamefont
  {Mankovsky}, \citenamefont {Polesya},\ and\ \citenamefont {Ebert}}]{MPE19}%
  \BibitemOpen
  \bibfield  {author} {\bibinfo {author} {\bibfnamefont {S.}~\bibnamefont
  {Mankovsky}}, \bibinfo {author} {\bibfnamefont {S.}~\bibnamefont {Polesya}},
  \ and\ \bibinfo {author} {\bibfnamefont {H.}~\bibnamefont {Ebert}},\ }\href
  {\doibase {10.1103/PhysRevB.99.104427}} {\bibfield  {journal} {\bibinfo
  {journal} {Phys. Rev. B}\ }\textbf {\bibinfo {volume} {{99}}} (\bibinfo
  {year} {{2019}}),\ {10.1103/PhysRevB.99.104427}}\BibitemShut {NoStop}%
\bibitem [{\citenamefont {Mankovsky}\ \emph {et~al.}(2020)\citenamefont
  {Mankovsky}, \citenamefont {Polesya},\ and\ \citenamefont {Ebert}}]{MPE20}%
  \BibitemOpen
  \bibfield  {author} {\bibinfo {author} {\bibfnamefont {S.}~\bibnamefont
  {Mankovsky}}, \bibinfo {author} {\bibfnamefont {S.}~\bibnamefont {Polesya}},
  \ and\ \bibinfo {author} {\bibfnamefont {H.}~\bibnamefont {Ebert}},\ }\href
  {\doibase 10.1103/PhysRevB.101.174401} {\bibfield  {journal} {\bibinfo
  {journal} {Phys. Rev. B}\ }\textbf {\bibinfo {volume} {101}},\ \bibinfo
  {pages} {174401} (\bibinfo {year} {2020})}\BibitemShut {NoStop}%
\bibitem [{\citenamefont {Ebert}\ \emph
  {et~al.}(2011{\natexlab{b}})\citenamefont {Ebert}, \citenamefont {Mankovsky},
  \citenamefont {K\"odderitzsch},\ and\ \citenamefont {Kelly}}]{EMKK11}%
  \BibitemOpen
  \bibfield  {author} {\bibinfo {author} {\bibfnamefont {H.}~\bibnamefont
  {Ebert}}, \bibinfo {author} {\bibfnamefont {S.}~\bibnamefont {Mankovsky}},
  \bibinfo {author} {\bibfnamefont {D.}~\bibnamefont {K\"odderitzsch}}, \ and\
  \bibinfo {author} {\bibfnamefont {P.~J.}\ \bibnamefont {Kelly}},\ }\href
  {\doibase 10.1103/PhysRevLett.107.066603} {\bibfield  {journal} {\bibinfo
  {journal} {Phys. Rev. Lett.}\ }\textbf {\bibinfo {volume} {107}},\ \bibinfo
  {pages} {066603} (\bibinfo {year} {2011}{\natexlab{b}})},\ \Eprint
  {http://arxiv.org/abs/http://arxiv.org/abs/1102.4551v1}
  {http://arxiv.org/abs/1102.4551v1} \BibitemShut {NoStop}%
\bibitem [{\citenamefont {Mankovsky}\ \emph {et~al.}(2013)\citenamefont
  {Mankovsky}, \citenamefont {K\"odderitzsch}, \citenamefont {Woltersdorf},\
  and\ \citenamefont {Ebert}}]{MKWE13}%
  \BibitemOpen
  \bibfield  {author} {\bibinfo {author} {\bibfnamefont {S.}~\bibnamefont
  {Mankovsky}}, \bibinfo {author} {\bibfnamefont {D.}~\bibnamefont
  {K\"odderitzsch}}, \bibinfo {author} {\bibfnamefont {G.}~\bibnamefont
  {Woltersdorf}}, \ and\ \bibinfo {author} {\bibfnamefont {H.}~\bibnamefont
  {Ebert}},\ }\href {\doibase 10.1103/PhysRevB.87.014430} {\bibfield  {journal}
  {\bibinfo  {journal} {Phys. Rev. B}\ }\textbf {\bibinfo {volume} {87}},\
  \bibinfo {pages} {014430} (\bibinfo {year} {2013})}\BibitemShut {NoStop}%
\bibitem [{\citenamefont {Ebert}\ \emph {et~al.}(2015)\citenamefont {Ebert},
  \citenamefont {Mankovsky}, \citenamefont {Chadova}, \citenamefont {Polesya},
  \citenamefont {Min\'{a}r},\ and\ \citenamefont {K\"odderitzsch}}]{EMC+15}%
  \BibitemOpen
  \bibfield  {author} {\bibinfo {author} {\bibfnamefont {H.}~\bibnamefont
  {Ebert}}, \bibinfo {author} {\bibfnamefont {S.}~\bibnamefont {Mankovsky}},
  \bibinfo {author} {\bibfnamefont {K.}~\bibnamefont {Chadova}}, \bibinfo
  {author} {\bibfnamefont {S.}~\bibnamefont {Polesya}}, \bibinfo {author}
  {\bibfnamefont {J.}~\bibnamefont {Min\'{a}r}}, \ and\ \bibinfo {author}
  {\bibfnamefont {D.}~\bibnamefont {K\"odderitzsch}},\ }\href {\doibase
  http://dx.doi.org/10.1103/PhysRevB.91.165132} {\bibfield  {journal} {\bibinfo
   {journal} {Phys. Rev. B}\ }\textbf {\bibinfo {volume} {91}},\ \bibinfo
  {pages} {165132} (\bibinfo {year} {2015})}\BibitemShut {NoStop}%
\bibitem [{\citenamefont {Papanikolaou}\ \emph {et~al.}(1997)\citenamefont
  {Papanikolaou}, \citenamefont {Zeller}, \citenamefont {Dederichs},\ and\
  \citenamefont {Stefanou}}]{PZDS97}%
  \BibitemOpen
  \bibfield  {author} {\bibinfo {author} {\bibfnamefont {N.}~\bibnamefont
  {Papanikolaou}}, \bibinfo {author} {\bibfnamefont {R.}~\bibnamefont
  {Zeller}}, \bibinfo {author} {\bibfnamefont {P.~H.}\ \bibnamefont
  {Dederichs}}, \ and\ \bibinfo {author} {\bibfnamefont {N.}~\bibnamefont
  {Stefanou}},\ }\href {\doibase 10.1103/PhysRevB.55.4157} {\bibfield
  {journal} {\bibinfo  {journal} {Phys. Rev. B}\ }\textbf {\bibinfo {volume}
  {55}},\ \bibinfo {pages} {4157} (\bibinfo {year} {1997})}\BibitemShut
  {NoStop}%
\bibitem [{\citenamefont {Lodder}(1976)}]{Lod76}%
  \BibitemOpen
  \bibfield  {author} {\bibinfo {author} {\bibfnamefont {A.}~\bibnamefont
  {Lodder}},\ }\href {\doibase 10.1088/0305-4608/6/10/018} {\bibfield
  {journal} {\bibinfo  {journal} {J. Phys. F: Met. Phys.}\ }\textbf {\bibinfo
  {volume} {6}},\ \bibinfo {pages} {1885} (\bibinfo {year} {1976})}\BibitemShut
  {NoStop}%
\bibitem [{\citenamefont {Butler}(1985)}]{But85}%
  \BibitemOpen
  \bibfield  {author} {\bibinfo {author} {\bibfnamefont {W.~H.}\ \bibnamefont
  {Butler}},\ }\href {\doibase 10.1103/PhysRevB.31.3260} {\bibfield  {journal}
  {\bibinfo  {journal} {Phys. Rev. B}\ }\textbf {\bibinfo {volume} {31}},\
  \bibinfo {pages} {3260} (\bibinfo {year} {1985})}\BibitemShut {NoStop}%
\bibitem [{\citenamefont {Harris}\ and\ \citenamefont {Owen}(1963)}]{HO63}%
  \BibitemOpen
  \bibfield  {author} {\bibinfo {author} {\bibfnamefont {E.~A.}\ \bibnamefont
  {Harris}}\ and\ \bibinfo {author} {\bibfnamefont {J.}~\bibnamefont {Owen}},\
  }\href {\doibase 10.1103/PhysRevLett.11.9} {\bibfield  {journal} {\bibinfo
  {journal} {Phys. Rev. Lett.}\ }\textbf {\bibinfo {volume} {11}},\ \bibinfo
  {pages} {9} (\bibinfo {year} {1963})}\BibitemShut {NoStop}%
\bibitem [{\citenamefont {Huang}\ and\ \citenamefont {Orbach}(1964)}]{HO64}%
  \BibitemOpen
  \bibfield  {author} {\bibinfo {author} {\bibfnamefont {N.~L.}\ \bibnamefont
  {Huang}}\ and\ \bibinfo {author} {\bibfnamefont {R.}~\bibnamefont {Orbach}},\
  }\href {\doibase 10.1103/PhysRevLett.12.275} {\bibfield  {journal} {\bibinfo
  {journal} {Phys. Rev. Lett.}\ }\textbf {\bibinfo {volume} {12}},\ \bibinfo
  {pages} {275} (\bibinfo {year} {1964})}\BibitemShut {NoStop}%
\bibitem [{\citenamefont {Allan}\ and\ \citenamefont {Betts}(1967)}]{AB67}%
  \BibitemOpen
  \bibfield  {author} {\bibinfo {author} {\bibfnamefont {G.~A.~T.}\
  \bibnamefont {Allan}}\ and\ \bibinfo {author} {\bibfnamefont {D.~D.}\
  \bibnamefont {Betts}},\ }\href {\doibase 10.1088/0370-1328/91/2/311}
  {\bibfield  {journal} {\bibinfo  {journal} {Proceedings of the Physical
  Society}\ }\textbf {\bibinfo {volume} {91}},\ \bibinfo {pages} {341}
  (\bibinfo {year} {1967})}\BibitemShut {NoStop}%
\bibitem [{\citenamefont {Iwashita}\ and\ \citenamefont
  {Ury\^a}(1974)}]{IU74}%
  \BibitemOpen
  \bibfield  {author} {\bibinfo {author} {\bibfnamefont {T.}~\bibnamefont
  {Iwashita}}\ and\ \bibinfo {author} {\bibfnamefont {N.}~\bibnamefont
  {Ury\^a}},\ }\href {\doibase 10.1143/JPSJ.36.48} {\bibfield  {journal}
  {\bibinfo  {journal} {Journal of the Physical Society of Japan}\ }\textbf
  {\bibinfo {volume} {36}},\ \bibinfo {pages} {48} (\bibinfo {year} {1974})},\
  \Eprint {http://arxiv.org/abs/https://doi.org/10.1143/JPSJ.36.48}
  {https://doi.org/10.1143/JPSJ.36.48} \BibitemShut {NoStop}%
\bibitem [{\citenamefont {Iwashita}\ and\ \citenamefont {Ury\^u}(1976)}]{IU76}%
  \BibitemOpen
  \bibfield  {author} {\bibinfo {author} {\bibfnamefont {T.}~\bibnamefont
  {Iwashita}}\ and\ \bibinfo {author} {\bibfnamefont {N.}~\bibnamefont
  {Ury\^u}},\ }\href {\doibase 10.1103/PhysRevB.14.3090} {\bibfield  {journal}
  {\bibinfo  {journal} {Phys. Rev. B}\ }\textbf {\bibinfo {volume} {14}},\
  \bibinfo {pages} {3090} (\bibinfo {year} {1976})}\BibitemShut {NoStop}%
\bibitem [{\citenamefont {Aksamit}(1980)}]{Aks80}%
  \BibitemOpen
  \bibfield  {author} {\bibinfo {author} {\bibfnamefont {J.}~\bibnamefont
  {Aksamit}},\ }\href {\doibase 10.1088/0022-3719/13/30/010} {\bibfield
  {journal} {\bibinfo  {journal} {Journal of Physics C: Solid State Physics}\
  }\textbf {\bibinfo {volume} {13}},\ \bibinfo {pages} {L871} (\bibinfo {year}
  {1980})}\BibitemShut {NoStop}%
\bibitem [{\citenamefont {Brown}(1984)}]{Bro84a}%
  \BibitemOpen
  \bibfield  {author} {\bibinfo {author} {\bibfnamefont {H.}~\bibnamefont
  {Brown}},\ }\href {\doibase https://doi.org/10.1016/0304-8853(84)90267-1}
  {\bibfield  {journal} {\bibinfo  {journal} {Journal of Magnetism and Magnetic
  Materials}\ }\textbf {\bibinfo {volume} {43}},\ \bibinfo {pages} {L1 }
  (\bibinfo {year} {1984})}\BibitemShut {NoStop}%
\bibitem [{\citenamefont {Ivanov}\ \emph {et~al.}(2014)\citenamefont {Ivanov},
  \citenamefont {Ummethum},\ and\ \citenamefont {Schnack}}]{IUS14}%
  \BibitemOpen
  \bibfield  {author} {\bibinfo {author} {\bibfnamefont {N.~B.}\ \bibnamefont
  {Ivanov}}, \bibinfo {author} {\bibfnamefont {J.}~\bibnamefont {Ummethum}}, \
  and\ \bibinfo {author} {\bibfnamefont {J.}~\bibnamefont {Schnack}},\ }\href
  {\doibase 10.1140/epjb/e2014-50423-7} {\bibfield  {journal} {\bibinfo
  {journal} {The European Physical Journal B}\ }\textbf {\bibinfo {volume}
  {87}},\ \bibinfo {pages} {226} (\bibinfo {year} {2014})}\BibitemShut
  {NoStop}%
\bibitem [{\citenamefont {Antal}\ \emph {et~al.}(2008)\citenamefont {Antal},
  \citenamefont {Lazarovits}, \citenamefont {Udvardi}, \citenamefont
  {Szunyogh}, \citenamefont {\'Ujfalussy},\ and\ \citenamefont
  {Weinberger}}]{ALU+08}%
  \BibitemOpen
  \bibfield  {author} {\bibinfo {author} {\bibfnamefont {A.}~\bibnamefont
  {Antal}}, \bibinfo {author} {\bibfnamefont {B.}~\bibnamefont {Lazarovits}},
  \bibinfo {author} {\bibfnamefont {L.}~\bibnamefont {Udvardi}}, \bibinfo
  {author} {\bibfnamefont {L.}~\bibnamefont {Szunyogh}}, \bibinfo {author}
  {\bibfnamefont {B.}~\bibnamefont {\'Ujfalussy}}, \ and\ \bibinfo {author}
  {\bibfnamefont {P.}~\bibnamefont {Weinberger}},\ }\href {\doibase
  10.1103/PhysRevB.77.174429} {\bibfield  {journal} {\bibinfo  {journal} {Phys.
  Rev. B}\ }\textbf {\bibinfo {volume} {77}},\ \bibinfo {pages} {174429}
  (\bibinfo {year} {2008})}\BibitemShut {NoStop}%
\bibitem [{\citenamefont {M\"uller-Hartmann}\ \emph {et~al.}(1997)\citenamefont
  {M\"uller-Hartmann}, \citenamefont {K\"obler},\ and\ \citenamefont
  {Smardz}}]{MKS97}%
  \BibitemOpen
  \bibfield  {author} {\bibinfo {author} {\bibfnamefont {E.}~\bibnamefont
  {M\"uller-Hartmann}}, \bibinfo {author} {\bibfnamefont {U.}~\bibnamefont
  {K\"obler}}, \ and\ \bibinfo {author} {\bibfnamefont {L.}~\bibnamefont
  {Smardz}},\ }\href {\doibase http://dx.doi.org/10.1016/S0304-8853(97)00166-2}
  {\bibfield  {journal} {\bibinfo  {journal} {Journal of Magnetism and Magnetic
  Materials}\ }\textbf {\bibinfo {volume} {173}},\ \bibinfo {pages} {133 }
  (\bibinfo {year} {1997})}\BibitemShut {NoStop}%
\bibitem [{\citenamefont {Greiter}\ and\ \citenamefont {Thomale}(2009)}]{GT09}%
  \BibitemOpen
  \bibfield  {author} {\bibinfo {author} {\bibfnamefont {M.}~\bibnamefont
  {Greiter}}\ and\ \bibinfo {author} {\bibfnamefont {R.}~\bibnamefont
  {Thomale}},\ }\href {\doibase 10.1103/PhysRevLett.102.207203} {\bibfield
  {journal} {\bibinfo  {journal} {Phys. Rev. Lett.}\ }\textbf {\bibinfo
  {volume} {102}},\ \bibinfo {pages} {207203} (\bibinfo {year}
  {2009})}\BibitemShut {NoStop}%
\bibitem [{\citenamefont {Greiter}\ \emph {et~al.}(2014)\citenamefont
  {Greiter}, \citenamefont {Schroeter},\ and\ \citenamefont {Thomale}}]{GST14}%
  \BibitemOpen
  \bibfield  {author} {\bibinfo {author} {\bibfnamefont {M.}~\bibnamefont
  {Greiter}}, \bibinfo {author} {\bibfnamefont {D.~F.}\ \bibnamefont
  {Schroeter}}, \ and\ \bibinfo {author} {\bibfnamefont {R.}~\bibnamefont
  {Thomale}},\ }\href {\doibase 10.1103/PhysRevB.89.165125} {\bibfield
  {journal} {\bibinfo  {journal} {Phys. Rev. B}\ }\textbf {\bibinfo {volume}
  {89}},\ \bibinfo {pages} {165125} (\bibinfo {year} {2014})}\BibitemShut
  {NoStop}%
\bibitem [{\citenamefont {Fedorova}\ \emph {et~al.}(2015)\citenamefont
  {Fedorova}, \citenamefont {Ederer}, \citenamefont {Spaldin},\ and\
  \citenamefont {Scaramucci}}]{FESS15}%
  \BibitemOpen
  \bibfield  {author} {\bibinfo {author} {\bibfnamefont {N.~S.}\ \bibnamefont
  {Fedorova}}, \bibinfo {author} {\bibfnamefont {C.}~\bibnamefont {Ederer}},
  \bibinfo {author} {\bibfnamefont {N.~A.}\ \bibnamefont {Spaldin}}, \ and\
  \bibinfo {author} {\bibfnamefont {A.}~\bibnamefont {Scaramucci}},\ }\href
  {\doibase 10.1103/PhysRevB.91.165122} {\bibfield  {journal} {\bibinfo
  {journal} {Phys. Rev. B}\ }\textbf {\bibinfo {volume} {91}},\ \bibinfo
  {pages} {165122} (\bibinfo {year} {2015})}\BibitemShut {NoStop}%
\bibitem [{\citenamefont {Mendive-Tapia}\ \emph {et~al.}(2021)\citenamefont
  {Mendive-Tapia}, \citenamefont {dos Santos~Dias}, \citenamefont {Grytsiuk},
  \citenamefont {Staunton}, \citenamefont {Bl\"ugel},\ and\ \citenamefont
  {Lounis}}]{MdSDG+21}%
  \BibitemOpen
  \bibfield  {author} {\bibinfo {author} {\bibfnamefont {E.}~\bibnamefont
  {Mendive-Tapia}}, \bibinfo {author} {\bibfnamefont {M.}~\bibnamefont {dos
  Santos~Dias}}, \bibinfo {author} {\bibfnamefont {S.}~\bibnamefont
  {Grytsiuk}}, \bibinfo {author} {\bibfnamefont {J.~B.}\ \bibnamefont
  {Staunton}}, \bibinfo {author} {\bibfnamefont {S.}~\bibnamefont {Bl\"ugel}},
  \ and\ \bibinfo {author} {\bibfnamefont {S.}~\bibnamefont {Lounis}},\ }\href
  {\doibase 10.1103/PhysRevB.103.024410} {\bibfield  {journal} {\bibinfo
  {journal} {Phys. Rev. B}\ }\textbf {\bibinfo {volume} {103}},\ \bibinfo
  {pages} {024410} (\bibinfo {year} {2021})}\BibitemShut {NoStop}%
\bibitem [{\citenamefont {Gutzeit}\ \emph {et~al.}(2021)\citenamefont
  {Gutzeit}, \citenamefont {Haldar}, \citenamefont {Meyer},\ and\ \citenamefont
  {Heinze}}]{GHMH21}%
  \BibitemOpen
  \bibfield  {author} {\bibinfo {author} {\bibfnamefont {M.}~\bibnamefont
  {Gutzeit}}, \bibinfo {author} {\bibfnamefont {S.}~\bibnamefont {Haldar}},
  \bibinfo {author} {\bibfnamefont {S.}~\bibnamefont {Meyer}}, \ and\ \bibinfo
  {author} {\bibfnamefont {S.}~\bibnamefont {Heinze}},\ }\href {\doibase
  10.1103/PhysRevB.104.024420} {\bibfield  {journal} {\bibinfo  {journal}
  {Phys. Rev. B}\ }\textbf {\bibinfo {volume} {104}},\ \bibinfo {pages}
  {024420} (\bibinfo {year} {2021})}\BibitemShut {NoStop}%
\bibitem [{\citenamefont {Hayami}(2022)}]{Hay22}%
  \BibitemOpen
  \bibfield  {author} {\bibinfo {author} {\bibfnamefont {S.}~\bibnamefont
  {Hayami}},\ }\href {\doibase 10.1103/PhysRevB.105.024413} {\bibfield
  {journal} {\bibinfo  {journal} {Phys. Rev. B}\ }\textbf {\bibinfo {volume}
  {105}},\ \bibinfo {pages} {024413} (\bibinfo {year} {2022})}\BibitemShut
  {NoStop}%
\bibitem [{\citenamefont {L\'aszl\'offy}\ \emph {et~al.}(2019)\citenamefont
  {L\'aszl\'offy}, \citenamefont {R\'ozsa}, \citenamefont {Palot\'as},
  \citenamefont {Udvardi},\ and\ \citenamefont {Szunyogh}}]{LRP+19a}%
  \BibitemOpen
  \bibfield  {author} {\bibinfo {author} {\bibfnamefont {A.}~\bibnamefont
  {L\'aszl\'offy}}, \bibinfo {author} {\bibfnamefont {L.}~\bibnamefont
  {R\'ozsa}}, \bibinfo {author} {\bibfnamefont {K.}~\bibnamefont {Palot\'as}},
  \bibinfo {author} {\bibfnamefont {L.}~\bibnamefont {Udvardi}}, \ and\
  \bibinfo {author} {\bibfnamefont {L.}~\bibnamefont {Szunyogh}},\ }\href
  {\doibase 10.1103/PhysRevB.99.184430} {\bibfield  {journal} {\bibinfo
  {journal} {Phys. Rev. B}\ }\textbf {\bibinfo {volume} {99}},\ \bibinfo
  {pages} {184430} (\bibinfo {year} {2019})}\BibitemShut {NoStop}%
\bibitem [{\citenamefont {Paul}\ \emph {et~al.}(2020)\citenamefont {Paul},
  \citenamefont {Haldar}, \citenamefont {von Malottki},\ and\ \citenamefont
  {Heinze}}]{PHMH20}%
  \BibitemOpen
  \bibfield  {author} {\bibinfo {author} {\bibfnamefont {S.}~\bibnamefont
  {Paul}}, \bibinfo {author} {\bibfnamefont {S.}~\bibnamefont {Haldar}},
  \bibinfo {author} {\bibfnamefont {S.}~\bibnamefont {von Malottki}}, \ and\
  \bibinfo {author} {\bibfnamefont {S.}~\bibnamefont {Heinze}},\ }\href
  {\doibase 10.1038/s41467-020-18473-x} {\bibfield  {journal} {\bibinfo
  {journal} {Nature Communications}\ }\textbf {\bibinfo {volume} {1}},\
  \bibinfo {pages} {475} (\bibinfo {year} {2020})}\BibitemShut {NoStop}%
\bibitem [{\citenamefont {Brinker}\ \emph {et~al.}(2019)\citenamefont
  {Brinker}, \citenamefont {dos Santos~Dias},\ and\ \citenamefont
  {Lounis}}]{BSL19}%
  \BibitemOpen
  \bibfield  {author} {\bibinfo {author} {\bibfnamefont {S.}~\bibnamefont
  {Brinker}}, \bibinfo {author} {\bibfnamefont {M.}~\bibnamefont {dos
  Santos~Dias}}, \ and\ \bibinfo {author} {\bibfnamefont {S.}~\bibnamefont
  {Lounis}},\ }\href {\doibase 10.1088/1367-2630/ab35c9} {\bibfield  {journal}
  {\bibinfo  {journal} {New Journal of Physics}\ }\textbf {\bibinfo {volume}
  {21}},\ \bibinfo {pages} {083015} (\bibinfo {year} {2019})}\BibitemShut
  {NoStop}%
\bibitem [{\citenamefont {Brinker}\ \emph {et~al.}(2020)\citenamefont
  {Brinker}, \citenamefont {dos Santos~Dias},\ and\ \citenamefont
  {Lounis}}]{BSL20}%
  \BibitemOpen
  \bibfield  {author} {\bibinfo {author} {\bibfnamefont {S.}~\bibnamefont
  {Brinker}}, \bibinfo {author} {\bibfnamefont {M.}~\bibnamefont {dos
  Santos~Dias}}, \ and\ \bibinfo {author} {\bibfnamefont {S.}~\bibnamefont
  {Lounis}},\ }\href {\doibase 10.1103/PhysRevResearch.2.033240} {\bibfield
  {journal} {\bibinfo  {journal} {Phys. Rev. Research}\ }\textbf {\bibinfo
  {volume} {2}},\ \bibinfo {pages} {033240} (\bibinfo {year}
  {2020})}\BibitemShut {NoStop}%
\bibitem [{\citenamefont {Mankovsky}\ and\ \citenamefont {Ebert}(2017)}]{ME17}%
  \BibitemOpen
  \bibfield  {author} {\bibinfo {author} {\bibfnamefont {S.}~\bibnamefont
  {Mankovsky}}\ and\ \bibinfo {author} {\bibfnamefont {H.}~\bibnamefont
  {Ebert}},\ }\href {\doibase 10.1103/PhysRevB.96.104416} {\bibfield  {journal}
  {\bibinfo  {journal} {Phys. Rev. B}\ }\textbf {\bibinfo {volume} {96}},\
  \bibinfo {pages} {104416} (\bibinfo {year} {2017})}\BibitemShut {NoStop}%
\bibitem [{\citenamefont {Solenov}\ \emph {et~al.}(2012)\citenamefont
  {Solenov}, \citenamefont {Mozyrsky},\ and\ \citenamefont {Martin}}]{SMM12}%
  \BibitemOpen
  \bibfield  {author} {\bibinfo {author} {\bibfnamefont {D.}~\bibnamefont
  {Solenov}}, \bibinfo {author} {\bibfnamefont {D.}~\bibnamefont {Mozyrsky}}, \
  and\ \bibinfo {author} {\bibfnamefont {I.}~\bibnamefont {Martin}},\ }\href
  {\doibase 10.1103/PhysRevLett.108.096403} {\bibfield  {journal} {\bibinfo
  {journal} {Phys. Rev. Lett.}\ }\textbf {\bibinfo {volume} {108}},\ \bibinfo
  {pages} {096403} (\bibinfo {year} {2012})}\BibitemShut {NoStop}%
\bibitem [{\citenamefont {Okubo}\ \emph {et~al.}(2012)\citenamefont {Okubo},
  \citenamefont {Chung},\ and\ \citenamefont {Kawamura}}]{OCK12}%
  \BibitemOpen
  \bibfield  {author} {\bibinfo {author} {\bibfnamefont {T.}~\bibnamefont
  {Okubo}}, \bibinfo {author} {\bibfnamefont {S.}~\bibnamefont {Chung}}, \ and\
  \bibinfo {author} {\bibfnamefont {H.}~\bibnamefont {Kawamura}},\ }\href
  {\doibase 10.1103/PhysRevLett.108.017206} {\bibfield  {journal} {\bibinfo
  {journal} {Phys. Rev. Lett.}\ }\textbf {\bibinfo {volume} {108}},\ \bibinfo
  {pages} {017206} (\bibinfo {year} {2012})}\BibitemShut {NoStop}%
\bibitem [{\citenamefont {Batista}\ \emph {et~al.}(2016)\citenamefont
  {Batista}, \citenamefont {Lin}, \citenamefont {Hayami},\ and\ \citenamefont
  {Kamiya}}]{BLHK16}%
  \BibitemOpen
  \bibfield  {author} {\bibinfo {author} {\bibfnamefont {C.~D.}\ \bibnamefont
  {Batista}}, \bibinfo {author} {\bibfnamefont {S.-Z.}\ \bibnamefont {Lin}},
  \bibinfo {author} {\bibfnamefont {S.}~\bibnamefont {Hayami}}, \ and\ \bibinfo
  {author} {\bibfnamefont {Y.}~\bibnamefont {Kamiya}},\ }\href {\doibase
  10.1088/0034-4885/79/8/084504} {\bibfield  {journal} {\bibinfo  {journal}
  {Reports on Progress in Physics}\ }\textbf {\bibinfo {volume} {79}},\
  \bibinfo {pages} {084504} (\bibinfo {year} {2016})}\BibitemShut {NoStop}%
\bibitem [{\citenamefont {Mankovsky}\ \emph {et~al.}(2021)\citenamefont
  {Mankovsky}, \citenamefont {Polesya},\ and\ \citenamefont {Ebert}}]{MPE21}%
  \BibitemOpen
  \bibfield  {author} {\bibinfo {author} {\bibfnamefont {S.}~\bibnamefont
  {Mankovsky}}, \bibinfo {author} {\bibfnamefont {S.}~\bibnamefont {Polesya}},
  \ and\ \bibinfo {author} {\bibfnamefont {H.}~\bibnamefont {Ebert}},\ }\href
  {\doibase 10.1103/PhysRevB.104.054418} {\bibfield  {journal} {\bibinfo
  {journal} {Phys. Rev. B}\ }\textbf {\bibinfo {volume} {104}},\ \bibinfo
  {pages} {054418} (\bibinfo {year} {2021})}\BibitemShut {NoStop}%
\bibitem [{\citenamefont {Grytsiuk}\ \emph {et~al.}(2020)\citenamefont
  {Grytsiuk}, \citenamefont {Hanke}, \citenamefont {Hoffmann}, \citenamefont
  {Bouaziz}, \citenamefont {Gomonay}, \citenamefont {Bihlmayer}, \citenamefont
  {Lounis}, \citenamefont {Mokrousov},\ and\ \citenamefont
  {Bl\"ugel}}]{GHH+20}%
  \BibitemOpen
  \bibfield  {author} {\bibinfo {author} {\bibfnamefont {S.}~\bibnamefont
  {Grytsiuk}}, \bibinfo {author} {\bibfnamefont {J.-P.}\ \bibnamefont {Hanke}},
  \bibinfo {author} {\bibfnamefont {M.}~\bibnamefont {Hoffmann}}, \bibinfo
  {author} {\bibfnamefont {J.}~\bibnamefont {Bouaziz}}, \bibinfo {author}
  {\bibfnamefont {O.}~\bibnamefont {Gomonay}}, \bibinfo {author} {\bibfnamefont
  {G.}~\bibnamefont {Bihlmayer}}, \bibinfo {author} {\bibfnamefont
  {S.}~\bibnamefont {Lounis}}, \bibinfo {author} {\bibfnamefont
  {Y.}~\bibnamefont {Mokrousov}}, \ and\ \bibinfo {author} {\bibfnamefont
  {S.}~\bibnamefont {Bl\"ugel}},\ }\href {\doibase
  https://doi.org/10.1038/s41467-019-14030-3} {\bibfield  {journal} {\bibinfo
  {journal} {Nature Communications}\ }\textbf {\bibinfo {volume} {11}},\
  \bibinfo {pages} {511} (\bibinfo {year} {2020})}\BibitemShut {NoStop}%
\bibitem [{\citenamefont {dos Santos~Dias}\ \emph {et~al.}(2021)\citenamefont
  {dos Santos~Dias}, \citenamefont {Brinker}, \citenamefont {L\'aszl\'offy},
  \citenamefont {Ny\'ari}, \citenamefont {Bl\"ugel}, \citenamefont {Szunyogh},\
  and\ \citenamefont {Lounis}}]{SBL+21a}%
  \BibitemOpen
  \bibfield  {author} {\bibinfo {author} {\bibfnamefont {M.}~\bibnamefont {dos
  Santos~Dias}}, \bibinfo {author} {\bibfnamefont {S.}~\bibnamefont {Brinker}},
  \bibinfo {author} {\bibfnamefont {A.}~\bibnamefont {L\'aszl\'offy}}, \bibinfo
  {author} {\bibfnamefont {B.}~\bibnamefont {Ny\'ari}}, \bibinfo {author}
  {\bibfnamefont {S.}~\bibnamefont {Bl\"ugel}}, \bibinfo {author}
  {\bibfnamefont {L.}~\bibnamefont {Szunyogh}}, \ and\ \bibinfo {author}
  {\bibfnamefont {S.}~\bibnamefont {Lounis}},\ }\href {\doibase
  10.1103/PhysRevB.103.L140408} {\bibfield  {journal} {\bibinfo  {journal}
  {Phys. Rev. B}\ }\textbf {\bibinfo {volume} {103}},\ \bibinfo {pages}
  {L140408} (\bibinfo {year} {2021})}\BibitemShut {NoStop}%
\bibitem [{\citenamefont {Lounis}(2020)}]{Lou20}%
  \BibitemOpen
  \bibfield  {author} {\bibinfo {author} {\bibfnamefont {S.}~\bibnamefont
  {Lounis}},\ }\href {\doibase 10.1088/1367-2630/abb514} {\bibfield  {journal}
  {\bibinfo  {journal} {New Journal of Physics}\ }\textbf {\bibinfo {volume}
  {22}},\ \bibinfo {pages} {103003} (\bibinfo {year} {2020})}\BibitemShut
  {NoStop}%
\bibitem [{\citenamefont {Streib}\ \emph {et~al.}(2021)\citenamefont {Streib},
  \citenamefont {Szilva}, \citenamefont {Borisov}, \citenamefont {Pereiro},
  \citenamefont {Bergman}, \citenamefont {Sj\"oqvist}, \citenamefont {Delin},
  \citenamefont {Katsnelson}, \citenamefont {Eriksson},\ and\ \citenamefont
  {Thonig}}]{SSB+21}%
  \BibitemOpen
  \bibfield  {author} {\bibinfo {author} {\bibfnamefont {S.}~\bibnamefont
  {Streib}}, \bibinfo {author} {\bibfnamefont {A.}~\bibnamefont {Szilva}},
  \bibinfo {author} {\bibfnamefont {V.}~\bibnamefont {Borisov}}, \bibinfo
  {author} {\bibfnamefont {M.}~\bibnamefont {Pereiro}}, \bibinfo {author}
  {\bibfnamefont {A.}~\bibnamefont {Bergman}}, \bibinfo {author} {\bibfnamefont
  {E.}~\bibnamefont {Sj\"oqvist}}, \bibinfo {author} {\bibfnamefont
  {A.}~\bibnamefont {Delin}}, \bibinfo {author} {\bibfnamefont {M.~I.}\
  \bibnamefont {Katsnelson}}, \bibinfo {author} {\bibfnamefont
  {O.}~\bibnamefont {Eriksson}}, \ and\ \bibinfo {author} {\bibfnamefont
  {D.}~\bibnamefont {Thonig}},\ }\href {\doibase 10.1103/PhysRevB.103.224413}
  {\bibfield  {journal} {\bibinfo  {journal} {Phys. Rev. B}\ }\textbf {\bibinfo
  {volume} {103}},\ \bibinfo {pages} {224413} (\bibinfo {year}
  {2021})}\BibitemShut {NoStop}%
\bibitem [{\citenamefont {Streib}\ \emph {et~al.}(2022)\citenamefont {Streib},
  \citenamefont {Cardias}, \citenamefont {Pereiro}, \citenamefont {Bergman},
  \citenamefont {Sjöqvist}, \citenamefont {Barreteau}, \citenamefont {Delin},
  \citenamefont {Eriksson},\ and\ \citenamefont {Thonig}}]{SCP+22}%
  \BibitemOpen
  \bibfield  {author} {\bibinfo {author} {\bibfnamefont {S.}~\bibnamefont
  {Streib}}, \bibinfo {author} {\bibfnamefont {R.}~\bibnamefont {Cardias}},
  \bibinfo {author} {\bibfnamefont {M.}~\bibnamefont {Pereiro}}, \bibinfo
  {author} {\bibfnamefont {A.}~\bibnamefont {Bergman}}, \bibinfo {author}
  {\bibfnamefont {E.}~\bibnamefont {Sjöqvist}}, \bibinfo {author}
  {\bibfnamefont {C.}~\bibnamefont {Barreteau}}, \bibinfo {author}
  {\bibfnamefont {A.}~\bibnamefont {Delin}}, \bibinfo {author} {\bibfnamefont
  {O.}~\bibnamefont {Eriksson}}, \ and\ \bibinfo {author} {\bibfnamefont
  {D.}~\bibnamefont {Thonig}},\ }\href {\doibase 10.48550/ARXIV.2203.11759}
  {\enquote {\bibinfo {title} {Adiabatic spin dynamics and effective exchange
  interactions from constrained tight-binding electronic structure theory:
  beyond the heisenberg regime},}\ } (\bibinfo {year} {2022})\BibitemShut
  {NoStop}%
\bibitem [{\citenamefont {Kambersky}(1970)}]{Kam70}%
  \BibitemOpen
  \bibfield  {author} {\bibinfo {author} {\bibfnamefont {V.}~\bibnamefont
  {Kambersky}},\ }\href
  {http://www.nrcresearchpress.com/doi/abs/10.1139/p70-361} {\bibfield
  {journal} {\bibinfo  {journal} {Can. J. Phys.}\ }\textbf {\bibinfo {volume}
  {48}},\ \bibinfo {pages} {2906} (\bibinfo {year} {1970})}\BibitemShut
  {NoStop}%
\bibitem [{\citenamefont {F\"ahnle}\ and\ \citenamefont
  {Steiauf}(2006)}]{FS06}%
  \BibitemOpen
  \bibfield  {author} {\bibinfo {author} {\bibfnamefont {M.}~\bibnamefont
  {F\"ahnle}}\ and\ \bibinfo {author} {\bibfnamefont {D.}~\bibnamefont
  {Steiauf}},\ }\href@noop {} {\bibfield  {journal} {\bibinfo  {journal} {Phys.
  Rev. B}\ }\textbf {\bibinfo {volume} {73}},\ \bibinfo {pages} {184427}
  (\bibinfo {year} {2006})}\BibitemShut {NoStop}%
\bibitem [{\citenamefont {Kambersky}(1976)}]{Kam76}%
  \BibitemOpen
  \bibfield  {author} {\bibinfo {author} {\bibfnamefont {V.}~\bibnamefont
  {Kambersky}},\ }\href {http://dx.doi.org/10.1007/BF01587621} {\bibfield
  {journal} {\bibinfo  {journal} {Czech. J. Phys.}\ }\textbf {\bibinfo {volume}
  {26}},\ \bibinfo {pages} {1366} (\bibinfo {year} {1976})}\BibitemShut
  {NoStop}%
\bibitem [{\citenamefont {Gilmore}\ \emph {et~al.}(2007)\citenamefont
  {Gilmore}, \citenamefont {Idzerda},\ and\ \citenamefont {Stiles}}]{GIS07}%
  \BibitemOpen
  \bibfield  {author} {\bibinfo {author} {\bibfnamefont {K.}~\bibnamefont
  {Gilmore}}, \bibinfo {author} {\bibfnamefont {Y.~U.}\ \bibnamefont
  {Idzerda}}, \ and\ \bibinfo {author} {\bibfnamefont {M.~D.}\ \bibnamefont
  {Stiles}},\ }\href {\doibase 10.1103/PhysRevLett.99.027204} {\bibfield
  {journal} {\bibinfo  {journal} {Phys. Rev. Lett.}\ }\textbf {\bibinfo
  {volume} {99}},\ \bibinfo {pages} {027204} (\bibinfo {year}
  {2007})}\BibitemShut {NoStop}%
\bibitem [{\citenamefont {Brataas}\ \emph {et~al.}(2008)\citenamefont
  {Brataas}, \citenamefont {Tserkovnyak},\ and\ \citenamefont {Bauer}}]{BTB08}%
  \BibitemOpen
  \bibfield  {author} {\bibinfo {author} {\bibfnamefont {A.}~\bibnamefont
  {Brataas}}, \bibinfo {author} {\bibfnamefont {Y.}~\bibnamefont
  {Tserkovnyak}}, \ and\ \bibinfo {author} {\bibfnamefont {G.~E.~W.}\
  \bibnamefont {Bauer}},\ }\href {\doibase 10.1103/PhysRevLett.101.037207}
  {\bibfield  {journal} {\bibinfo  {journal} {Phys. Rev. Lett.}\ }\textbf
  {\bibinfo {volume} {101}},\ \bibinfo {pages} {037207} (\bibinfo {year}
  {2008})}\BibitemShut {NoStop}%
\bibitem [{\citenamefont {Starikov}\ \emph {et~al.}(2010)\citenamefont
  {Starikov}, \citenamefont {Kelly}, \citenamefont {Brataas}, \citenamefont
  {Tserkovnyak},\ and\ \citenamefont {Bauer}}]{SKB+10}%
  \BibitemOpen
  \bibfield  {author} {\bibinfo {author} {\bibfnamefont {A.~A.}\ \bibnamefont
  {Starikov}}, \bibinfo {author} {\bibfnamefont {P.~J.}\ \bibnamefont {Kelly}},
  \bibinfo {author} {\bibfnamefont {A.}~\bibnamefont {Brataas}}, \bibinfo
  {author} {\bibfnamefont {Y.}~\bibnamefont {Tserkovnyak}}, \ and\ \bibinfo
  {author} {\bibfnamefont {G.~E.~W.}\ \bibnamefont {Bauer}},\ }\href {\doibase
  10.1103/PhysRevLett.105.236601} {\bibfield  {journal} {\bibinfo  {journal}
  {Phys. Rev. Lett.}\ }\textbf {\bibinfo {volume} {105}},\ \bibinfo {pages}
  {236601} (\bibinfo {year} {2010})}\BibitemShut {NoStop}%
\bibitem [{\citenamefont {Carva}\ \emph {et~al.}(2013)\citenamefont {Carva},
  \citenamefont {Battiato}, \citenamefont {Legut},\ and\ \citenamefont
  {Oppeneer}}]{CBLO13}%
  \BibitemOpen
  \bibfield  {author} {\bibinfo {author} {\bibfnamefont {K.}~\bibnamefont
  {Carva}}, \bibinfo {author} {\bibfnamefont {M.}~\bibnamefont {Battiato}},
  \bibinfo {author} {\bibfnamefont {D.}~\bibnamefont {Legut}}, \ and\ \bibinfo
  {author} {\bibfnamefont {P.~M.}\ \bibnamefont {Oppeneer}},\ }\href {\doibase
  10.1103/PhysRevB.87.184425} {\bibfield  {journal} {\bibinfo  {journal} {Phys.
  Rev. B}\ }\textbf {\bibinfo {volume} {87}},\ \bibinfo {pages} {184425}
  (\bibinfo {year} {2013})}\BibitemShut {NoStop}%
\bibitem [{\citenamefont {Carva}\ \emph {et~al.}(2011)\citenamefont {Carva},
  \citenamefont {Battiato},\ and\ \citenamefont {Oppeneer}}]{CBO11}%
  \BibitemOpen
  \bibfield  {author} {\bibinfo {author} {\bibfnamefont {K.}~\bibnamefont
  {Carva}}, \bibinfo {author} {\bibfnamefont {M.}~\bibnamefont {Battiato}}, \
  and\ \bibinfo {author} {\bibfnamefont {P.~M.}\ \bibnamefont {Oppeneer}},\
  }\href {\doibase 10.1103/PhysRevLett.107.207201} {\bibfield  {journal}
  {\bibinfo  {journal} {Phys. Rev. Lett.}\ }\textbf {\bibinfo {volume} {107}},\
  \bibinfo {pages} {207201} (\bibinfo {year} {2011})}\BibitemShut {NoStop}%
\bibitem [{\citenamefont {Thonig}\ \emph {et~al.}(2018)\citenamefont {Thonig},
  \citenamefont {Kvashnin}, \citenamefont {Eriksson},\ and\ \citenamefont
  {Pereiro}}]{TKEP18}%
  \BibitemOpen
  \bibfield  {author} {\bibinfo {author} {\bibfnamefont {D.}~\bibnamefont
  {Thonig}}, \bibinfo {author} {\bibfnamefont {Y.}~\bibnamefont {Kvashnin}},
  \bibinfo {author} {\bibfnamefont {O.}~\bibnamefont {Eriksson}}, \ and\
  \bibinfo {author} {\bibfnamefont {M.}~\bibnamefont {Pereiro}},\ }\href
  {\doibase 10.1103/PhysRevMaterials.2.013801} {\bibfield  {journal} {\bibinfo
  {journal} {Phys. Rev. Materials}\ }\textbf {\bibinfo {volume} {2}},\ \bibinfo
  {pages} {013801} (\bibinfo {year} {2018})}\BibitemShut {NoStop}%
\bibitem [{\citenamefont {Mankovsky}\ \emph {et~al.}(2018)\citenamefont
  {Mankovsky}, \citenamefont {Wimmer},\ and\ \citenamefont {Ebert}}]{MWE18}%
  \BibitemOpen
  \bibfield  {author} {\bibinfo {author} {\bibfnamefont {S.}~\bibnamefont
  {Mankovsky}}, \bibinfo {author} {\bibfnamefont {S.}~\bibnamefont {Wimmer}}, \
  and\ \bibinfo {author} {\bibfnamefont {H.}~\bibnamefont {Ebert}},\ }\href
  {\doibase 10.1103/PhysRevB.98.104406} {\bibfield  {journal} {\bibinfo
  {journal} {Phys. Rev. B}\ }\textbf {\bibinfo {volume} {98}},\ \bibinfo
  {pages} {104406} (\bibinfo {year} {2018})}\BibitemShut {NoStop}%
\bibitem [{\citenamefont {Qian}\ and\ \citenamefont {Vignale}(2002)}]{QV02}%
  \BibitemOpen
  \bibfield  {author} {\bibinfo {author} {\bibfnamefont {Z.}~\bibnamefont
  {Qian}}\ and\ \bibinfo {author} {\bibfnamefont {G.}~\bibnamefont {Vignale}},\
  }\href {\doibase 10.1103/PhysRevLett.88.056404} {\bibfield  {journal}
  {\bibinfo  {journal} {Phys. Rev. Lett.}\ }\textbf {\bibinfo {volume} {88}},\
  \bibinfo {pages} {056404} (\bibinfo {year} {2002})}\BibitemShut {NoStop}%
\bibitem [{\citenamefont {Hankiewicz}\ \emph {et~al.}(2007)\citenamefont
  {Hankiewicz}, \citenamefont {Vignale},\ and\ \citenamefont
  {Tserkovnyak}}]{HVT07}%
  \BibitemOpen
  \bibfield  {author} {\bibinfo {author} {\bibfnamefont {E.~M.}\ \bibnamefont
  {Hankiewicz}}, \bibinfo {author} {\bibfnamefont {G.}~\bibnamefont {Vignale}},
  \ and\ \bibinfo {author} {\bibfnamefont {Y.}~\bibnamefont {Tserkovnyak}},\
  }\href {\doibase 10.1103/PhysRevB.75.174434} {\bibfield  {journal} {\bibinfo
  {journal} {Phys. Rev. B}\ }\textbf {\bibinfo {volume} {75}},\ \bibinfo
  {pages} {174434} (\bibinfo {year} {2007})}\BibitemShut {NoStop}%
\bibitem [{\citenamefont {Brinker}\ \emph {et~al.}(2022)\citenamefont
  {Brinker}, \citenamefont {dos Santos~Dias},\ and\ \citenamefont
  {Lounis}}]{BSL22}%
  \BibitemOpen
  \bibfield  {author} {\bibinfo {author} {\bibfnamefont {S.}~\bibnamefont
  {Brinker}}, \bibinfo {author} {\bibfnamefont {M.}~\bibnamefont {dos
  Santos~Dias}}, \ and\ \bibinfo {author} {\bibfnamefont {S.}~\bibnamefont
  {Lounis}},\ }\href {\doibase 10.1088/1361-648x/ac699d} {\bibfield  {journal}
  {\bibinfo  {journal} {Journal of Physics: Condensed Matter}\ }\textbf
  {\bibinfo {volume} {34}},\ \bibinfo {pages} {285802} (\bibinfo {year}
  {2022})}\BibitemShut {NoStop}%
\bibitem [{\citenamefont {Aßmann}\ and\ \citenamefont {Nowak}(2019)}]{AN19}%
  \BibitemOpen
  \bibfield  {author} {\bibinfo {author} {\bibfnamefont {M.}~\bibnamefont
  {Aßmann}}\ and\ \bibinfo {author} {\bibfnamefont {U.}~\bibnamefont
  {Nowak}},\ }\href {\doibase https://doi.org/10.1016/j.jmmm.2018.08.034}
  {\bibfield  {journal} {\bibinfo  {journal} {Journal of Magnetism and Magnetic
  Materials}\ }\textbf {\bibinfo {volume} {469}},\ \bibinfo {pages} {217}
  (\bibinfo {year} {2019})}\BibitemShut {NoStop}%
\bibitem [{\citenamefont {Hellsvik}\ \emph {et~al.}(2019)\citenamefont
  {Hellsvik}, \citenamefont {Thonig}, \citenamefont {Modin}, \citenamefont
  {Iu\ifmmode~\mbox{\c{s}}\else \c{s}\fi{}an}, \citenamefont {Bergman},
  \citenamefont {Eriksson}, \citenamefont {Bergqvist},\ and\ \citenamefont
  {Delin}}]{HTM+19}%
  \BibitemOpen
  \bibfield  {author} {\bibinfo {author} {\bibfnamefont {J.}~\bibnamefont
  {Hellsvik}}, \bibinfo {author} {\bibfnamefont {D.}~\bibnamefont {Thonig}},
  \bibinfo {author} {\bibfnamefont {K.}~\bibnamefont {Modin}}, \bibinfo
  {author} {\bibfnamefont {D.}~\bibnamefont {Iu\ifmmode~\mbox{\c{s}}\else
  \c{s}\fi{}an}}, \bibinfo {author} {\bibfnamefont {A.}~\bibnamefont
  {Bergman}}, \bibinfo {author} {\bibfnamefont {O.}~\bibnamefont {Eriksson}},
  \bibinfo {author} {\bibfnamefont {L.}~\bibnamefont {Bergqvist}}, \ and\
  \bibinfo {author} {\bibfnamefont {A.}~\bibnamefont {Delin}},\ }\href
  {\doibase 10.1103/PhysRevB.99.104302} {\bibfield  {journal} {\bibinfo
  {journal} {Phys. Rev. B}\ }\textbf {\bibinfo {volume} {99}},\ \bibinfo
  {pages} {104302} (\bibinfo {year} {2019})}\BibitemShut {NoStop}%
\bibitem [{\citenamefont {Sadhukhan}\ \emph {et~al.}(2022)\citenamefont
  {Sadhukhan}, \citenamefont {Bergman}, \citenamefont {Kvashnin}, \citenamefont
  {Hellsvik},\ and\ \citenamefont {Delin}}]{SBK+22}%
  \BibitemOpen
  \bibfield  {author} {\bibinfo {author} {\bibfnamefont {B.}~\bibnamefont
  {Sadhukhan}}, \bibinfo {author} {\bibfnamefont {A.}~\bibnamefont {Bergman}},
  \bibinfo {author} {\bibfnamefont {Y.~O.}\ \bibnamefont {Kvashnin}}, \bibinfo
  {author} {\bibfnamefont {J.}~\bibnamefont {Hellsvik}}, \ and\ \bibinfo
  {author} {\bibfnamefont {A.}~\bibnamefont {Delin}},\ }\href {\doibase
  10.1103/PhysRevB.105.104418} {\bibfield  {journal} {\bibinfo  {journal}
  {Phys. Rev. B}\ }\textbf {\bibinfo {volume} {105}},\ \bibinfo {pages}
  {104418} (\bibinfo {year} {2022})}\BibitemShut {NoStop}%
\bibitem [{\citenamefont {Mankovsky}\ \emph {et~al.}(2022)\citenamefont
  {Mankovsky}, \citenamefont {Polesya}, \citenamefont {Lange}, \citenamefont
  {Wei{\ss}enhofer}, \citenamefont {Nowak},\ and\ \citenamefont
  {Ebert}}]{MPL+22}%
  \BibitemOpen
  \bibfield  {author} {\bibinfo {author} {\bibfnamefont {S.}~\bibnamefont
  {Mankovsky}}, \bibinfo {author} {\bibfnamefont {S.}~\bibnamefont {Polesya}},
  \bibinfo {author} {\bibfnamefont {H.}~\bibnamefont {Lange}}, \bibinfo
  {author} {\bibfnamefont {M.}~\bibnamefont {Wei{\ss}enhofer}}, \bibinfo
  {author} {\bibfnamefont {U.}~\bibnamefont {Nowak}}, \ and\ \bibinfo {author}
  {\bibfnamefont {H.}~\bibnamefont {Ebert}},\ }\href {\doibase
  10.48550/ARXIV.2203.16144} {\bibfield  {journal} {\bibinfo  {journal}
  {arXiv:2203.16144v1}\ } (\bibinfo {year} {2022}),\
  10.48550/ARXIV.2203.16144}\BibitemShut {NoStop}%
\end{thebibliography}

%

\end{document}